\documentclass[11pt,a4paper]{article}
\usepackage[english]{babel}
\usepackage[utf8]{inputenc}
\usepackage{epsf}
\usepackage{cite}
\usepackage{graphicx}
\usepackage{subcaption} 

\usepackage{verbatim} 

\usepackage{amsmath}
\usepackage{amssymb}
\usepackage{mathrsfs}

\usepackage{mathbbol}

\usepackage{amsfonts}
\usepackage{bbding}

\usepackage[
colorlinks=true
,urlcolor=black
,anchorcolor=black
,citecolor=black
,filecolor=black
,linkcolor=black
,menucolor=black
,linktocpage=true
,pdfproducer=medialab
,pdfa=true
]{hyperref}

\usepackage[noabbrev]{cleveref}
\usepackage{color} 

\usepackage[left=2cm,right=2cm,top=2cm,bottom=2cm]{geometry}
\usepackage{tikz}
\usetikzlibrary{shapes,arrows,positioning,automata,backgrounds,calc,er,patterns}
\usepackage[compat=1.1.0]{tikz-feynman}

\vspace*{1cm}
\allowdisplaybreaks
\begin{document}

\begin{center}
\vspace*{1mm}
\mathversion{bold}
{\Large\bf
Prospects for a flavour violating $Z^\prime$ explanation of $\Delta a_{\mu,e}$
}
\mathversion{normal}
\vspace*{1.2cm}

{\bf J.~Kriewald, J. Orloff, E.~Pinsard and A.~M.~Teixeira}

\vspace*{.5cm}

Laboratoire de Physique de Clermont (UMR 6533), CNRS/IN2P3,\\
Univ. Clermont Auvergne, 4 Av. Blaise Pascal, 63178 Aubi\`ere Cedex,
France

\end{center}

\vspace*{5mm}
\begin{abstract}
\noindent
The apparent tensions emerging from the comparison of experimental data of the anomalous magnetic moments of the muon and electron to the Standard Model predictions ($\Delta a_{\mu,e}$) could be interpreted as a potential signal of  New Physics. 
Models encompassing a light vector boson have been known to offer a satisfactory explanation to $\Delta a_{\mu}$, albeit subject to stringent experimental constraints. Here we explore a minimal extension of the Standard Model via a leptophilic vector boson $Z^\prime$, under the hypothesis of strictly flavour-violating couplings of the latter to leptons. The most constraining observables to this ad-hoc construction emerge from lepton flavour universality violation (in $Z$ and $\tau$ decays) and from rare charged lepton flavour violating transitions. Once these are accommodated, one can saturate the tensions in $\Delta a_{\mu}$, but $\Delta a_{e}$ is predicted to be Standard Model-like. 
We infer prospects for several observables, including leptonic $Z$ decays and several charged lepton flavour violating processes.
We also discuss potential signatures of the considered $Z^\prime$ at a future muon collider, emphasising the role of the $\mu^+\mu^- \to\tau^+\tau^- $ forward-backward asymmetry as a key probe of the model.  
\end{abstract}

\newpage
\section{Introduction}
In recent years, numerous tensions between the Standard Model 
(SM) expectations and observation have emerged, many (if not most) in relation with high-intensity flavour observables. In addition to the various ``anomalous'' behaviours associated with $B$-meson decays, the anomalous magnetic moment of charged leptons (both muons and electrons, albeit the latter to a smaller degree) have been the object of intensive dedicated studies. The  anomalous magnetic moment of a charged lepton $\ell$, defined as 
\begin{equation}\label{eq:aell:def}
    a_\ell \, =\, \frac{1}{2}\left(g_\ell-2\right)\,,
\end{equation}
allows probing numerous aspects of the SM, and is also instrumental in determining some of its fundamental quantities. 

Following the disclosed results from the ``g-2'' E989 experiment at FNAL~\cite{Muong-2:2021ojo}, which are in good agreement with the previous findings of the BNL E821 experiment~\cite{Muong-2:2006rrc}, the current experimental average for the muon anomalous magnetic moment~\cite{Muong-2:2021ojo} is given by
\begin{equation}\label{eq:amu:exp}
    a_\mu^\text{exp}\, =\, 116\, 592\, 061\, (41) \times 10^{-11}\,,
\end{equation}
which should be compared to its SM expectation. Prior to the most recent lattice QCD based computation of the hadronic vacuum polarisation contribution by the BMW collaboration~\cite{Borsanyi:2020mff}, the SM prediction - as
compiled by the ``Muon $g-2$ Theory Inititative''~\cite{Aoyama:2020ynm} (see also~\cite{Aoyama:2012wk,Aoyama:2019ryr,Czarnecki:2002nt,Gnendiger:2013pva,Davier:2017zfy,Keshavarzi:2018mgv,Colangelo:2018mtw,Hoferichter:2019mqg,Davier:2019can,Keshavarzi:2019abf,Kurz:2014wya,Melnikov:2003xd,Masjuan:2017tvw,Colangelo:2017fiz,Hoferichter:2018kwz,Gerardin:2019vio,Bijnens:2019ghy,Colangelo:2019uex,Blum:2019ugy,Colangelo:2014qya} - was found to be
\begin{equation}\label{eq:amu:SMwhite}
    a_\mu^\text{SM}\, =\, 116\, 591\, 810 \, (43) \times 10^{-11}\,.
\end{equation}
When compared, the experimental average and the latter SM prediction lead to the following $4.2\,\sigma$ tension between  
theory and observation\footnote{In what follows we will rely on  $\Delta a_\mu$ as obtained from the SM value as given in Eq.~(\ref{eq:amu:SMwhite}); the value obtained taking into account the BMW collaboration computation ($a_\mu^\text{SM}=116\,591\,954\, (57)\times 10^{-11}$) would suggest $\Delta a_\mu= 107\, (70)\times 10^{-11}$, corresponding to a $1.5\,\sigma$ tension between  
theory and observation.}
\begin{equation}\label{eq:amu:delta}
    \Delta a_\mu\, \equiv a_\mu^\text{SM}\,- \, 
    a_\mu^\text{exp}\,
    \, =\, 251 \, (59) \times 10^{-11}\,.
\end{equation}
Under the assumption of a significant tension between theory and observation, as given by Eq.~(\ref{eq:amu:delta}), the need for new physics (NP) capable of accounting for such a sizeable discrepancy is manifest; 
several minimal, as well as more complete NP models, have been thoroughly explored in the light of recent experimental results (for a review see, for example,~\cite{Athron:2021iuf} and references therein).  

The magnetic moment of the electron has also been at the origin of possible new tensions, upon comparison of the experimental value
\begin{equation}\label{eq:ae:exp}
    a_e^\text{exp}\, =\, 1 \,159\,652\,180.73\,(28) \times 10^{-12}\,,
\end{equation}
to the SM prediction (depending on the value of $\alpha_e$ that is used for the computation of the latter): using $\alpha_e$ as extracted from  measurements using Cs atoms~\cite{Parker:2018vye,Yu:2019gdh},  one is led to
\begin{equation}\label{eq:ae:deltaCs}
   \Delta a_e^\text{Cs}\, \equiv a_e^\text{SM,Cs}\,- \, 
    a_e^\text{exp}\,    \, =\, -0.88 \, (0.36)\times 10^{-12}\,,
    \end{equation}
corresponding to a $-2.5\, \sigma$ deviation; in~\cite{Morel:2020dww}, a more recent estimation of $\alpha_e$ was obtained, this time relying on Rubidium atoms, and the new determination of $\alpha_e$ (implying an overall deviation above the $5\, \sigma$ level for $\alpha_e$) now suggests milder discrepancies between observation and theory prediction,  $   \Delta a_e^\text{Rb} \, =\, 0.48 \, (0.30)\times 10^{-12},$
corresponding to $\mathcal{O}(1.7\, \sigma)$ deviation. 

Other than (possibly) signalling deviations from the SM expectation, 
it is interesting to notice the potential impact of {\it both}
$\Delta a_e^{\mathrm{Cs}}$ and $\Delta a_\mu$: other than having an opposite sign, the ratio $\Delta a_\mu/\Delta a_e$ does not exhibit the 
na\"ive scaling $\sim m^2_\mu/ m^2_e$ (expected  from the magnetic dipole operator, in which a mass insertion of the SM lepton is responsible for the required chirality flip~\cite{Giudice:2012ms}). This behaviour renders a common explanation of both tensions quite challenging, calling upon a departure from a minimal flavour violation (MFV) hypothesis, or from single new particle extensions of the SM (coupling to charged leptons~\cite{Davoudiasl:2018fbb,Kahn:2016vjr,Crivellin:2018qmi}). Notice that the pattern in both $\Delta a_e^{\mathrm{Cs}}$ and $\Delta a_\mu$ can be also suggestive of a violation of lepton flavour universality (LFU).

Several new physics constructions have  been put forward to simultaneously explain the tensions in $a_e$ and $a_\mu$ (see, for example,~\cite{Davoudiasl:2018fbb,Kahn:2016vjr,Crivellin:2018qmi,Liu:2018xkx,Dutta:2018fge,Han:2018znu,Endo:2019bcj,Kawamura:2019rth,Abdullah:2019ofw,Badziak:2019gaf,CarcamoHernandez:2019ydc,Hiller:2019mou,Cornella:2019uxs,CarcamoHernandez:2020pxw,Haba:2020gkr,Bigaran:2020jil,Jana:2020pxx,Calibbi:2020emz,Chen:2020jvl,Yang:2020bmh,Dorsner:2020aaz,Hati:2020fzp,DelleRose:2020oaa, Hernandez:2021tii, Bodas:2021fsy, Han:2021gfu, Hernandez:2021iss, Borah:2021khc, Hue:2021xzl}). Among the most minimal models, extensions of the SM gauge group via additional $U(1)$ groups have been intensively explored, as these offer several appealing features. The new (vector) boson (as well as other potentially present state) and new associated neutral currents can open the door to 
extensive implications both for particle and astroparticle physics, across vast energy scales~\cite{Langacker:2008yv}. In addition to their potential in what concerns well-motivated dark matter mediators see, for instance~\cite{Altmannshofer:2016jzy,Correia:2016xcs,Correia:2019pnn,Correia:2019woz}, $Z^\prime$ extensions of the SM have been considered in the context of flavour physics, in particular in what concerns $B$-meson decay  observables (especially $b\to s\ell\ell$ transitions)~\cite{Altmannshofer:2014cfa,Crivellin:2015mga,Crivellin:2015lwa,Sierra:2015fma,Crivellin:2015era,Celis:2015ara,Altmannshofer:2016jzy,Bhatia:2017tgo,Kamenik:2017tnu,Chen:2017usq,Camargo-Molina:2018cwu,Darme:2018hqg,Baek:2018aru,Biswas:2019twf,Allanach:2019iiy,Crivellin:2020oup,Crivellin:2022obd}.

The most minimal SM extensions leading to a (light) $Z^\prime$ boson have mostly been oriented towards constructions featuring flavour-conserving couplings of the new vector boson to matter (quarks and leptons). Models in which the  $Z^\prime$ couples to hadrons have a severely constrained parameter space due to conflict with numerous bounds from meson decays and oscillations, as well as from atomic parity violation experiments~\cite{Raggi:2015noa, Anastasi:2015qla, Androic:2018kni}. Likewise, and despite their potential to address the $B$-meson decay ``anomalies'' (and/or tensions in $a_\mu$), lepton flavour conserving (although not necessarily lepton flavour universal) leptophilic $Z^\prime$ extensions have also been strongly constrained, as a consequence of the extensive implications for (lepton) flavour observables, high- and low-energy neutrino observables, LHC phenomenology, as well as dark matter direct detection experiments~\cite{Baek:2001kca,Ma:2001md,Altmannshofer:2014cfa,Amaral:2021rzw,Heeck:2011wj,Foldenauer:2018zrz,Gninenko:2018tlp,Holst:2021lzm,Huang:2021nkl,Banerjee:2020zvi,Biswas:2019twf,Kamada:2018zxi,Araki:2017wyg,Araki:2014ona,Patra:2016shz,Harigaya:2013twa,Frank:2021nkq,Bhattacharyya:2018evo,Arcadi:2017jqd,Arcadi:2013qia,Dudas:2009uq,Chun:2010ve,Frandsen:2011cg,Alves:2013tqa}.

\bigskip
A NP construction relying on the addition of a new vector boson usually calls for the extension of the SM gauge group by at least an additional $U(1)$. 
Likewise further states must be added, for instance to explain the origin of a massive vector or to provide a rationale for the peculiar flavour structure.
However, as a first step to evaluate the phenomenological viability of SM extensions via leptophilic flavour violating 
(light) $Z^\prime$ bosons, one can rely in a minimal, ad-hoc approach. 
In this work, we consider a toy-model construction in which only new 
interactions with the leptons are added to the SM Lagrangian, allowing for left-handed and right-handed couplings ($g_{L,R}^{\alpha \beta} \bar \ell_\alpha  \ell_\beta Z^\prime$). No assumption is made on the underlying gauge group, nor on the actual mechanism leading to its breaking. 
Motivated by an explanation to $\Delta a_{e,\mu}$, we only hypothesise the existence of a massive (albeit light) vector boson, and investigate the constraints on its 
$g_{L,R}^{\alpha \beta}$ couplings: in addition to addressing whether or not it can simultaneously saturate the tensions for both the electron and the muon anomalous magnetic moments, we consider numerous constraints arising from electroweak precision observables (EWPO), rare (lepton) flavour violating transitions and decays, as well as LFU tests in $Z$ and tau-lepton decays, providing in most cases a full computation (beyond renormalisation group (RG) approximation) of the relevant quantities (see also~\cite{Foot:1994vd,Heeck:2016xkh,Altmannshofer:2016brv,Buras:2021btx} and~\cite{Crivellin:2013hpa}, albeit in the context of effective theory studies). 
As we proceed to discuss, the latter bounds lead to very stringent 
constraints on the $Z^\prime$ parameter space: perturbativity of the couplings allows inferring an upper limit for its mass ($m_{Z^\prime}$), and the flavour constraints on the couplings suggest that 
$\Delta a_{e}$ (and $\Delta a_{\tau}$) should be SM-like. Interestingly, several charged lepton flavour violating (cLFV) observables could be within future experimental sensitivity.

Our analysis is further complemented by investigating the prospects of such a leptophilic $Z^\prime$ in what concerns a future muon collider. Muon colliders have received increasing attention in recent years~\cite{EuropeanStrategyforParticlePhysicsPreparatoryGroup:2019qin,Delahaye:2019omf,Long:2020wfp,MuonCollider:2022xlm} and offer promising testing grounds for beyond the SM (BSM) constructions with preferred couplings to leptons. 
In line with recent studies (see, e.g.~\cite{Huang:2021nkl}), in  this work 
we also discuss the $t$-channel $Z^\prime$ contribution to 
di-tau pair production, $\mu^+ \mu^- \to \tau^+ \tau^- $, comparing the deviations with respect to the SM expectation regarding the production cross section, and the associated forward-backward asymmetry ($A_{\text{FB}}$). As discussed here, our findings suggest that a muon collider could indeed offer good testing (and discovery) grounds for a leptophilic $Z^\prime$ capable of explaining the tension in $\Delta a_\mu$, or conversely being capable of falsifying the proposed explanation for the tension in the anomalous magnetic moment of the muon.

We emphasise that in this work we have focused on  studying the low-energy effects 
induced by the presence of a new neutral vector boson exhibiting distinctive couplings exclusively to the lepton sector.
This is but a first step towards the model-building of ultraviolet (UV)-complete SM extensions 
whose particle content would feature such a neutral vector boson, in addition to other new states (scalars, fermions, ...).  

The manuscript is organised as follows:
in Section~\ref{sec:model}, we describe the minimal (ad-hoc) construction we will explore, subsequently describing the most stringent constraints on flavour violating 
$Z^\prime$ couplings to leptons in Section~\ref{sec:cLFV-LFU},  stemming from complying with relevant bounds - from both cLFV transitions and from observables sensitive to lepton flavour universality violation (LFUV). In Section~\ref{sec:amuae-impact}
we address the possibility of 
saturating $\Delta a_{e,\mu}$ while complying with the considered 
bounds. 
In Section~\ref{sec:cLFVprospects} we discuss the impact that the latter constraints might have on the prospects for the maximal predictions of a number of charged lepton flavour violating observables. This is followed by a discussion of the prospects of such a state at a future muon collider (Section~\ref{sec:ZpMuon}). We then present our summarising conclusions and a brief overview.

\mathversion{bold}
\section{Leptophilic cLFV $Z^\prime$: a minimal model}\label{sec:model}
\mathversion{normal}
As mentioned in the Introduction, we consider here a minimal extension of the SM via a single neutral vector boson $Z^\prime$ (without specifying the underlying gauge group). 
Generic $Z^\prime$ extensions of the SM, especially those in which the new mediator is lighter than the electroweak (EW) scale, $\Lambda_\text{EW}$,  
are subject to stringent constraints, both from direct searches and from electroweak precision observables.

Beyond the SM constructions in which the $Z^\prime$  couples to hadrons are constrained from light meson decays, $\pi^0 \to \gamma Z^\prime(Z^\prime\to ee)$ and $K^+ \to \pi^+ Z^\prime(Z^\prime\to ee)$, as searched for 
at the NA48/2~\cite{Raggi:2015noa} experiment,
as well as by searches for 
$\phi^+ \to \eta^+ Z^\prime(Z^\prime\to
ee)$ at KLOE-2~\cite{Anastasi:2015qla}. Likewise, rare meson decays (e.g. $B_{(s)}\to \ell^+\ell^-$), neutral meson oscillations ($K^0-\bar K^0$ and $B_{(s)}-\bar B_{(s)}$ mixing), and atomic parity violation~\cite{Androic:2018kni}, all play an important role in constraining 
hadrophilic $Z^\prime$ extensions of the SM.
Most of these constraints can be avoided by considering specific 
extensions in which the $Z^\prime$ only couples to the lepton sector (neutrinos and charged leptons)\footnote{Notice that higher order processes - typically at the one-loop level -, arising from gauge kinetic mixing can still lead to new contributions to the ``hadronic" observables. Furthermore, as we proceed to discuss, we will only consider flavour-violating couplings to leptons and thus gauge-kinetic mixing between the photon and the $Z^\prime$ only arises at the two-loop level. Kinetic mixing ($\epsilon_\text{kin}$) is therefore strongly suppressed, of the order $\epsilon_\text{kin}\simeq e g^{e\mu}g^{e\tau}g^{\mu\tau}/(256\pi^4)\log(\mu^2/m_{Z'}^2)$. In what follows, these effects will not be taken into account.}. Nevertheless, flavour conserving 
$Z^\prime$ couplings to leptons, i.e. $Z^\prime \ell_\alpha \ell_\alpha$, also lead to NP contributions that are potentially in conflict with numerous observables: notice that the non-observation of a $Z^\prime$ at electron beam dump experiments (SLAC E141, Orsay, NA64~\cite{NA64:2018lsq, NA64:2019auh}), in dark photon production searches (KLOE-2 experiment~\cite{Anastasi:2015qla}, BaBar~\cite{BaBar:2014zli}) or at parity-violation experiments (SLAC E158~\cite{SLACE158:2005uay}), set severe bounds on $Z^\prime ee$ couplings. The other diagonal couplings (second and third generation, as well as to left-handed neutrinos) are also constrained by $\bar\nu_e$-electron and $\nu_\mu$-electron scattering, in particular from the data of the TEXONO~\cite{Deniz:2009mu} and CHARM-II~\cite{Vilain:1993kd} experiments, respectively.
Diagonal couplings to muons are further severely constrained from neutrino trident production ($\nu_\mu N\to \nu_\mu N \mu^+\mu^-$)~\cite{CCFR:1991lpl}.

In order to evade these severe bounds one can envisage NP models in which the new $Z^\prime$ has no flavour conserving couplings to leptons; in our analysis, we thus 
consider a strictly leptophilic flavour violating (light) $Z^\prime$, leading to the presence of the following new terms in the interaction Lagrangian:
\begin{eqnarray}\label{eq:lagrangian0}
  \mathcal{L}_{Z^\prime}^\text{int} &=&  \sum_{\alpha ,\beta} Z^\prime_\mu  \, \bar{L }_\alpha \gamma^\mu \left(g_{X}^{\alpha \beta} \, P_X  \right)L_\beta  + \text{H.c.} \,.
\end{eqnarray}
In the above, the indices run over $\alpha ,\beta = e, \mu, \tau $, with $\alpha \neq \beta$; $P_X=P_{L,R}$ are the chiral projectors and $g_X^{\alpha \beta} = g_{L,R}^{\alpha \beta}$ are the new couplings hermitian matrices.
Notice that throughout the study, and for simplicity, we will always consider real couplings.
In the ``toy-model" here considered, we do not extend the lepton sector to account for massive neutrinos; only left-handed (LH) neutrinos are present and, due to $SU(2)_L$ gauge invariance, they couple to the $Z^\prime$ through the same LH charged lepton couplings, leading to  
\begin{eqnarray}\label{eq:lagrangian}
  \mathcal{L}_{Z^\prime}^\text{int} &=&  \sum_{\alpha ,\beta} Z^\prime_\mu \left[ \bar{\ell }_\alpha \gamma^\mu \left(g_{X}^{\alpha \beta} \, P_X  \right)\ell_\beta + \bar{\nu}_\alpha \gamma^\mu \left(g_{L}^{\alpha \beta} \, P_L \right)\nu_\beta \right] + \text{H.c.} \,.
\end{eqnarray}
Despite the simplicity of this BSM realisation, it is clear from the structure of the Lagrangian in Eq.~(\ref{eq:lagrangian}) that the leptonic couplings of the new neutral gauge boson can potentially lead to extensive contributions to several leptonic and EW precision observables - including the desired sizeable contributions to the charged lepton anomalous magnetic moments.

Before proceeding to discuss this in the following sections, 
a few comments are in order concerning formal aspects related to possible UV completions for this ad-hoc phenomenological construction, especially regarding anomaly cancellations.
Without attempting a full discussion, let us briefly consider the new gauge anomalies emerging in this context, possibly in association with the presence of a new $U(1)^\prime$. First, the simplest ones are the $Z^\prime-W^{1,2}-W^{1,2}$ and $Z^\prime-B-B$ one-loop triangle diagrams involving $SU(2)_L$ and $U(1)_Y$ gauge fields. The vanishing of these anomalies respectively requires $\mathrm{Tr}(g_L)=0$ and $\mathrm{Tr}(g_R)=0$. Our peculiar hypothesis of purely off-diagonal leptonic $Z^\prime$ couplings provides the simplest solution, which in turn also partly justifies this assumption. Concerning the vanishing of the  $Z^\prime-Z^\prime-Z^\prime$ triangle anomaly, this requires $2\mathrm{Tr}(g_L^3)=\mathrm{Tr}(g_R^3)$. 
This can be achieved by setting one of the three off-diagonal couplings of each chirality to zero, as this trivially ensures the equality of the traces (identical to zero). This choice can also further prevent the appearance of flavour-conserving $Z^\prime$ vertices at the one-loop level. Finally, the $Z^\prime-Z^\prime-W^3$ triangle is proportional to  $\mathrm{Tr}(g_L^2)$ (a positive definite quantity), and also to the vanishing sum of the lepton and neutrino isospins. The only difficulty then lies in association with the triangle $Z^\prime-Z^\prime-B$ anomaly (or equivalently $Z^\prime-Z^\prime-\gamma$), whose vanishing would require $\mathrm{Tr}(g_L^2)=\mathrm{Tr}(g_R^2)$, in conflict
with the results of the subsequent phenomenological analysis\footnote{This will be extensively discussed in Section~\ref{sec:amuae-impact}; nevertheless, we quickly draw the attention to, for example, Eqs.~(\ref{eq:summ:gLRmutau}, \ref{eq:summ:gLRmutau:cLFV}), which encode some key-features of the $\Delta a_\mu$ preferred regimes for the cLFV $Z^\prime$ couplings.}.
New fields should thus be added: on top of a new scalar responsible for symmetry breaking and for giving a mass to the $Z^\prime$ boson, additional (heavy) charged fermions will be needed to properly cancel the above gauge anomaly,  and to provide a realistic origin of the (strictly) flavour violating couplings. 

\mathversion{bold}
\section{Constraining flavour violating $Z^\prime$ couplings to leptons}\label{sec:cLFV-LFU}
\mathversion{normal}
In addition to the stringent constraints that have been obtained for the lepton flavour-conserving (LFC) couplings of a new 
$Z^\prime$ boson~\cite{NA64:2018lsq, NA64:2019auh,Anastasi:2015qla,BaBar:2014zli,SLACE158:2005uay,Deniz:2009mu,Vilain:1993kd,CCFR:1991lpl}, there is an extensive  array  of bounds on its flavour-violating couplings  
to leptons, as a consequence of the associated experimental limits. 
In particular, the bounds on the $g_{L,R}^{\alpha \beta}$  couplings  (with $\alpha \neq  \beta$) stem  from both lepton flavour violating transitions and decays, as well as processes sensitive to the breaking of LFU (the  latter being an indirect consequence of the former in the present case).
Thus, and before addressing the ultimate requirement that should be fulfilled by this class of simplified leptophilic $Z^\prime$ extensions (i.e. the tensions in $(g-2)_\ell$), in this section we address the constraints on 
$g_{L,R}^{\alpha \beta}$ arising from cLFV and LFUV limits, including $Z$ 
decays, as well as several leptonic processes.

\mathversion{bold}
\subsection{Constraints from $Z$ decays}
\mathversion{normal}
The presence of the new $Z^\prime$ can lead to new (higher order) contributions to both lepton flavour conserving and lepton flavour violating (LFV)  $Z$ decays\footnote{From a formal point of view, in both cases of  lepton flavour conserving and lepton flavour violating $Z$ decays, one has to consider loop diagrams which are UV-divergent and must thus be renormalised. We have carried out the self-energy renormalisations (for diagonal and off-diagonal contributions) following~\cite{Aoki:1982ed, Espriu:2002xv, Denner:1991kt}. The remaining divergences are then taken into account through the vertex counterterms in the $\overline{\text{MS}}$ scheme.}. 

Concerning flavour conserving $Z$ decays, the new $Z^\prime$-mediated loops (and the non-negligible interferences between the SM-like processes and the NP ones) will lead to  
new contributions to the individual decay widths and, as expected, to modifications of the effective $Z\ell \ell$ couplings. 
In order to assess the impact of the current experimental measurements, 
we consider the following ratios of decay widths, which further have the advantage of allowing the cancellation of quantum electrodynamics (QED) corrections in the theoretical predictions. Moreover, these ratios are also probes of the lepton flavour universality of $Z$-boson couplings. 
We thus consider the ratios
\begin{equation}
    R^Z_{\alpha\beta} = \dfrac{\Gamma(Z \to \ell_\alpha^+\ell_\alpha^-)}{\Gamma(Z \to \ell_\beta^+\ell_\beta^-)}\,, \quad \text{with } \alpha \neq \beta \, =\, e, \, \mu, \, \tau\,.
\end{equation}
The SM prediction for these ratios (at 2-loop accuracy) are~\cite{Freitas:2014hra}
\begin{eqnarray}
  \dfrac{\Gamma(Z \to \mu^+\mu^-)^{\text{SM}}}{\Gamma(Z \to e^+e^-)^{\text{SM}}} &=& 1 \,, \nonumber \\
    \dfrac{\Gamma(Z \to \tau^+\tau^-)^{\text{SM}}}{\Gamma(Z \to \mu^+\mu^-)^{\text{SM}}} &=& 0.9977\,,\nonumber\\
     \dfrac{\Gamma(Z \to \tau^+\tau^-)^{\text{SM}}}{\Gamma(Z \to e^+e^-)^{\text{SM}}} &=& 0.9977\,,
\end{eqnarray}
with negligible associated uncertainties.
These should be compared with the corresponding experimental values~\cite{ParticleDataGroup:2020ssz},
\begin{eqnarray}
    \dfrac{\Gamma(Z \to \mu^+\mu^-)^{\text{exp}}}{\Gamma(Z \to e^+e^-)^{\text{exp}}} &=& 1.0001 \pm 0.0024 \,, \nonumber \\
    \dfrac{\Gamma(Z \to \tau^+\tau^-)^{\text{exp}}}{\Gamma(Z \to \mu^+\mu^-)^{\text{exp}}} &=& 1.0010 \pm 0.0026\,,\nonumber\\
     \dfrac{\Gamma(Z \to \tau^+\tau^-)^{\text{exp}}}{\Gamma(Z \to e^+e^-)^{\text{exp}}} &=& 1.0020 \pm 0.0032\,.
\label{eq::ZdecayLFU}
\end{eqnarray} 
As can be seen, the experimental measurements are in good agreement with their respective SM predictions, thus placing strong bounds on any NP contribution.
In what concerns the NP contributions, we estimate the modified individual partial widths as
\begin{equation}
    \Gamma (Z\to \ell^+\ell^-) \simeq \Gamma^{\text{SM}_\text{full}} + \Gamma^{\text{SM}_\text{tree}-Z^\prime} + \Gamma^{Z^\prime}\,,
\end{equation}
where $\Gamma^{\text{SM}_\text{full}}$ is given in~\cite{Freitas:2014hra} at 2-loop accuracy, $\Gamma^{\text{SM}_\text{tree}-Z^\prime}$ is the interference term between the SM tree-level contribution and the $Z^\prime$-mediated 1-loop diagrams and $\Gamma^{Z^\prime}$ purely consists of the $Z^\prime$-mediated 1-loop diagrams.
In particular, the interference term between the SM tree-level diagrams and the $Z^\prime$ contributions is at the source of important corrections to the $Z$-boson leptonic partial widths, which in turn leads to stringent constraints on the $Z^\prime$ couplings to SM leptons.
The amplitude of the $Z^\prime$ contributions is given by
\begin{eqnarray}
  \mathcal{M}^{Z^\prime}_{Z \to \ell_\alpha^+\ell_\alpha^-} &=& u_\alpha \left[ \mathcal{C}_X^V\,  \gamma^\mu \, P_X + \mathcal{C}_X^T \, \sigma^{\mu \nu}\, q_\nu \, P_X \right] v_\alpha \, \epsilon_\mu^*(q) \, , 
\end{eqnarray} 
where $u_\alpha, \, v_\alpha$ denote the $\ell_\alpha$ spinors, $\epsilon_\mu^*(q)$ the $Z$-boson polarisation vector, $q$ is the $Z$ momentum and $\mathcal{C}_X^{V, T}$ are the vector and tensor coefficients computed from the decay ``triangle" diagram (i.e. one-loop vertex correction), which are given in Appendix~\ref{app:lfcZ}. (These calculations were done relying on \texttt{Package-X}~\cite{Patel:2015tea}.)

In view of the structure of the couplings of the $Z^\prime$ to leptons - cf. Eq.~(\ref{eq:lagrangian})  -, even in the absence of a specific mechanism of neutrino mass generation, new contributions to the so-called invisible $Z$ decay width are expected: these are fuelled by the non-vanishing $Z^\prime \bar \nu_\alpha \nu_\beta$ couplings ($\propto g_L^{\alpha\beta}$). 
However, due to the observed smallness of $m_{\nu_i}$ (or zero, in the strict SM limit of massless neutrinos) the $Z^\prime$-mediated 1-loop contributions turn out to be always negligible. Moreover, notice that 
the $Z^\prime$ can also mediate loop decays leading to cLFV $Z$-decays\footnote{We also estimated effects of the $Z^\prime$ on lepton flavour universality violation and lepton flavour violation in Higgs decays; due to the smallness of the lepton Yukawa couplings these effects are however negligible and hence we neglect them in our analysis.}, $Z \to \ell_\alpha^\pm \ell_\beta^\mp$. The relevant expressions of the associated cLFV $Z$ decay amplitude are collected in Appendix~\ref{app:LFV_Zdecays}.
The current limits from ATLAS on $Z \to e^\pm\mu^\mp $~\cite{ATLAS:2014vur}, OPAL on $Z \to e^\pm\tau^\mp $ and  on $Z \to \mu^\pm\tau^\mp $~\cite{OPAL:1995grn} (see Table~\ref{tab:cLFVdata}) are not sufficiently strong to constrain the different $Z^\prime$ couplings; in fact, it turns out that  
the NP contributions to the decay rate are very small: at most one has $\text{BR}(Z \to \ell_\alpha^\pm \ell_\beta^\mp)\sim \mathcal{O}(10^{-14})$ (throughout the model's parameter space which will be numerically explored in subsequent sections).

\renewcommand{\arraystretch}{1.3}
\begin{table}[h!]
    \centering
    \hspace*{-7mm}{\small\begin{tabular}{|c|c|c|}
    \hline
    Observable & Current bound & Future sensitivity  \\
    \hline\hline
    $\text{BR}(\mu\to e \gamma)$    &
    \quad $<4.2\times 10^{-13}$ \quad (MEG~\cite{TheMEG:2016wtm})   &
    \quad $6\times 10^{-14}$ \quad (MEG II~\cite{Baldini:2018nnn}) \\
    $\text{BR}(\tau \to e \gamma)$  &
    \quad $<3.3\times 10^{-8}$ \quad (BaBar~\cite{Aubert:2009ag})    &
    \quad $3\times10^{-9}$ \quad (Belle II~\cite{Kou:2018nap})      \\
    $\text{BR}(\tau \to \mu \gamma)$    &
     \quad $ <4.4\times 10^{-8}$ \quad (BaBar~\cite{Aubert:2009ag})  &
    \quad $10^{-9}$ \quad (Belle II~\cite{Kou:2018nap})     \\
    \hline
    $\text{BR}(\mu \to 3 e)$    &
     \quad $<1.0\times 10^{-12}$ \quad (SINDRUM~\cite{Bellgardt:1987du})    &
     \quad $10^{-15(-16)}$ \quad (Mu3e~\cite{Blondel:2013ia})   \\
    $\text{BR}(\tau \to 3 e)$   &
    \quad $<2.7\times 10^{-8}$ \quad (Belle~\cite{Hayasaka:2010np})&
    \quad $5\times10^{-10}$ \quad (Belle II~\cite{Kou:2018nap})     \\
    $\text{BR}(\tau \to 3 \mu )$    &
    \quad $<3.3\times 10^{-8}$ \quad (Belle~\cite{Hayasaka:2010np})  &
    \quad $5\times10^{-10}$ \quad (Belle II~\cite{Kou:2018nap})     \\
    & & \quad$5\times 10^{-11}$\quad (FCC-ee~\cite{Abada:2019lih})\\
        $\text{BR}(\tau^- \to e^-\mu^+\mu^-)$   &
    \quad $<2.7\times 10^{-8}$ \quad (Belle~\cite{Hayasaka:2010np})&
    \quad $5\times10^{-10}$ \quad (Belle II~\cite{Kou:2018nap})     \\
    $\text{BR}(\tau^- \to \mu^-e^+e^-)$ &
    \quad $<1.8\times 10^{-8}$ \quad (Belle~\cite{Hayasaka:2010np})&
    \quad $5\times10^{-10}$ \quad (Belle II~\cite{Kou:2018nap})     \\
    $\text{BR}(\tau^- \to e^-\mu^+e^-)$ &
    \quad $<1.5\times 10^{-8}$ \quad (Belle~\cite{Hayasaka:2010np})&
    \quad $3\times10^{-10}$ \quad (Belle II~\cite{Kou:2018nap})     \\
    $\text{BR}(\tau^- \to \mu^-e^+\mu^-)$   &
    \quad $<1.7\times 10^{-8}$ \quad (Belle~\cite{Hayasaka:2010np})&
    \quad $4\times10^{-10}$ \quad (Belle II~\cite{Kou:2018nap})     \\
    \hline
    $\text{CR}(\mu- e, \text{N})$ &
     \quad $<7 \times 10^{-13}$ \quad  (Au, SINDRUM~\cite{Bertl:2006up}) &
    \quad $10^{-14}$  \quad (SiC, DeeMe~\cite{Nguyen:2015vkk})    \\
    & &  \quad $2.6\times 10^{-17}$  \quad (Al, COMET~\cite{Krikler:2015msn,Adamov:2018vin,KunoESPP19})  \\
    & &  \quad $8 \times 10^{-17}$  \quad (Al, Mu2e~\cite{Bartoszek:2014mya})\\
    \hline
    \hline 
    $\mathrm{BR}(Z\to e^\pm\mu^\mp)$ & \quad$< 4.2\times 10^{-7}$\quad (ATLAS~\cite{Aad:2014bca}) & \quad$\mathcal O (10^{-10})$\quad (FCC-ee~\cite{Abada:2019lih})\\
    $\mathrm{BR}(Z\to e^\pm\tau^\mp)$ & \quad$< 5.2\times 10^{-6}$\quad (OPAL~\cite{Akers:1995gz}) & \quad$\mathcal O (10^{-10})$\quad (FCC-ee~\cite{Abada:2019lih})\\
    $\mathrm{BR}(Z\to \mu^\pm\tau^\mp)$ & \quad$< 5.4\times 10^{-6}$\quad (OPAL~\cite{Akers:1995gz}) & \quad $\mathcal O (10^{-10})$\quad (FCC-ee~\cite{Abada:2019lih})\\
    \hline
    \end{tabular}}
    \caption{Current experimental bounds and future sensitivities on cLFV observables considered in this work. All limits are given at $90\%\:\mathrm{C.L.}$ (notice that the Belle II sensitivities correspond to an integrated luminosity of $50\:\mathrm{ab}^{-1}$).}
    \label{tab:cLFVdata}
\end{table}
\renewcommand{\arraystretch}{1.}

\subsection{Leptonic processes: rare decays and transitions }
Lepton decays offer numerous probes of the potential contribution of a light leptophilic $Z^\prime$, and lead to extensive constraints on its flavour-violating couplings. 
Among these processes one finds cLFV muon and tau three-body decays ($\mu \to 3e$, $\tau \to 3e$, $\tau \to 3\mu$, $\tau^- \to \mu^- e^+ e^-$, $\tau^- \to \mu^- e^+ \mu^-$, $\tau^- \to e^- \mu^+ \mu^-$ and $\tau^- \to e^- \mu^+ e^-$), most of them mediated by $Z$ and photon-penguin, and radiative decays ($\mu \to e \gamma$, $\tau \to e \gamma$ and $\tau \to \mu \gamma$). 
As mentioned in the Introduction, we present here a full calculation of the one-loop contributions to LFV operators beyond leading-log approximations (see e.g.~\cite{Buras:2021btx})\footnote{Notice that in certain scenarios (depending on the symmetry breaking mechanism) there can be sizeable $Z-Z^\prime$ mixing, leading to  effectively flavour violating $Z$-vertices already at the tree-level.}.
In the muon sector, the $Z^\prime$ can also mediate Muonium oscillations as well as $\mu-e$ conversion. Finally, one must also consider the impact of the $Z^\prime$ on $\tau \to \ell \nu \bar \nu$ decays, and on the associated ratios of decay rates.

\subsubsection{cLFV 3 body decays: $\ell_\alpha \to \ell_\beta \, \overline{\ell\,}_{\!\gamma}\, \ell_\delta$}
\begin{figure}
     \centering
     \begin{subfigure}[b]{0.24\textwidth}
         \centering
\begin{tikzpicture}
\begin{feynman}
\vertex (b) at (0,0);
\vertex (l1) at (-0.65,0);
\vertex (l2) at (0.65,0);
\vertex (a) at (-1.5,0) {\(\ell_\alpha\)};
\vertex (f1) at (1.5,0) {\(\ell_\beta\)};
\vertex (c) at (0,-1.5);
\vertex (f2) at (1.5,-1) {\(\overline{\ell\,}_{\!\beta}\)};
\vertex (f3) at (1.5,-2) {\(\ell_\beta\)};
\diagram* {
(a) -- [fermion] (l1) -- [fermion] (b) -- [fermion] (l2) -- [fermion] (f1),
(b) -- [boson, edge label'=\(Z/\gamma\)] (c),
(c) -- [anti fermion] (f2),
(c) -- [fermion] (f3),
(l1) -- [boson, out=90, in=90, looseness=1.5, edge label=\(Z^\prime\)] (l2)
};
\end{feynman}
\end{tikzpicture}
         \caption{$\ell_\alpha \to \ell_\beta \overline{\ell\,}_{\!\beta} \ell_\beta$}
         \label{fig:3bd:a_bbb}
     \end{subfigure}
     \hfill
     \begin{subfigure}[b]{0.48\textwidth}
         \centering
\begin{tikzpicture}
\begin{feynman}
\vertex (b) at (0,0);
\vertex (l1) at (-0.65,0);
\vertex (l2) at (0.65,0);
\vertex (a) at (-1.5,0) {\(\ell_\alpha\)};
\vertex (f1) at (1.5,0) {\(\ell_\beta\)};
\vertex (c) at (0,-1.5);
\vertex (f2) at (1.5,-1) {\(\overline{\ell\,}_{\!\gamma}\)};
\vertex (f3) at (1.5,-2) {\(\ell_\gamma\)};
\diagram* {
(a) -- [fermion] (l1) -- [fermion] (b) -- [fermion] (l2) -- [fermion] (f1),
(b) -- [boson, edge label'=\(Z/\gamma\)] (c),
(c) -- [anti fermion] (f2),
(c) -- [fermion] (f3),
(l1) -- [boson, out=90, in=90, looseness=1.5, edge label=\(Z^\prime\)] (l2)
};
\end{feynman}
\end{tikzpicture}
\hspace{4mm}
\begin{tikzpicture}
\begin{feynman}
\vertex (b) at (0,0);
\vertex (a) at (-1.5,0) {\(\ell_\alpha\)};
\vertex (f1) at (1.5,0) {\(\ell_\gamma\)};
\vertex (c) at (0,-1.5);
\vertex (f2) at (1.5,-1) {\(\overline{\ell\,}_{\!\gamma}\)};
\vertex (f3) at (1.5,-2) {\(\ell_\beta\)};
\diagram* {
(a) -- [fermion] (b) -- [fermion] (f1),
(b) -- [boson, edge label'=\(Z^\prime\)] (c),
(c) -- [anti fermion] (f2),
(c) -- [fermion] (f3),
};
\end{feynman}
\end{tikzpicture}
         \caption{$\ell_\alpha \to \ell_\beta \overline{\ell\,}_{\!\gamma} \ell_\gamma$}
         \label{fig:3bd:a_bgg}
     \end{subfigure}
          \hfill
     \begin{subfigure}[b]{0.24\textwidth}
         \centering
    \begin{tikzpicture}
\begin{feynman}
\vertex  (b) at (0,0);
\vertex (a) at (-1.5,0) {\(\ell_\alpha\)};
\vertex (f1) at (1.5,0) {\(\ell_\gamma\)};
\vertex (c) at (0,-1.5);
\vertex (f2) at (1.5,-1) {\(\overline{\ell\,}_{\!\beta}\)};
\vertex (f3) at (1.5,-2) {\(\ell_\gamma\)};
\diagram* {
(a) -- [fermion] (b) -- [fermion] (f1),
(b) -- [boson, edge label'=\(Z^\prime\)] (c),
(c) -- [anti fermion] (f2),
(c) -- [fermion] (f3),
};
\end{feynman}
\end{tikzpicture}
         \caption{$\ell_\alpha \to \ell_\gamma \overline{\ell\,}_{\!\beta} \ell_\gamma$}
         \label{fig:3bd:a_gbg}
     \end{subfigure}
        \caption{Feynman diagrams contributing to cLFV three-body decays: from left to right, photon and $Z$ penguins, and tree level $Z^\prime$ exchange.}
        \label{fig:3_body_decay}
\end{figure}
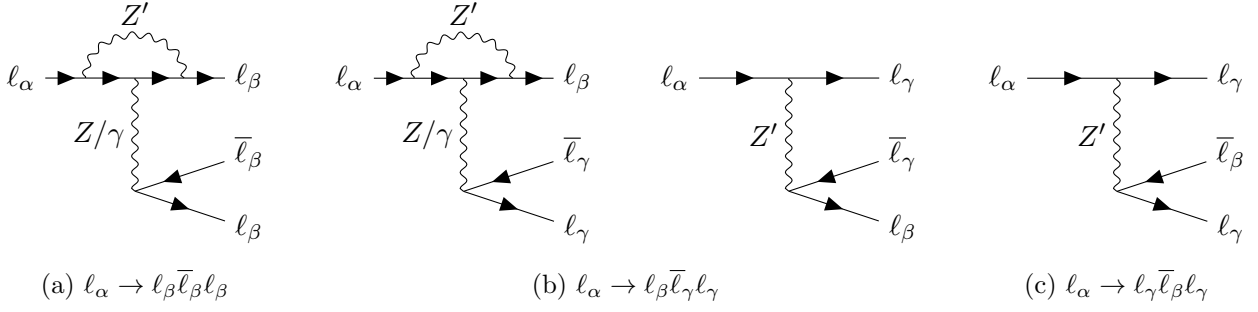

Depending on the mass of the decaying lepton, and as illustrated in Fig.~\ref{fig:3_body_decay}, 
several final-state flavour configurations are possible.
The (effective) Lagrangian governing these transitions can be written as~\cite{Abada:2014kba}
\begin{eqnarray}
     \mathcal{L}_{\text{cLFV}} \,= \,\mathcal{L}_{\ell \ell \gamma} \,+ \,\mathcal{L}_{4\ell}\,, \quad\text{with}
     \quad \mathcal{L}_{4\ell}\, = \,\mathcal{L}_{Z} + \mathcal{L}_{\mathrm{tree}}\,,
\end{eqnarray}
in which $\mathcal{L}_{\ell \ell \gamma}$ includes photon exchanges (dipole and anapole), and 
$\mathcal{L}_{4\ell}$ encodes the interactions relevant for
$Z$-penguins and tree-level decays. The former ($\mathcal{L}_{Z}$) can be cast as
\begin{eqnarray}\label{eq:lagrangian_zpenguin:def}
     \mathcal{L}_{Z} \,= \,F^{Z, \alpha\beta}_{XY} \,\bar{\ell}_\beta \left[\gamma^\mu P_X \right] \ell_\alpha \,
     \times \,\bar{\ell}_\gamma \left[\gamma_\mu P_Y \right] \ell_\gamma + \text{H.c.}\,,
\end{eqnarray}
with $\{X,Y\} = \{L,R\}$ and 
\begin{equation}
F^{Z, \alpha\beta}_{XY} \,= \,
-\dfrac{F^{Z, \alpha\beta}_X\, g_Y^Z}{m_{Z}^2}\, = \,
\dfrac{(F^{V, \alpha\beta}_X + i m_\alpha \,F^{T, \alpha\beta}_{\bar{X}})\, g_Y^Z}{m_{Z}^2}\,.
\end{equation}
In the above, the vector and tensor form factors, 
$F_X^{V,T}$, are computed from cLFV $Z$ decays 
in the limit of vanishing external fermion masses ($m_\alpha, m_\beta \to 0$) and of vanishing momentum transfer 
(i.e. $q.q \to 0$); moreover one has $\bar{X}=R,L$ (for $X=L,R$). One thus finds
\begin{eqnarray}
     F_L^{T, \alpha\beta} \,&=&\,  - \sum_{i} \frac{i g_L^{i\alpha} g_R^{i\beta *}  (g^Z_L+g^Z_R)}{16 \pi^2 m_{Z^\prime}} \left[ \frac{ \sqrt{x_i} \, \left(x_i^3+3 x_i-6 x_i \log (x_i)-4\right)}{4(x_i-1)^3}\right] \, ,
\\
        F_{L}^{V, \alpha\beta} \,&=&\,   \sum_{i} \frac{g_L^{i\alpha} g_L^{i \beta *}  (g^Z_L-g^Z_R)}{16 \pi^2 } \left[ \frac{x_i}{2}\left(\log \left(\frac{\mu ^2}{m_{Z^\prime}^2 x_i}\right)+\Delta_\varepsilon 
        \right)
-\frac{ x_i \left(x_i^2-8 x_i+6 \log (x_i)+7\right)}{4 (x_i-1)^2}\right] \, ,
\label{eq:vector_formfactor_L_R}
\end{eqnarray}
in which $\ell_i$ is the internal lepton ($\ell_i \neq \ell_\alpha \, , \ell_\beta$),  $x_i= m_i^2/m_{Z^\prime}^2$ and the right-handed (RH) coefficients are given by the exchange $(L \leftrightarrow R)$.
The divergence $\Delta_\varepsilon$ with
\begin{equation}
    \Delta_\varepsilon = \frac{1}{\varepsilon} - \gamma_E + \log(4\pi)\,,
\end{equation}
is minimally substracted ($\overline{\text{MS}}$-scheme) and the remnant 't Hooft scale set to $\mu^2 = m_\alpha^2$.
The tree-level interaction (encoded in $\mathcal{L}_{\mathrm{tree}}$) is given by
\begin{eqnarray}\label{eq:treelagragian:def}
     \mathcal{L}_{\mathrm{tree}} \, =\,  F_{XY}^{\alpha\beta\gamma\delta} \, \bar{\ell}_\beta \left[\gamma^\mu P_X \right] \ell_\alpha\,  \times\,  \bar{\ell}_\delta \left[\gamma_\mu P_Y \right] \ell_\gamma + \text{H.c.}\,,
     \quad \text{with} \quad 
     F_{XY}^{\alpha\beta\gamma\delta} \,= \,-\dfrac{g_X^{\alpha\beta *} \,g_Y^{\gamma\delta}}{m_{Z^\prime}^2}\,.
\end{eqnarray}
One thus finds for the ``effective" 4-lepton interactions:
\begin{eqnarray}
     \mathcal{L}_{4\ell}\, =\, \mathcal{C}_{XY}^{\alpha\beta\gamma\delta}\,\bar{\ell}_\beta \left[\gamma^\mu P_X \right] \ell_\alpha \,\times\, \bar{\ell}_\delta \left[\gamma_\mu P_Y \right] \ell_\gamma + \text{H.c.}\,, 
          \quad \text{with} \quad 
\mathcal{C}_{XY}^{\alpha\beta\gamma\delta} \,= \,F^{Z, \alpha\beta}_{XY} \,+ \,F_{XY}^{\alpha\beta\gamma\delta} \,;
\end{eqnarray}
in the above notice that $F^{Z, \alpha\beta}_{XY}$ only contributes in the case of $\ell_\gamma = \ell_\delta$ see Eq.~\ref{eq:lagrangian_zpenguin:def}.
Concerning the photon-exchange contributions to the cLFV 3-body decays, one has
\begin{eqnarray}
     \mathcal{L}_{\ell \ell \gamma} \,&=&\, e \,\bar{\ell}_\beta \left[\gamma^\mu (K_{1L}^{\alpha\beta} P_L + K_{1R}^{\alpha\beta} P_R) + i m_\alpha \sigma^{\mu \nu} \,q_\nu \,(K_{2L}^{\alpha\beta} P_L + K_{2R}^{\alpha\beta} P_R) \right] \ell_\alpha\, A_\mu + \text{H.c.}\,,
\end{eqnarray}
in which $e$ is the electric charge, $q$ denotes the photon 4-momentum and the anapole coupling $K_{1X}^{\alpha\beta}= q^2 F_{1X}^{\alpha\beta}$ with $F_{1X}^{\alpha\beta}$ defined below. In turn, this leads to the following photon-penguin amplitude:
\begin{eqnarray}\label{eq:gammapenguin:def}
     \mathcal{M}_{\gamma - \text{penguin}} =& - \dfrac{e^2}{q^2}\, \bar{u}(p_\beta)\left[ q^2 \gamma^\mu (F_{1L}^{\alpha\beta} P_L + F_{1R}^{\alpha\beta} P_R) + i m_\alpha \sigma^{\mu \nu} q_\nu (K_{2L}^{\alpha\beta} P_L + K_{2R}^{\alpha\beta} P_R) \right]u(p_\alpha) \nonumber \\
     &\times \bar{u}(p_\delta)\,Q_\delta\, \gamma_\mu\, v(p_\gamma) - ( p_\beta \leftrightarrow p_\delta)\,,
\end{eqnarray}
with $Q_\delta$ the electric charge of $\ell_\delta$.
Taking for simplicity (and clarity) the limit of vanishing external lepton masses,  
$m_\alpha, m_\beta \to 0$ (notice however that we 
keep all terms in the numerical computation of the decay rates), 
the anapole form factors entering the above equation are given by
\begin{eqnarray}\label{eq:anapole:coefficients}
     F_{1L}^{\alpha\beta} \,= \,\sum_{i}\, \dfrac{g_L^{i\alpha}\,g_L^{i\beta *}Q_i}{16\pi^2 \,m_{Z^\prime}^2} \,f(x_i), \quad
     F_{1R}^{\alpha\beta} \,=\, \sum_{i}\, \dfrac{g_R^{i\alpha}\,g_R^{i\beta *}Q_i}{16\pi^2 \,m_{Z^\prime}^2} \,f(x_i)\,,
\end{eqnarray}
in which $\ell_i$ denotes the internal lepton, 
$x_i = m_i^2/m_{Z^\prime}^2$, and with $f(x)$ defined as  
\begin{eqnarray}
    f(x) \,= \,-\dfrac{4 + 38 x - 63 x^2 + 14 x^3 + 7 x^4 - 6(4 - 16 x + 9 x^2) \log(x)}{36 (x -1 )^4}\,.
\end{eqnarray}
In the limit $m_\beta, \,q^2 \to 0$ (again, we emphasise that in the numerical computations we do take into account $m_\beta \neq 0$), the dipole coefficients are given by
\begin{eqnarray}\label{eq:dipole:coefficients}
K_{2L}^{\alpha \beta} &=& \sum_{i}\, \frac{g_R^{i \beta *} Q_i  }{16\pi^2 m_\alpha m_{Z^\prime}^2}\left\{
2 m_{Z^\prime}^2 \left[g_R^{i \alpha} m_\alpha-2 g_L^{i \alpha} m_i\right] C_0
\right. \nonumber \\
&+& \left[g_L^{i \alpha} m_i^3 -g_L^{i \alpha} m_\alpha^2 m_i +4 g_R^{i \alpha} m_\alpha m_{Z^\prime}^2  -4 g_L^{i \alpha} m_i m_{Z^\prime}^2 \right] C_1 \nonumber \\
&+&\left[g_L^{i \alpha} m_i^3-4 g_L^{i \alpha} m_i m_{Z^\prime}^2-g_R^{i \alpha} m_\alpha m_i^2+2 g_R^{i \alpha} m_\alpha m_{Z^\prime}^2\right] C_2  \nonumber \\
&+& \left. \left[2 g_R^{i \alpha} m_\alpha m_{Z^\prime}^2 +g_R^{i \alpha} m_\alpha m_i^2 -g_L^{i \alpha} m_\alpha^2 m_i \right] \left[ C_{11} + C_{12} \right]
\right\} \,,
\end{eqnarray}
where $C_{PV} \equiv C_{PV}(m_{\alpha}^2 ,0, 0 ,m_{Z^\prime}^2,m_{i}^2,m_{i}^2) $ 
are the Passarino-Veltman functions\footnote{In our study we adopt the LoopTools convention and notation, see~\cite{Hahn:1998yk,Passarino:1978jh}.} (with $PV={0,1,2,11,12}$), and $K_{2R}^{\alpha \beta} = K_{2L}^{\alpha \beta} (L\leftrightarrow R)$ (see Eq.~(\ref{eq:gammapenguin:def})).

\medskip
After considering the most general case for the 3-body decays, we proceed to address some simpler realisations, $\ell_\alpha \to \ell_\beta \overline{\ell\,}_{\!\beta} \ell_\beta$ and $\ell_\alpha \to \ell_\beta \overline{\ell\,}_{\!\gamma} \ell_\gamma$, $\ell_\alpha \to \ell_\gamma \overline{\ell\,}_{\!\beta} \ell_\gamma$.

\subsubsection{cLFV 3-body decays: $\ell_\alpha \to \ell_\beta \overline{\ell\,}_{\!\beta} \ell_\beta$}
This subset of processes receives contributions from $Z$- and photon-penguins, as can be seen in Fig.~\ref{fig:3bd:a_bbb}. The associated branching ratio is given by:
\begin{eqnarray}
     \hspace*{-3mm}\mathrm{BR}(\ell_\alpha \to 3\ell_\beta) &=& \dfrac{m_\alpha^5 \,\tau_\alpha}{512 \pi^3} \left[ e^4 \,\left(|K_{2L}^{\alpha \beta}|^2+ |K_{2R}^{\alpha \beta}|^2 \right)\left(\dfrac{16}{3}\log\dfrac{m_\alpha}{m_\beta} - \dfrac{22}{3} \right) \right. \nonumber \\
     &+& \left. \dfrac{2}{3}\left( |A^V_{LL}|^2 + |A^V_{RR}|^2 \right) + \dfrac{1}{3}\left( |A^V_{LR}|^2 + |A^V_{RL}|^2 \right) \right. \nonumber \\
     &-& \left. \dfrac{4e^2}{3}\left( K_{2L}^{\alpha\beta}\,A_{RR}^{V*} + K_{2R}^{\alpha\beta}\,A_{LL}^{V*} + \text{c.c.} \right)
     - \dfrac{2e^2}{3}\left( K_{2L}^{\alpha\beta}\,A_{RL}^{V*} + K_{2R}^{\alpha\beta}\,A_{LR}^{V*} + \text{c.c.} \right)\right],
\end{eqnarray}
in which $\tau_\ell$ denotes the lifetime of the decaying fermion, and where we have introduced the vector contribution, $A^V_{XY} = F^{Z,\alpha\beta}_{XY} + e^2 \, Q_\beta F_{1X}^{\alpha\beta}$. The relevant form factors have been defined in the previous subsection.

\subsubsection{cLFV 3-body decays: $\ell_\alpha \to \ell_\beta \overline{\ell\,}_{\!\gamma} \ell_\gamma$}
Another relevant case is that of $\ell_\alpha \to \ell_\beta \overline{\ell\,}_{\!\gamma} \ell_\gamma$ (with $\beta \neq \gamma$), corresponding to the processes $\tau \to e \bar\mu \mu $ and $\tau \to \mu \bar{e} e$. In the model under consideration these decays can occur at loop-level (penguin diagrams) and at tree-level, see Fig.~\ref{fig:3bd:a_bgg}. The corresponding branching ratios are given by:
\begin{eqnarray}
     \mathrm{BR}(\ell_\alpha \to \ell_\beta \overline{\ell\,}_{\!\gamma} \ell_\gamma) &\,=\,& \dfrac{m_\alpha^5 \,\tau_\alpha}{512 \pi^3} \left[ e^4 \left(|K_{2L}^{\alpha \beta}|^2+ |K_{2R}^{\alpha \beta}|^2 \right)\left(\dfrac{16}{3}\log\dfrac{m_\alpha}{m_\beta} - 8 \right) \right. \nonumber \\
     &+& \left.  \dfrac{1}{12}\left( |A^S_{LR}|^2 + |A^S_{RL}|^2 \right) \right. \nonumber \\
     &+& \left. \dfrac{1}{3}\left( |A^V_{LL}|^2 + |A^V_{RR}|^2 \right) + \dfrac{1}{3}\left( |A^V_{LR}|^2 + |A^V_{RL}|^2 \right) \right. \nonumber \\
     &-& \left. \dfrac{2e^2}{3}\left( K_{2L}^{\alpha\beta}\,A_{RR}^{V*} + K_{2R}^{\alpha\beta}\,A_{LL}^{V*} + K_{2L}^{\alpha\beta}\,A_{RL}^{V*} + K_{2R}^{\alpha\beta}\,A_{LR}^{V*} + \text{c.c.} \right)\right] \, .
\end{eqnarray}
As before we define distinct vector contributions 
(for $XX=LL,RR$): 
\begin{equation}
    A^V_{XX}\, = \,F_{XX}^{\alpha\gamma\gamma\beta} + F_{XX}^{Z, \alpha\beta} + e^2 Q_\gamma\, F_{1X}^{\alpha\beta}\,,
\end{equation} 
that include tree-level, $Z$- and $\gamma$-penguin mediated exchanges, and 
\begin{equation}
A^V_{XY} \,=\, F_{XY}^{Z, \alpha\beta} + e^2 Q_\gamma\, F_{1X}^{\alpha\beta}\,,
\end{equation} 
which does not include any tree-level contributions\footnote{The latter will be taken into account in the scalar contributions of the type $A^S_{XY} = -2 F_{XY}^{\alpha\gamma\gamma\beta}$ that stem from the Fierz-transformed tree-level operators to match the penguin ones.}.

\subsubsection{cLFV 3-body decays: $\ell_\alpha \to \ell_\gamma \overline{\ell\,}_{\!\beta} \ell_\gamma$}
Finally we consider the decay $\ell_\alpha \to \ell_\gamma \overline{\ell\,}_{\!\beta} \ell_\gamma$ (with $\beta \neq \gamma$), that can only arise from tree-level $Z^\prime$ exchanges, see Fig.~\ref{fig:3bd:a_gbg}. The physical processes corresponding to this type of decay are $\tau \to e \bar\mu e$ and $\tau \to \mu \bar{e} \mu$, and the corresponding branching ratio is given by:
\begin{eqnarray}
     \mathrm{BR}(\ell_\alpha \to \ell_\gamma \overline{\ell\,}_{\!\beta} \ell_\gamma) &=& \dfrac{m_\alpha^5 \, \tau_\alpha}{512 \pi^3} \left[ \dfrac{2}{3}\left( |A^V_{LL}|^2 + |A^V_{RR}|^2 \right) + \dfrac{1}{3}\left( |A^V_{LR}|^2 + |A^V_{RL}|^2 \right)\right] \, ,
\end{eqnarray}
with $A^V_{XY} = F_{XY}^{\alpha\gamma\beta\gamma}$ defined in Eq.~(\ref{eq:treelagragian:def}).

\subsubsection{cLFV radiative decays: $\ell_\alpha \to \ell_\beta \gamma$}
These correspond to higher order processes, mediated by the light $Z^\prime$ in the loop, and the associated branching ratios are given by~\cite{Hisano:1995cp}
\begin{eqnarray}
     \mathrm{BR}(\ell_\alpha \to \ell_\beta \gamma) \,=\, \dfrac{m_\alpha^5 \,\tau_\alpha}{16 \pi}\,e^2\,\left(|K_{2L}^{\alpha \beta}|^2+ |K_{2R}^{\alpha \beta}|^2 \right)\,,
\end{eqnarray}
in which all the quantities have been already defined.

\subsubsection{Neutrinoless muon-electron conversion in nuclei}
The cLFV $Z^\prime$ interactions can also lead to contributions to 
$\mu-e$ transitions in muonic atoms. 
Following~\cite{Kitano:2002mt}, the (coherent) conversion rate can be cast as 
\begin{eqnarray}\label{eq:CR:mu-e}
    \text{CR}(\mu - e, \text{ N}) \, &=&\, \dfrac{2\,G_F^2\,m_\mu^5}{\Gamma_{\text{capt}}} \left|A_R^* \,D + (2g_{LV(u)} + g_{LV(d)} )\,V^{(p)} + (g_{LV(u)} + 2g_{LV(d)} )V^{(n)} \right|^2  \nonumber \\
    && \hspace*{15mm}+ \, \{L,R \leftrightarrow R,L\}\,,
\end{eqnarray}
with $D, V^{(p)}, V^{(n)}$ the overlap integrals (see~\cite{Kitano:2002mt}). Here $\Gamma_{\text{capt}}$ is the muon capture rate, and $G_F$ is the Fermi constant. The relevant quantities entering the above equation are given by 
\begin{equation}
    A_{L/R} \, = \, -\frac{\sqrt{2}}{4 \, G_F} \, e \,  K_{2L/R}^{\mu e} \, ,
\end{equation}
with $K_{2L/R}^{\mu e}$ defined in Eq.~(\ref{eq:dipole:coefficients}) and 
\begin{equation}
g_{XV(q)} \,= \,-\frac{\sqrt{2}}{G_F}\,(e^2\,  Q_q \,F_{1X}^{\mu e} + F_{VX}^{Z, \mu e} \,g_V^{Z,q}) \quad (X=L,R)\,.
\end{equation}
In the above, $Q_q$ is the electric charge of the quark $q$, and 
\begin{equation}
F_{VX}^{Z, \mu e} \,=\, \dfrac{F_{VX}^{V,\mu e} + i\,m_\mu\, F_{V\bar{X}}^{T,\mu e}}{m_Z^2} \quad (X=L,R \quad \text{and} \quad \bar{X}= R,L)\,,
\end{equation}
with $F_{VX}^{V,T}$ given in Eq.~(\ref{eq:vector_formfactor_L_R}) and we clarify that in the above $g_V^{Z,q}$ denotes the {\it vector} coupling of the $Z$-boson to a quark $q$. 

\subsubsection{Muonium oscillations}
Another relevant observable to consider are spontaneous Muonium-antimuonium oscillations. Muonium is a bound state of an electron and an antimuon, $\mathrm{Mu} = \mu^+ e^-$. In presence of cLFV interactions, this system can oscillate into antimuonium $\mathrm{\overline{Mu}} = \mu^-e^+$. 
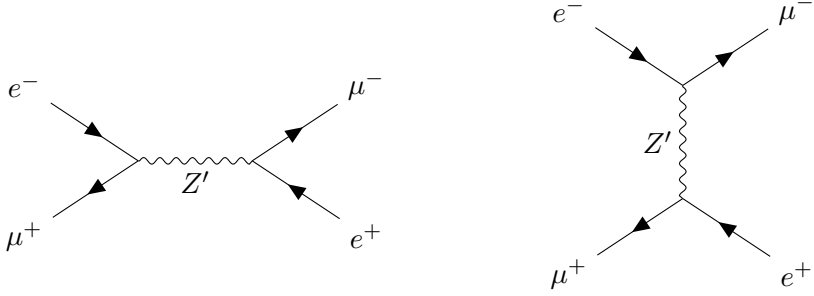
\begin{figure}
    \centering
\raisebox{5mm}{\begin{subfigure}[b]{0.49\textwidth}
\centering
\begin{tikzpicture}
\begin{feynman}
\vertex (a) at (0,0);
\vertex (l1) at (-1.5,1)  {\(e^-\)};
\vertex (l2) at (-1.5,-1) {\( \mu^+\)};
\vertex (b) at (1.5,0);
\vertex (f1) at (3, 1) {\(\mu^-\)};
\vertex (f2) at (3, -1) {\(e^+\)};
\diagram* {
(l1) -- [fermion] (a) -- [fermion] (l2),
(f2) -- [fermion] (b) -- [fermion] (f1),
(a) -- [boson, edge label'=\(Z^\prime\)] (b),
};
\end{feynman}
\end{tikzpicture}      
\end{subfigure} }
\hspace*{-23mm}
\begin{subfigure}[b]{0.49\textwidth}
\centering
\begin{tikzpicture}
\begin{feynman}
\vertex (a) at (0,0);
\vertex (l1) at (-1.5,1)  {\(e^-\)};
\vertex (l2) at (1.5,1) {\( \mu^-\)};
\vertex (b) at (0,-1.5);
\vertex (f1) at (-1.5, -2.5) {\(\mu^+\)};
\vertex (f2) at (1.5, -2.5) {\(e^+\)};
\diagram* {
(l1) -- [fermion] (a) -- [fermion] (l2),
(f2) -- [fermion] (b) -- [fermion] (f1),
(a) -- [boson, edge label'=\(Z^\prime\)] (b),
};
\end{feynman}
\end{tikzpicture}     
\end{subfigure} 
    \caption{Feynman diagrams for Muonium-antimuonium oscillations (tree-level $Z^\prime$ mediated, $s$ and $t$ channel exchanges).}
    \label{fig:muonium}
\end{figure}
In the model under consideration $\mathrm{Mu-\overline{Mu}}$ conversion can occur at tree-level (as shown in Fig.~\ref{fig:muonium}), from both $s-$ and $t$-channels $Z^\prime$ exchange.
The relevant effective Lagrangian for the oscillations can be cast as~\cite{Fukuyama:2021iyw}:
\begin{equation}
   - \mathcal{L}_{\mathrm{Mu-\overline{Mu}}} \,=\, \sum_{i=1,...,5}\dfrac{G_i}{\sqrt{2}}\, Q_i\,,
\label{eq::MuoniumLagrangian}
\end{equation}
where $Q_i$ are the four-fermion operators responsible for the $\mathrm{Mu-\overline{Mu}}$ transitions
\begin{eqnarray}
        &Q_{1(2)} &=\hspace{3mm} 4 \left(\overline{\mu}\, \gamma_\alpha \,P_{L(R)} \,e\right)\,\left(\overline{\mu}\, \gamma^\alpha\, P_{L(R)} \,e\right), \nonumber\\
        &Q_3 &=\hspace{3mm} 4 \left(\overline{\mu} \,\gamma_\alpha\, P_R \,e\right)\,\left(\overline{\mu}\, \gamma^\alpha \,P_L\, e\right), \nonumber\\
       & Q_{4(5)} &=\hspace{3mm} 4 \left(\overline{\mu}\, P_{L(R)} \,e\right)\,\left(\overline{\mu} \,P_{L(R)} \,e\right)\,.
\label{eq::MuoniumOperators}
\end{eqnarray}
Moreover, notice that the presence of an external magnetic field also induces an energy splitting, further contributing to $\mathrm{Mu-\overline{Mu}}$ mixing. 
In the present LFV leptophilic $Z^\prime$ model, the transition probability  can be written as
\begin{align}
    P &= \dfrac{2.57 \times 10^{-5}}{G_F^2}\left\{|c_{0,0}|^2 \left|-G_3 + \dfrac{G_1 +G_2 - \frac{1}{2}G_3}{\sqrt{1+X^2}} \right|^2 + |c_{1,0}|^2 \left|G_3 + \dfrac{G_1 +G_2 - \frac{1}{2}G_3}{\sqrt{1+X^2}} \right|^2 \right\}\,,  
\label{MuoniumProba}
\end{align}
where $|c_{F,m}|^2$ denote the population of Muonium states and the factor $X$ encodes the magnetic flux density~\cite{Fukuyama:2021iyw}. Notice that in the above we have  only considered vector couplings (so that $Q_4,Q_5=0$).
For Hermitian couplings $g^{e\mu}_{L,R}$ (we recall that in our study we work with real couplings), one finds
\begin{equation}
    \dfrac{G_{1}}{\sqrt{2}}\,=\,\dfrac{|g_{L}^{e\mu}|^2}{8 m_{Z^\prime}^2}, \quad \:
    \dfrac{G_{2}}{\sqrt{2}}\,=\,\dfrac{|g_{R}^{e\mu}|^2}{8 m_{Z^\prime}^2}, \quad \:
    \dfrac{G_3}{\sqrt{2}}\,=\,\dfrac{2 g_L^{e\mu}\,g_R^{e\mu,*}}{8 m_{Z^\prime}^2}\,.
\label{eq::MuoniumCouplings}
\end{equation}
The new contributions to the transition probability $P$ (see Eq.~(\ref{MuoniumProba})) should be compared with the bound 
set by the PSI experiment~\cite{Willmann:1998gd} 
\begin{eqnarray}
   P\,< \,8.3 \times 10^{-11}\,.
\end{eqnarray}

\mathversion{bold}
\subsection{Constraints from LFU in $\tau$ decays}
\mathversion{normal}
The comparison of the SM-allowed $\tau \to \ell \nu_{\tau} \overline{\nu}_{\ell}$ decays (mediated at tree-level by $W$-boson exchange) allows to construct the following ratio, which can be used as a probe of LFU in lepton decays:
\begin{equation}
        R^\tau_{\mu e} \,\equiv\, \dfrac{\Gamma(\tau^- \to \mu^- \nu_\tau \overline{\nu}_\mu)}{\Gamma(\tau^- \to e^- \nu_\tau \overline{\nu}_e)} \,.
\label{eq:def:tauratio}
\end{equation}
The measured $\tau$ leptonic decay ratio ${R^\tau_{\mu e}}|_\text{exp}$ from ARGUS~\cite{ARGUS:1991zhv}, CLEO~\cite{CLEO:1996oro} and BaBar~\cite{BaBar:2009lyd} is found to be consistent with the SM prediction ${R^\tau_{\mu e}}|_\text{SM} = 0.972564 \pm 0.00001$~\cite{Pich:2009zza}; 
the HFLAV collaboration reports the following global fit~\cite{HFLAV:2019otj}:
\begin{equation}
        {R^\tau_{\mu e}}|_\text{exp}\,\, \equiv\, \,\dfrac{\Gamma(\tau^- \to \mu^- \nu_\tau \overline{\nu}_\mu)}{\Gamma(\tau^- \to e^- \nu_\tau \overline{\nu}_e)}\,=\,0.9761\pm 0.0028\,.
\label{eq::tauratio}
\end{equation}
Experimentally, it is not possible to disentangle the different neutrino flavours, so any LFV tree-level decay ($\tau \to \ell \nu_\gamma \overline{\nu}_\delta$, with $\gamma,\delta = e, \mu, \tau$)
will contribute to this observable; tree-level contributions from the new light $Z^\prime$ can thus potentially compete with the SM processes, leading to deviations from the observed $ {R^\tau_{\mu e}}|_\text{exp}$.
The relevant contact interactions at the source of the  new contributions to the decays (generically cast for $ \ell_\alpha \to \ell_\beta \nu_\gamma \bar{\nu}_\delta$) can be written as:
\begin{equation}\label{eq:LFU-tau}
    \mathcal{L}\,=\, C_L^{\alpha\beta\nu_\gamma \nu_\delta} \,(\bar{\ell}_\beta\, \gamma_\mu \,P_L \,\ell_\alpha)(\bar{\nu}_\gamma \, \gamma^\mu \,P_L\, \nu_\delta) + C_R^{\alpha\beta\nu_\gamma \nu_\delta} (\bar{\ell}_\beta\,\gamma_\mu \,P_R\, \ell_\alpha)(\bar{\nu}_\gamma \, \gamma^\mu \,P_L \,\nu_\delta) \,,
\end{equation}
where we have introduced $C_L^{\alpha\beta\nu_\gamma \nu_\delta} = C_L^{\text{SM}} + C_L^{\text{NP}}$ and $C_R^{\alpha\beta\nu_\gamma \nu_\delta} = C_R^{\text{NP}}$; after Fierz-transforming the SM charged current operator, one has
\begin{equation}
    C_L^{\text{SM}}\,(\bar{\ell}_\beta \,\gamma_\mu\, P_L \,\nu_\beta)\,(\bar{\nu}_\alpha\, \gamma^\mu \,P_L\, \ell_\alpha) \,=\,  
    C_L^{\text{SM}} \,(\bar{\ell}_\beta \,\gamma_\mu \,P_L\, \ell_\alpha)\,(\bar{\nu}_\alpha \,\gamma^\mu \,P_L \,\nu_\beta)\,.
\end{equation}
The SM contribution, only defined for the process $ \ell_\alpha \to \ell_\beta \nu_\alpha \bar{\nu}_\beta $, reads
\begin{eqnarray}
  C_L^{\text{SM}} &=& - \dfrac{4 G_F^{\text{eff}}}{\sqrt{2}} \,,
\end{eqnarray}
with $G_F^\text{eff}$ the effective Fermi constant in the presence of NP in $\mu\to e\bar\nu_e\nu_\mu$ decays;
in fact, and as a consequence of the modification of the muon lifetime, 
a comparison of the new rate of $\mu\to e\bar\nu_e\nu_\mu$ decays with the SM prediction allows defining $G_F^\text{eff}$ as
 \begin{equation}
    G_F^\text{eff} \,\simeq\, G_F \,  \sqrt{\dfrac{\text{BR}^{\text{SM}}(\mu\to e\bar\nu_e\nu_\mu)}{\text{BR}^{\text{SM+NP}}(\mu\to e\bar\nu_e\nu_\mu)}}\,.
\end{equation}
The tree-level NP contribution to the decay $ \ell_\alpha \to \ell_\beta \nu_\gamma \bar{\nu}_\delta$ is given by
\begin{eqnarray}
  C_{L(R)}^{\text{NP}} &=&  \dfrac{g_{L(R)}^{\alpha \beta, *} \, g_L^{\gamma \delta}}{m_{Z^\prime}^2} \,,
\end{eqnarray}
thus leading to the following expression for the  $\tau \to \ell  \nu \bar{\nu}$ decay width
\begin{eqnarray}
     \Gamma(\tau \to \ell \nu \bar{\nu}) \,= \,\sum_{\gamma , \delta = e, \mu, \tau}\dfrac{m_\tau^5}{192\, (2\pi)^3}\left[ 4C_L^{\tau\ell\nu_\gamma \nu_\delta\,\dag}\, C_R^{\tau\ell\nu_\gamma \nu_\delta} \, g(x_\ell) - \left(|C_L^{\tau\ell\nu_\gamma \nu_\delta}|^2 + |C_R^{\tau\ell\nu_\gamma \nu_\delta}|^2\right) \,f(x_\ell) \right]\, r_{\text{RC}}^{\tau\ell}\,,
\end{eqnarray}
with $x_\ell = \dfrac{m_\ell^2}{m_\tau^2}$ and the associated functions defined as 
\begin{eqnarray}
  f(x) &=& -1 + 8x - 8x^3 + x^4 + 12x^2\log(x)\,,  \nonumber \\
  g(x)&=&\sqrt{x}\, [-1 -9x + 9x^2 + x^3 - 6x(1+x)\log(x)]\,.
\end{eqnarray}
Further QED corrections and effects from the non-local structure of the $W$-boson propagator are included in $r_\text{RC}^{\alpha\beta}$\footnote{One can also take higher-order EW corrections to the $W$-boson propagator~\cite{Sirlin:1980nh} into account, but since these are flavour universal, they will cancel in the ratio here considered.}:
\begin{eqnarray}
 r_{\text{RC}}^{\alpha\beta} \,=\, \left[ \vphantom{\dfrac{m_\beta^2}{m_W^2}} 1 + \dfrac{\alpha_e(m_\alpha)}{2\pi}\left(\dfrac{25}{4} - \pi^2\right)\right] \left[ 1 + \dfrac{3}{5}\dfrac{m_\alpha^2}{m_W^2} + \dfrac{9}{5}\dfrac{m_\beta^2}{m_W^2} \right] \, ,
\end{eqnarray}
where $\alpha_e(m_\alpha)$ is the running electromagnetic fine-structure ``constant" (at  $m_{\alpha}$).

\medskip
Generally, the above corrections to the Fermi constant $G_F$ due to the presence of New Physics in the Michel decay of the muon should be propagated to other electroweak observables such as the invisible $Z$-decay and the mass of the $W$-boson.
However, as discussed in detail in the following sections, experimental data constrains the $g^{e\mu}$ couplings to be very small and thus corrections to $G_F$ are negligible for all practical purposes.

\section{Implications for the anomalous magnetic moments}\label{sec:amuae-impact}
We now address whether or not the model under consideration (a light, leptophilic, cLFV-interacting $Z^\prime$ boson) allows to account for the observed $a_{e, \mu}$, while complying with the numerous bounds from boson decays as well as  rare leptonic processes presented in the previous section. 
We first discuss the new contributions mediated by the (light) $Z^\prime$, and then (numerically) explore the  model's parameter space (spanned by new boson's mass, and by its left-handed and right-handed cLFV couplings, $g_{L,R}^{\alpha \beta}$).
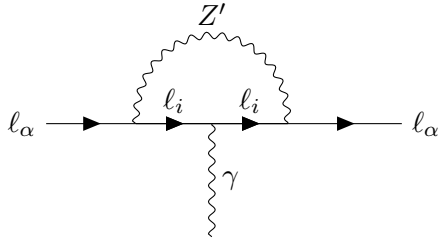
\begin{figure}[h!]
    \centering
\begin{tikzpicture}
\begin{feynman}
\vertex (a1) {\(\ell_\alpha\)};                    
\vertex[right=1.5cm of a1] (a2);
\vertex[right=1cm of a2] (a3);
\vertex[right=1cm of a3] (a4);
\vertex[right=1.5cm of a4] (a5) {\(\ell_\alpha\)};
\vertex[below=1.5cm of a3] (b);
\diagram* {
{[edges=fermion]
(a1) -- (a2) -- [fermion, edge label=\(\ell_i\)](a3) -- [fermion, edge label=\(\ell_i\)](a4) -- (a5),
},
(a3) -- [boson, edge label=\(\gamma\)] (b),
(a4) -- [boson, out=90, in=90, looseness=2.0, edge label'=\(Z^\prime\)] (a2)
};
\end{feynman}
\end{tikzpicture}
    \caption{New contributions to the anomalous magnetic moment of charged leptons, $a_{\ell_\alpha}$; notice that the internal lepton flavour is necessarily different from the external one ($\ell_i \neq \ell_\alpha$).}
    \label{fig:aell}
\end{figure}

The new physics contribution to $a_\ell$ from the additional $Z^\prime$ boson arises at one-loop (see Fig.~\ref{fig:aell}) and, following~\cite{Lindner:2016bgg, Jegerlehner:2009ry, Leveille:1977rc}, it can be expressed as: 
\begin{equation}
    \Delta a_\ell \,= \,\sum_i \left[\dfrac{|g_V^{\ell i}|^2}{4\pi^2}\,\dfrac{m_\ell^2}{m_{Z^\prime}^2}\,F(\lambda, \epsilon_i) + \dfrac{|g_A^{\ell i}|^2}{4\pi^2}\,\dfrac{m_\ell^2}{m_{Z^\prime}^2}\,F(\lambda, -\epsilon_i)\right]\,,
\label{eq::deltaaNP}
\end{equation}
with $g_V, g_A = (g_L \pm g_R)/2$ denoting  the 
vector and axial-vector couplings. The sum runs over the internal leptons (whose flavour is necessarily distinct from that of the external legs), and the loop function $F(\lambda, \epsilon_i)$ is defined as:
\begin{align}
       F(\lambda, \epsilon_i) &= \dfrac{1}{2}\int^1_0 dx \left[ \dfrac{2x(1-x)[x-2(1-\epsilon_i)] }{(1-x)(1-\lambda^2x)+\epsilon_i^2\lambda^2x} + \dfrac{\lambda^2x^2(1-\epsilon_i)^2(1+\epsilon_i-x)}{(1-x)(1-\lambda^2x)+\epsilon_i^2\lambda^2x}\right] , 
\label{eq::functionfordeltaa}
\end{align}
in which $\epsilon_i = m_i/m_\ell$, $m_i$ being the mass of the internal fermion and $\lambda = m_\ell/m_{Z^\prime}$.
Notice that the opposite sign between vector and axial-vector loop functions leads to partial cancellations, thus potentially opening the door for an explanation of both $\Delta a_\mu$ and $\Delta a_e$.

Throughout this section, in what concerns constraining the model's parameter space, we will fix the $Z^\prime$ mass to 
$m_{Z^\prime} = 10$~GeV (unless otherwise stated); this is a simplifying approach, which allows to evade the numerous constraints which would otherwise arise from the extensive searches at low-energy experiments such as indirect signals at $B$-factories~\cite{Iguro:2020rby}, and flavour violating $\tau\to\mu Z^\prime$ decays~\cite{Foot:1994vd,Heeck:2016xkh,Altmannshofer:2016brv}.
At the end of the discussion (Section~\ref{sec:amuae:heavier}), we will revisit this working hypothesis, and explore a wider range for $m_{Z^\prime}$.

\bigskip
\bigskip
Before carrying out the numerical study, let us briefly clarify some points regarding the associated statistical analysis.  
Leading to the (best-fit) contours presented in several of the plots of this section, we note here that contours of a single observable correspond to model predictions within the $1\,\sigma$ region of the experimental measurements; for combinations of observables (e.g. the $Z$-decay LFU ratios and the global contours), we construct a combined likelihood as a product of the experimental probability distribution functions (pdf) of the observables $\mathcal O_i$, evaluated at a point of the model's predictions given a set of input parameters $\vec p$:
\begin{equation}
    \mathcal L(\vec p) \,= \,\prod_i \mathrm{pdf}_i\,\left(\mathcal O_i^\text{exp}, \,\mathcal O_i^\text{th}(\vec p)\right)\,.
\end{equation}
For the observables here considered, theoretical uncertainties due to SM input parameters are negligible and we approximate the observables' probability density functions as uncorrelated gaussians.
Thus, the relative ($\log$)-likelihood function can be approximated as a simple $\chi^2$-function given by
\begin{equation}
    -2\Delta \log \mathcal L(\vec p) \,\simeq\, \chi^2(\vec p)\, =\, \sum_i \dfrac{(\mathcal O_i^\text{th}(\vec p) - \mathcal O_i^\text{exp})^2}{(\sigma_i^\text{exp})^2}\,,
\end{equation}
in which $\mathcal O_i^\text{exp}$ and $\sigma_i^{\text{exp}}$ respectively denote the mean value and the $1\,\sigma$ (gaussian) uncertainty of the $\mathcal O_i$ observable measurement.
The $\chi^2$-function is then minimised in terms of the model's parameters.

We further consider the test statistic $\Delta \chi^2 \equiv \chi^2(\vec p) - \chi^2_\text{min}$ in order to establish confidence intervals; in particular we assume that this quantity is distributed as a $\chi^2$ random variable with $n=2$ degrees of freedom.
The combined contours then correspond to the $k\,\sigma$ thresholds where the 2-dimensional cumulative $\chi^2$ distribution reaches the probability $P_{k\,\sigma}$ (for $1\,\sigma$ $\Delta \chi^2 \approx 2.3$, for $2\,\sigma$ $\Delta \chi^2\approx 6.2$), defined as the probability for a gaussian random variable to be measured within $k$ standard deviations from the mean.

\subsection{Accommodating the anomalous magnetic moment of the muon}\label{subsec:amu}
\begin{figure}[h!]
    \centering
    \mbox{\includegraphics[width=0.48\textwidth]{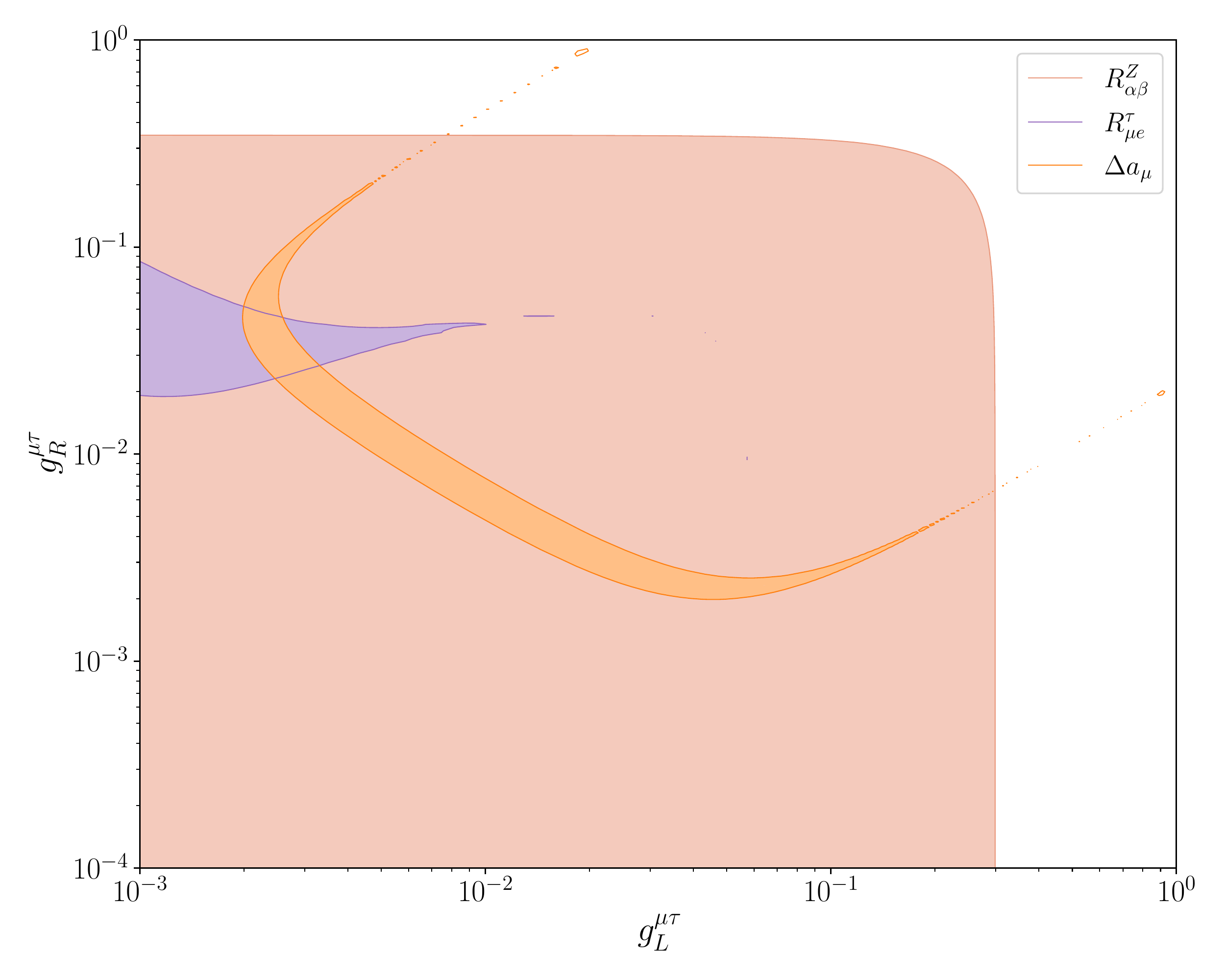}
    \includegraphics[width=0.48\textwidth]{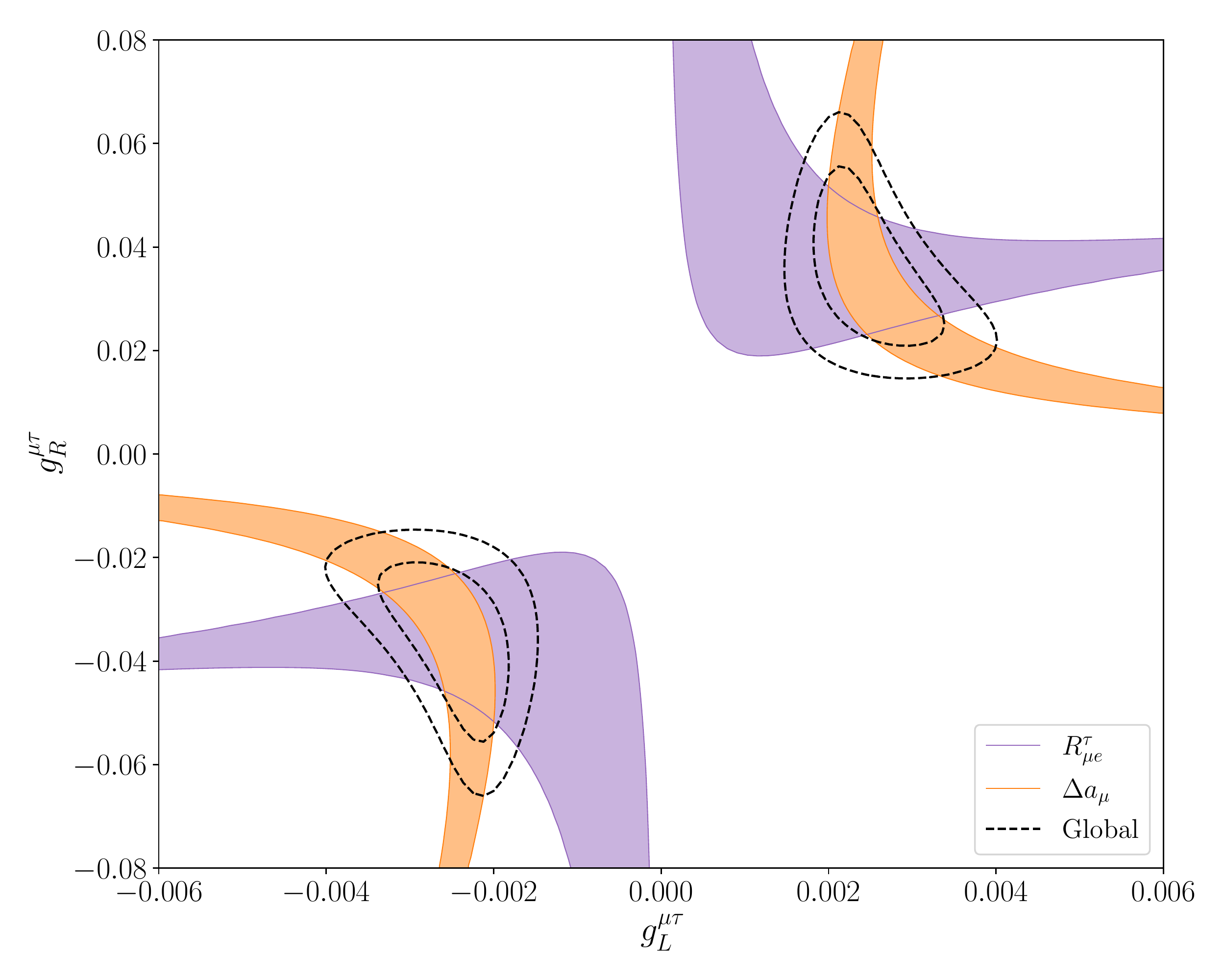}}
    \caption{Constraints on $Z^\prime$  
    couplings, $g_{L,R}^{\mu \tau}$, and prospects for $(g-2)_\mu$. On the left, allowed regimes for $\Delta a_\mu$ in the same-sign ($g_{L}^{\mu \tau}$-$g_{R}^{\mu \tau}$) parameter space, in agreement with  the experimental bounds on LFU $Z$ decay ratios (light orange) and $R^\tau_{\mu e}$ (purple); the dark orange region denotes the $1\,\sigma$ $\Delta a_\mu$ favoured regime.  
    On the right, detailed (linear) view of the viable  $g_{L}^{\mu \tau}$ vs. $g_{R}^{\mu \tau}$ parameter space, with the dashed curves now corresponding to the $1\, \sigma$ and $2\,\sigma$ global fits (see text).}
    \label{fig:g_mu_tauLR}
\end{figure}

We begin by discussing the regimes for the 
$Z^\prime \ell_\alpha \ell_\beta$ couplings
(i.e. $g_{L,R}^{\alpha \beta}$) which allow to alleviate the current tension in $\Delta a_\mu$,  see Eq.~(\ref{eq:amu:delta}), taking into account the impact regarding the numerous cLFV and LFU observables discussed in Section~\ref{sec:cLFV-LFU}. The NP contributions to 
$(g-2)_\mu$ can be associated with the exchange of either an electron or a tau lepton in the loop. 
The first possibility is strongly disfavoured, as it would lead to a negative shift in $\Delta a_\mu$.
Let us then consider the second possibility: in Fig.~\ref{fig:g_mu_tauLR}, we present two representations of the plane spanned by the $\mu-\tau$ couplings, 
($g_{L}^{\mu \tau}- g_{R}^{\mu \tau}$). As already mentioned, we set  $m_{Z^\prime} = 10$~GeV.
On the left panel of Fig.~\ref{fig:g_mu_tauLR} we display the $\Delta a_\mu$-favoured regimes relying on $\mu-\tau$ cLFV $Z^\prime$ couplings (all others being set to zero), as well as the most important associated constraints, which in this case arise from conflicts with the LFU-probing ratios, $R^\tau_{\mu e}$ and $R^Z_{\alpha\beta}$. 
Notice that the constraining role of $R^\tau_{\mu e}$ is much more important than that of  $R^Z_{\alpha\beta}$; the latter can be easily accommodated in wide regions of the parameter space (the only exception being large values of  $g_{L,R}^{\mu \tau} \gtrsim  3 \times 10^{-1}$). The other cLFV observables discussed in Section~\ref{sec:cLFV-LFU} play a far less restrictive role, and we do not discuss them here. For completeness, let us notice that corrections to the invisible $Z$ decay width due to one-loop contributions are negligible, $\Delta \Gamma (Z\to\text{inv.}) \lesssim 1$~keV, throughout the model's parameter space.

As can be seen, all experimental observations (i.e. $\Delta a_\mu$ and the LFU constraints) can be accommodated for $g_{L}^{\mu \tau} \sim \mathcal{O}(10^{-3})$, and $g_{R}^{\mu \tau} \sim \mathcal{O}(10^{-2})$. 
For clarity, this region has been expanded in the right panel of Fig.~\ref{fig:g_mu_tauLR}, where we have also displayed the 
corresponding regimes with negative values of the couplings (notice however that one must have same-sign couplings). In addition to $\Delta a_\mu$ and $R^\tau_{\mu e}$, we also display the $1\,\sigma$ and $2\,\sigma$ contours of the global fit to the observables. 
Finally, also visible from the right panel of Fig.~\ref{fig:g_mu_tauLR} is the apparent relation emerging for the left- and right-handed $Z^\prime$ couplings: the best-fit regions to saturate $\Delta a_\mu$ lie around $g_{R}^{\mu \tau} \sim \mathcal{O}(15) \times g_{L}^{\mu \tau}$. 

\subsection{Accounting for the anomalous magnetic moment of the electron}
\begin{figure}[h!]
    \centering
    \mbox{\includegraphics[width=0.48\textwidth]{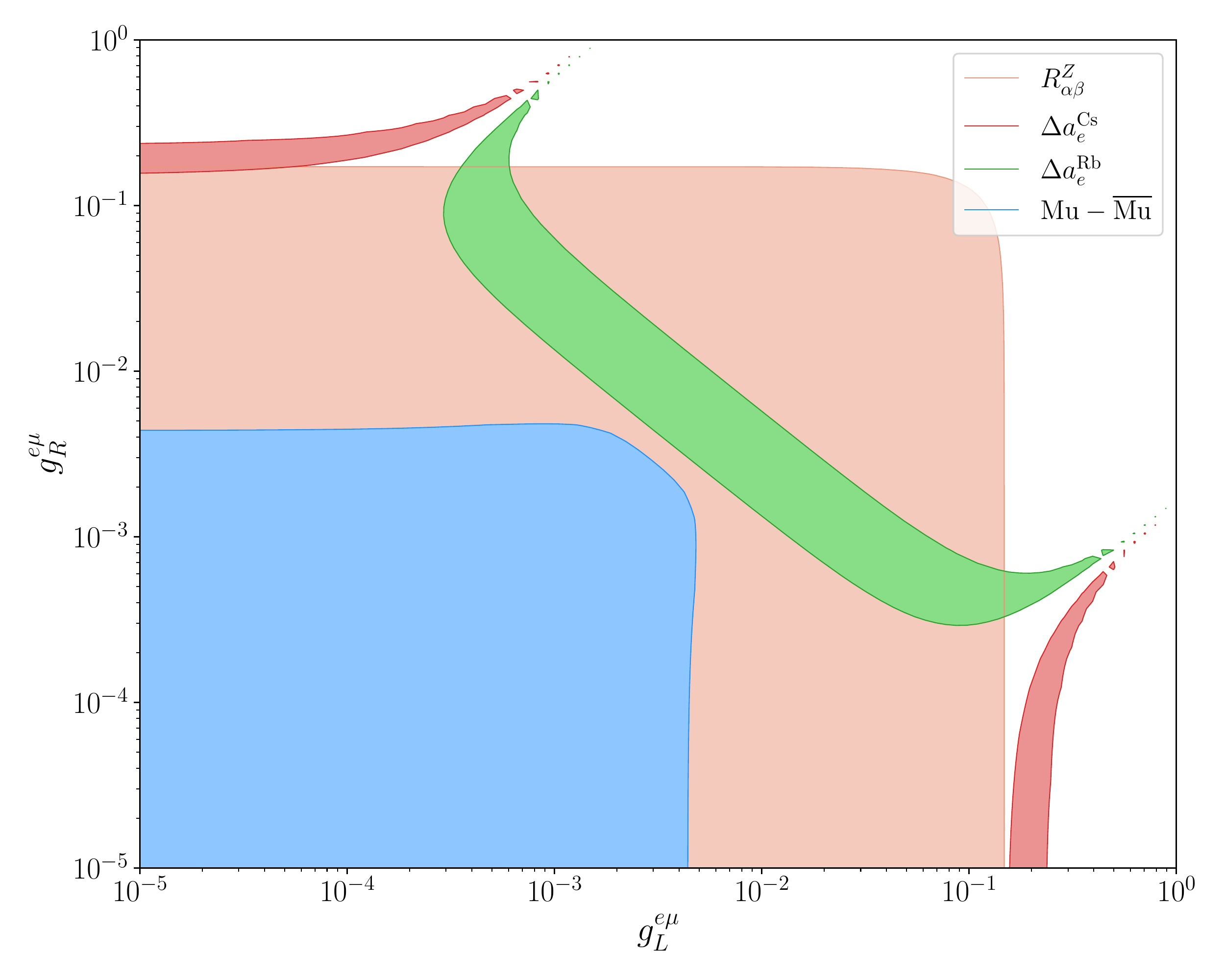}
    \includegraphics[width=0.48\textwidth]{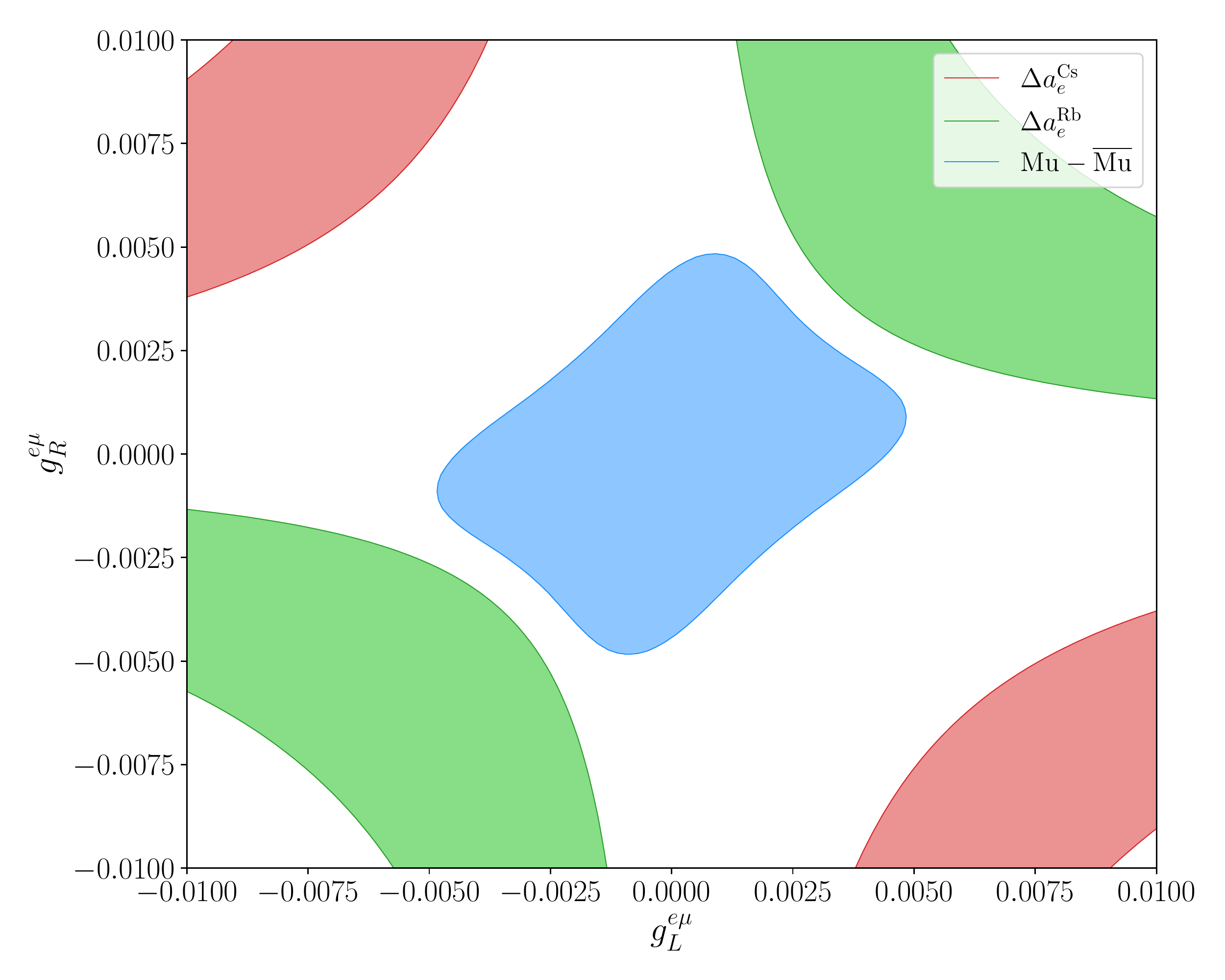}}
    \caption{Constraints on $g_{L,R}^{e \mu}$ couplings and prospects for $(g-2)_e$. On the left, allowed regions in the same-sign ($g_{L}^{e \mu }$-$g_{R}^{e \mu}$) parameter space, in agreement with  the experimental bounds on LFU $Z$ decay ratios (light orange) and Muonium oscillations (blue); the red and the green regions respectively denote the $1\,\sigma$ $(g-2)_e^\text{Cs}$ and $(g-2)_e^\text{Rb}$ favoured regimes. On the right, detailed (linear) view of the viable  $g_{L, R}^{e \mu }$ parameter space.}
    \label{fig:g_e_muLR}
\end{figure}

We now consider how the new $Z^\prime$ contribution(s) can further help addressing the recently identified tension in  
$\Delta a_e^\text{Cs}$ (and $\Delta a_e^\text{Rb}$), firstly studying it independently of $\Delta a_\mu$. 
In Fig.~\ref{fig:g_e_muLR} we thus display the viable regimes in the plane spanned by $e-\mu$ couplings, 
($g_{L}^{e\mu}- g_{R}^{e \mu}$), again setting $m_{Z^\prime} = 10$~GeV (and fixing the remaining couplings to zero). As can be readily seen from the left panel, the current tension arising from the determination of $\alpha_e^\text{Cs}$ cannot be accounted for with same-sign couplings due to conflicts with the experimental bounds on LFU $Z$ decays (i.e. $R^Z_{\alpha\beta}$). 
While the latter bounds still allow for a light $Z^\prime$ explanation of $\Delta a_e^\text{Rb}$, the preferred regime of 
$g_{L,R}^{e \mu}$ is incompatible with the current upper bounds on cLFV Muonium oscillations. The most promising regions identified in the left panel of Fig.~\ref{fig:g_e_muLR} (logarithmic scale) are presented in detail in a linear scale in its right panel, in which we omitted the  $R^Z_{\alpha\beta}$ constraints (but highlight that these are indeed satisfied in all the regions displayed). 

Although we will return to this in the following subsection, 
the regimes of $g_{L,R}^{e \mu}$ favoured by an explanation of $\Delta a_e$ would also contribute to the muon $(g-2)$, in fact leading to a negative shift in its value, and thus worsening the discrepancy with respect to the SM prediction.

\begin{figure}[h!]
    \centering
    \mbox{\includegraphics[width=0.48\textwidth]{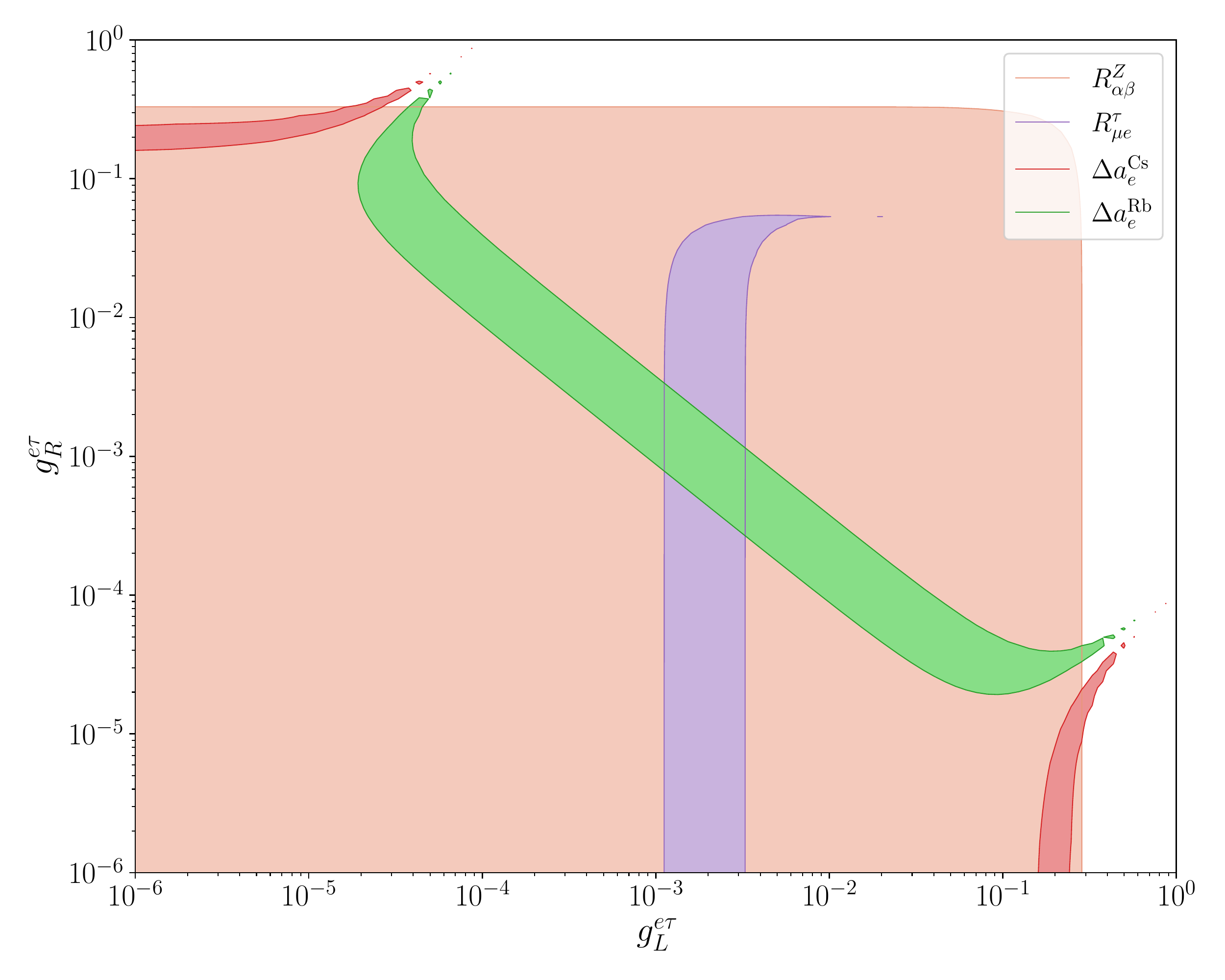}
    \includegraphics[width=0.48\textwidth]{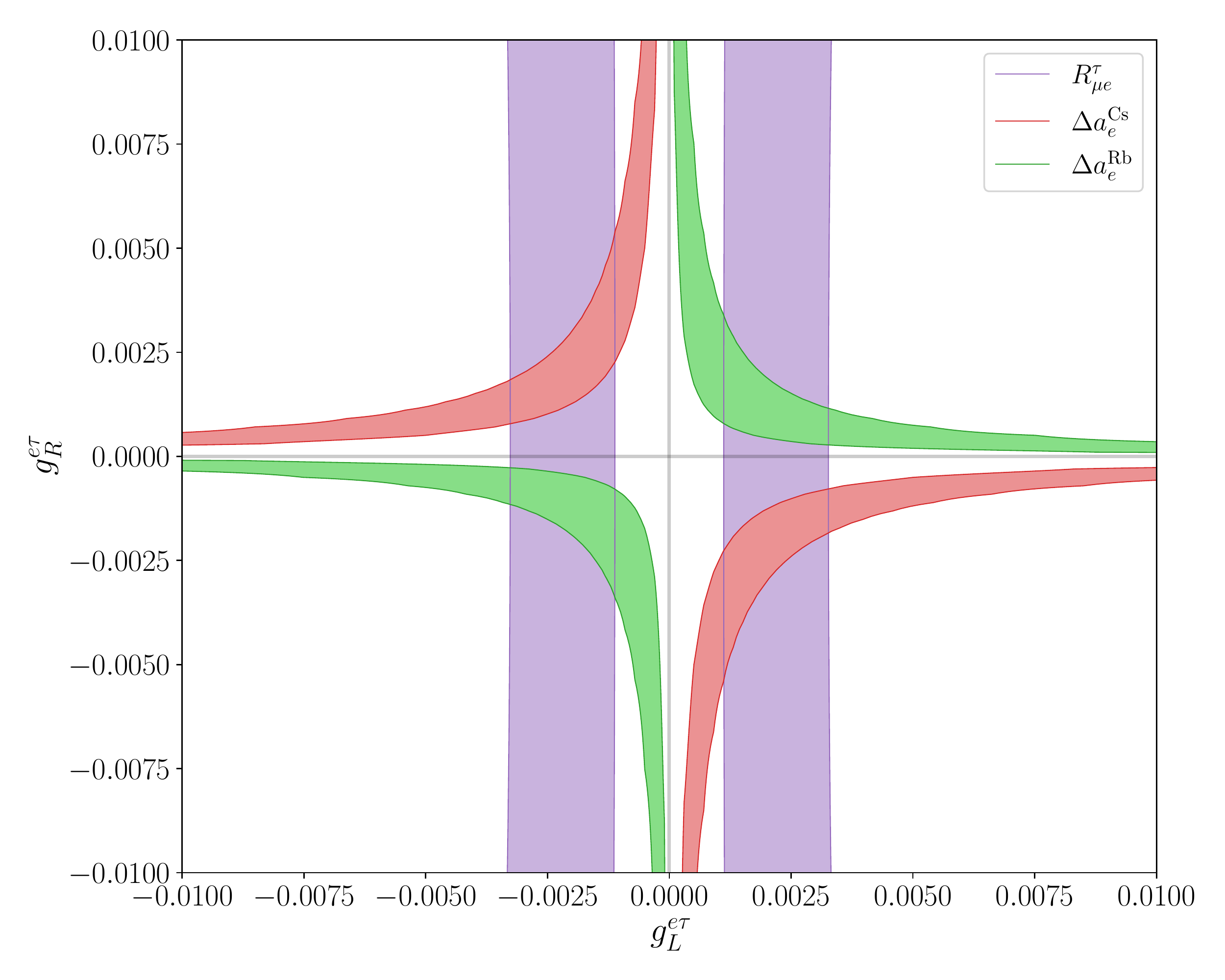}}
    \caption{Constraints on $g_{L,R}^{e \tau}$ couplings and prospects for $(g-2)_e$: the red and the green regions respectively denote the $1\,\sigma$ $(g-2)_e^\text{Cs}$ and $(g-2)_e^\text{Rb}$ favoured regimes.
    On the left, allowed regions in the same-sign ($g_{L}^{e \tau }$-$g_{R}^{e \tau}$) parameter space, in agreement with  the experimental bounds on LFU $Z$ decay ratios (light orange)
    and $R^\tau_{\mu e}$ (purple). 
    On the right, detailed (linear) view of the viable  $g_{L, R}^{e \tau }$ parameter space.
    }
    \label{fig:g_e_tauLR}
\end{figure}

\bigskip
\bigskip
Similar to what was done for $(g-2)_\mu$, one can also rely on 
 $Z^\prime - e\tau$ couplings  
(i.e. $g_{L,R}^{e \tau}$); the prospects for  $\Delta a_e^\text{Cs, Rb}$ are shown in Fig.~\ref{fig:g_e_tauLR}, which displays a view of the ($g_{L}^{e \tau}- g_{R}^{e\tau}$) plane, with all other couplings set to zero. As clearly seen from the plots, in order to saturate $\Delta a_e^\text{Cs}$ the couplings must have opposite signs (or one of them be zero); analogous to what was observed for $\Delta a_\mu$, 
$\Delta a_e^\text{Rb}$ can be explained relying on a combination of same-sign $g_{L,R}^{e \tau}$ couplings. Although all relevant cLFV and LFU observables were taken into account, the most constraining limits arise from $R^\tau_{\mu e}$. 

Interestingly, notice that such non-vanishing values of 
$g_{L,R}^{e \tau}$ also lead to contributions to $\Delta a_\tau$; these are negative, and typically $\mathcal{O}(10^{-7})$ (in fact much smaller than the uncertainty associated with the SM prediction for  $(g-2)_\tau$).

\mathversion{bold}
\subsection{A joint explanation for $\Delta a_\mu$ and $\Delta a_e$?}
\mathversion{normal}
After having independently considered the new contributions to the muon and to the electron anomalous magnetic moments (taking non-vanishing individual couplings at a time), 
one must now address the possibility of a joint explanation to both tensions. 

\begin{figure}[h!]
    \centering
    \mbox{\includegraphics[width=0.48\textwidth]{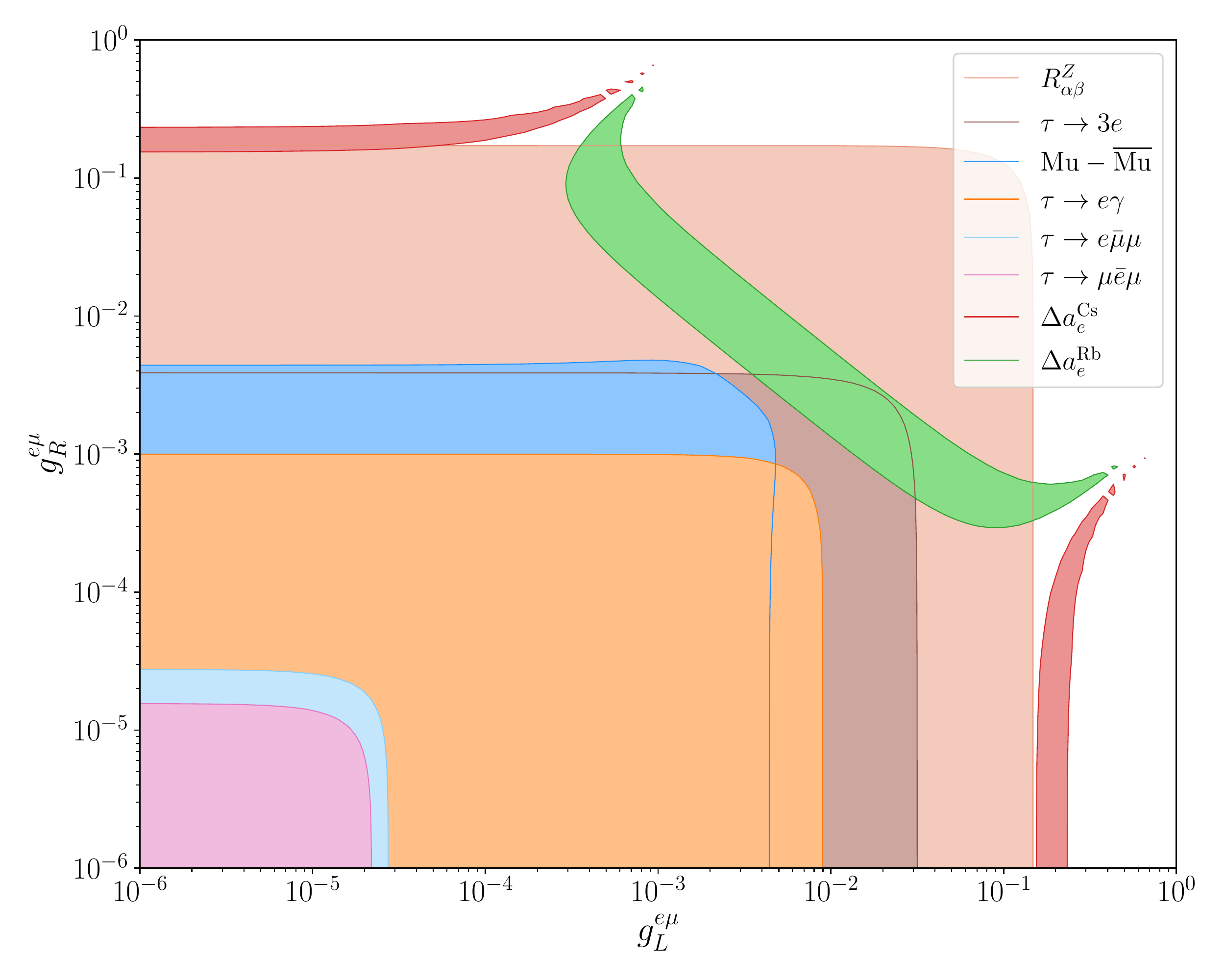}\includegraphics[width=0.48\textwidth]{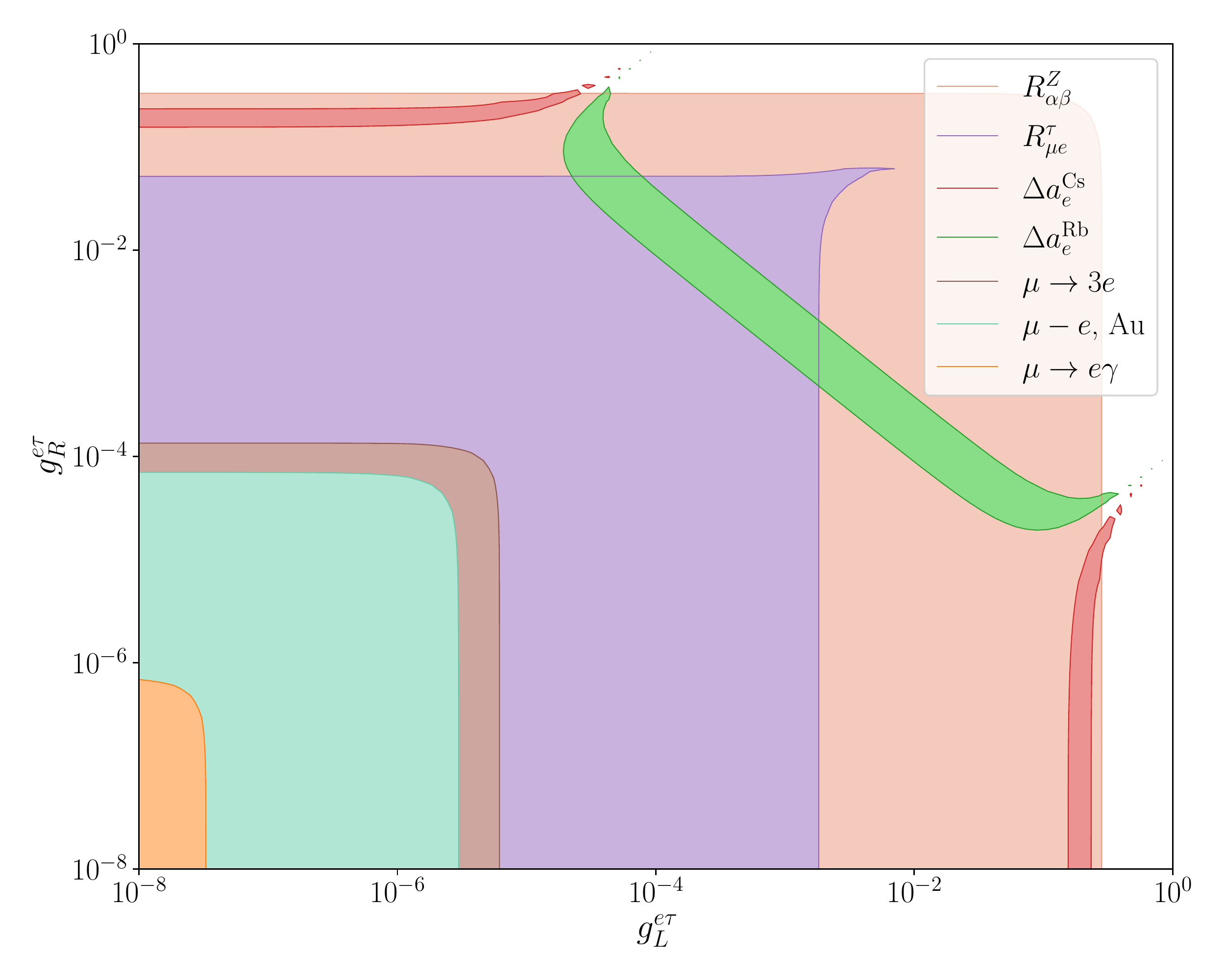}}
    \caption{Constraints on the $g_{L,R}^{e \mu}$ (left) and $g_{L,R}^{e \tau}$ (right) couplings and prospects for $(g-2)_e$ for fixed values of $g_{L,R}^{\mu \tau}$.
    On the left, allowed regions in the ($g_{L}^{e \mu }$-$g_{R}^{e \mu}$) parameter space, in agreement with constraints from cLFV $\tau$ decays ($\tau \to 3e$ in brown, $\tau \to e \gamma$ in orange, $\tau \to e \bar\mu \mu$ in light blue and $\tau \to \mu \bar{e} \mu$ in pink) and Muonium oscillations (blue); the red and the green regions respectively denote the $1\, \sigma$ $(g-2)_e^\text{Cs}$ and $(g-2)_e^\text{Rb}$ favoured regimes.
    On the right, allowed regions in the 
    ($g_{L}^{e \tau }$-$g_{R}^{e \tau}$) parameter space, in agreement with $R_{\mu e}(\tau \to \ell \nu \bar\nu)$ (purple), $\mu \to 3e$ (brown), $\mu \to e \gamma$ (orange) and $\mu-e$ conversion in Gold
    (turquoise); the red and the green regions respectively correspond to the $1\,\sigma$ $(g-2)_e^\text{Cs}$ and $(g-2)_e^\text{Rb}$ favoured regimes.}
    \label{fig:g_e_fix_mu_tau}
\end{figure}

In order to do so, and in view of the very restricted regimes for the
$g_{L,R}^{\mu \tau}$ couplings which allowed explaining 
$\Delta a_\mu$, we thus fix $g_{L,R}^{\mu \tau}$ as to comply with $(g-2)_\mu$ 
(we set $g_{L}^{\mu \tau}=0.0024$ and $g_{R}^{\mu \tau}=0.036$, see Fig.~\ref{fig:g_mu_tauLR}), and vary $g_{L,R}^{e \mu (\tau)}$. 
The prospects for a joint explanation of $\Delta a_\mu$ and $\Delta a_e$ are depicted in Fig.~\ref{fig:g_e_fix_mu_tau}: on the left (right) panel we explore the viable regimes for the
$g_{L,R}^{e \mu}$ ($g_{L,R}^{e \tau}$) couplings. 
In addition to the constraints already identified in Figs.~\ref{fig:g_e_muLR} and~\ref{fig:g_e_tauLR}, the simultaneously presence of $g_{L,R}^{e \mu} \neq 0$ 
(or $g_{L,R}^{e \tau} \neq 0$) and 
$g_{L,R}^{\mu \tau} \neq 0$ opens the door to sizeable contributions to cLFV processes: 
in the left panel of Fig.~\ref{fig:g_e_fix_mu_tau}, it is visible how the previously most stringent process (i.e. Muonium oscillations) is now strikingly superseded by various cLFV tau decays. 
In particular,   
$\tau \to e \bar\mu \mu$ and $\tau \to \mu \bar{e} \mu$ directly exclude any regime with 
$g_{L,R}^{e \mu} \gtrsim 10^{-5}$. 

Similar results are obtained when jointly considering the effects of $g_{L,R}^{e \tau} \neq 0$ and 
$g_{L,R}^{\mu \tau} \neq 0$: as displayed in the right panel of Fig.~\ref{fig:g_e_fix_mu_tau}, the constraints arising from rare cLFV muon decays are significantly more restrictive than those arising from  
$R^\tau_{\mu e}$. Three-body muon decays already constrain 
$g_{L(R)}^{e \tau} \lesssim 10^{-6 (-5) }$, 
with the muon cLFV radiative decay further imposing  $g_{L(R)}^{e \tau} \lesssim 10^{-8 (-7) }$.

In summary, for $m_{Z^\prime} = 10$~GeV, an excellent fit to the data (saturating $\Delta a_\mu$, and complying with $R_{\alpha\beta}^Z$, $R_{\mu e}^\tau$) can be found for 
\begin{equation}\label{eq:summ:gLRmutau}
    g_L^{\mu\tau} \simeq (2.4 \pm 0.5)\times 10^{-3}\,,
    \quad 
    g_R^{\mu\tau} \simeq 0.036 \pm 0.013\,;
\end{equation}
the bounds from $\tau\to\mu \bar{e}\mu$ and $\mu\to e \gamma $ then respectively imply the following upper limits for the combinations of couplings,
\begin{equation}\label{eq:summ:gLRmutau:cLFV}
    \sqrt{(g_L^{e\mu})^2 + (g_R^{e\mu})^2}\lesssim 10^{-5} \,,
    \quad 
    \sqrt{(g_L^{e\tau})^2 + (g_R^{e\tau})^2}\lesssim 4 \times 10^{-8}\,.
\end{equation}

The above discussion strongly suggests that a joint explanation\footnote{We have also considered complex couplings as a means to evade certain cLFV constraints. While in certain new physics models this is indeed possible (see e.g.~\cite{Abada:2021zcm}), the suppression/enhancement by CP violating phases relies on interference effects, and such effects are not present in this BSM construction.}  to the tensions in the light charged leptons anomalous magnetic moments is clearly precluded in view of the cLFV constraints. 

Should the current hypothetical model of a light $Z^\prime$ with flavour violating couplings to charged leptons be indeed an explanation to $\Delta a_\mu$, then one is led to mostly SM-like scenarios to both $(g-2)_e$ and $(g-2)_\tau$. 

\subsection{Heavier mediator regimes}\label{sec:amuae:heavier}
For simplicity, we have so far considered an illustrative value for the mass of the $Z^\prime$. In order to complete the study, it is important to discuss to which extent the conclusions of the previous subsections hold for other regimes of  $m_{Z^\prime}$. 

\begin{figure}[h!]
    \centering
    \mbox{\includegraphics[width=0.48\textwidth]{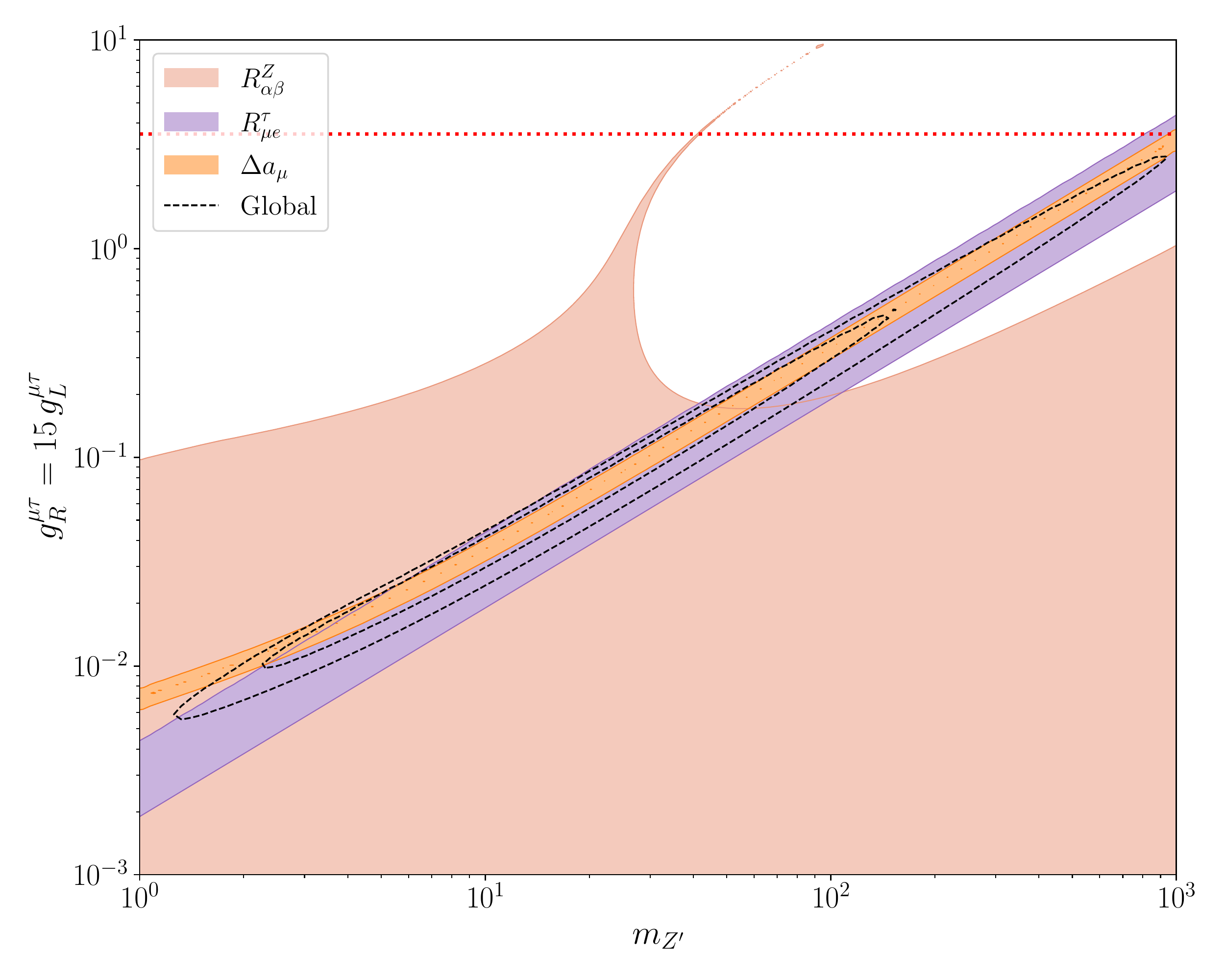}}
    \caption{Preferred regions by $(g-2)_\mu$ in the $(g_{R}^{\mu \tau}-m_{Z^\prime})$ parameter space, under the assumption 
    $g_{R}^{\mu \tau} \sim 15 \times g_{L}^{\mu \tau}$. The colour scheme denotes regions in agreement with  the experimental bounds on LFU $Z$ decay ratios (light orange) and $R^\tau_{\mu e}$ (purple). The dashed curves correspond to the $1\, \sigma$ and $2\,\sigma$ global fits (see text). The dotted horizontal line denotes the perturbativity limit for the $g_{R}^{\mu \tau}$ couplings. }
    \label{fig:g_mu_tau_masses}
\end{figure}

We thus explore regimes for the $Z^\prime$ couplings allowing to account for $\Delta a_\mu$ (i.e.  $g_{L,R}^{\mu \tau}$), for varying masses of the mediator. 
Relying on the findings of Section~\ref{subsec:amu}, we set  
$g_{R}^{\mu \tau} \sim 15 \times g_{L}^{\mu \tau}$, now for 
$m_{Z^\prime} \in [2\text{ GeV}, \, 1\text{ TeV}]$ (the lower bound allowing to escape the constraints that emerge for very light states below the $m_\tau$ threshold, especially $\tau \to \mu Z^\prime$~\cite{Foot:1994vd,Heeck:2016xkh,Altmannshofer:2016brv}, as discussed at the beginning of this section). In Fig.~\ref{fig:g_mu_tau_masses} we display information similar to that summarised in Fig.~\ref{fig:g_mu_tauLR}, presenting the best global fit regions ($1$ and $2\, \sigma$) in the model's parameter space now spanned by 
$g_{R}^{\mu \tau}$ and $m_{Z^\prime}$. As before, we colour-code the regimes in agreement with the most constraining bounds (for 
$g_{L,R}^{\mu \tau}$ couplings, the LFU ratios $R^\tau_{\mu e}$ and $R^Z_{\alpha \beta}$). 
The results displayed suggest that light $Z^\prime$ mediators, with masses $m_{Z^\prime} \in [10\text{ GeV}, \, 200\text{ GeV}]$, with associated couplings $0.01 \lesssim g_{R}^{\mu \tau} \lesssim 1$ offer the best prospects to account for the current 
tensions in $\Delta a_\mu$.

\begin{figure}[h!]
    \centering
    \includegraphics[width=0.48\textwidth]{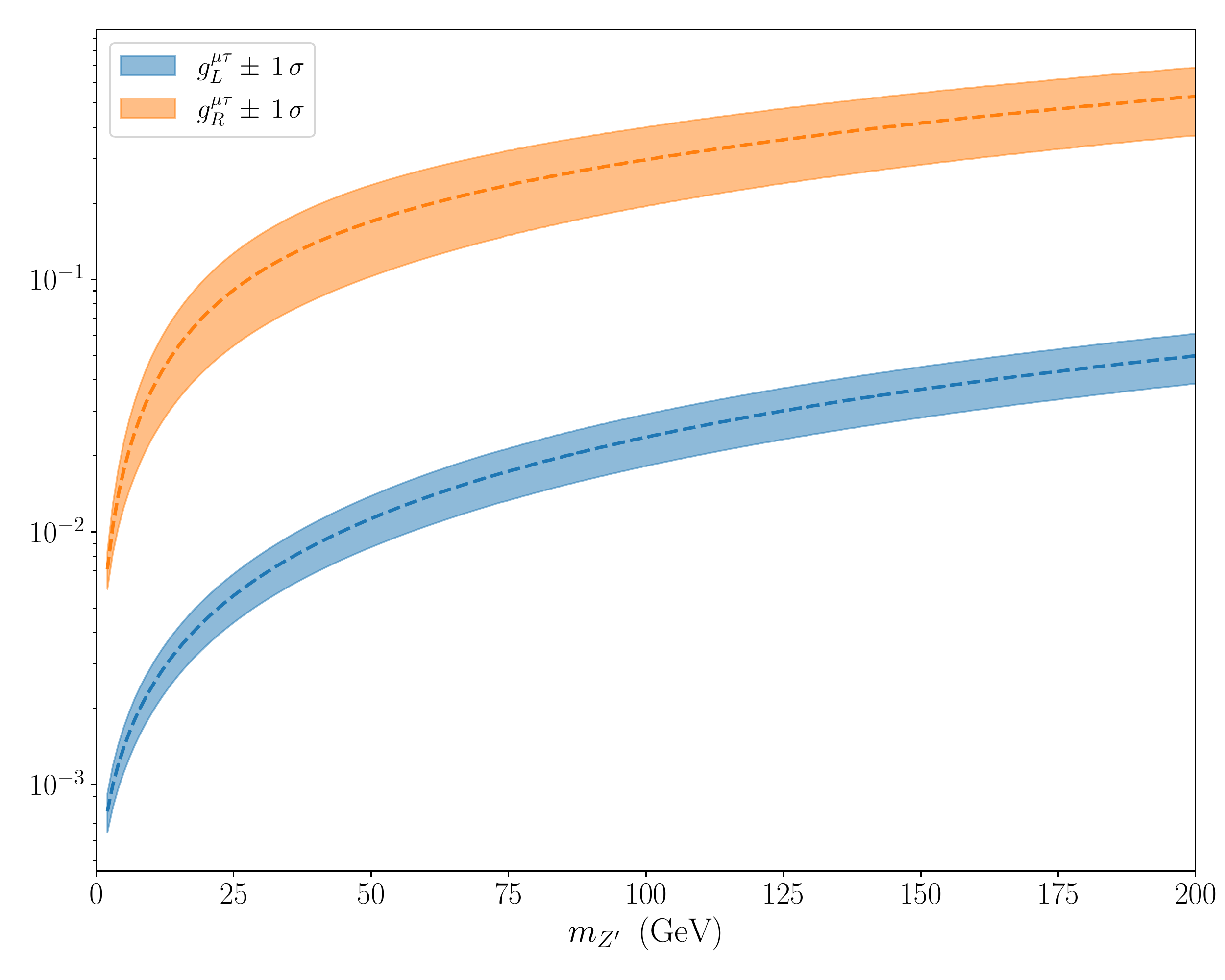}
    \includegraphics[width=0.48\textwidth]{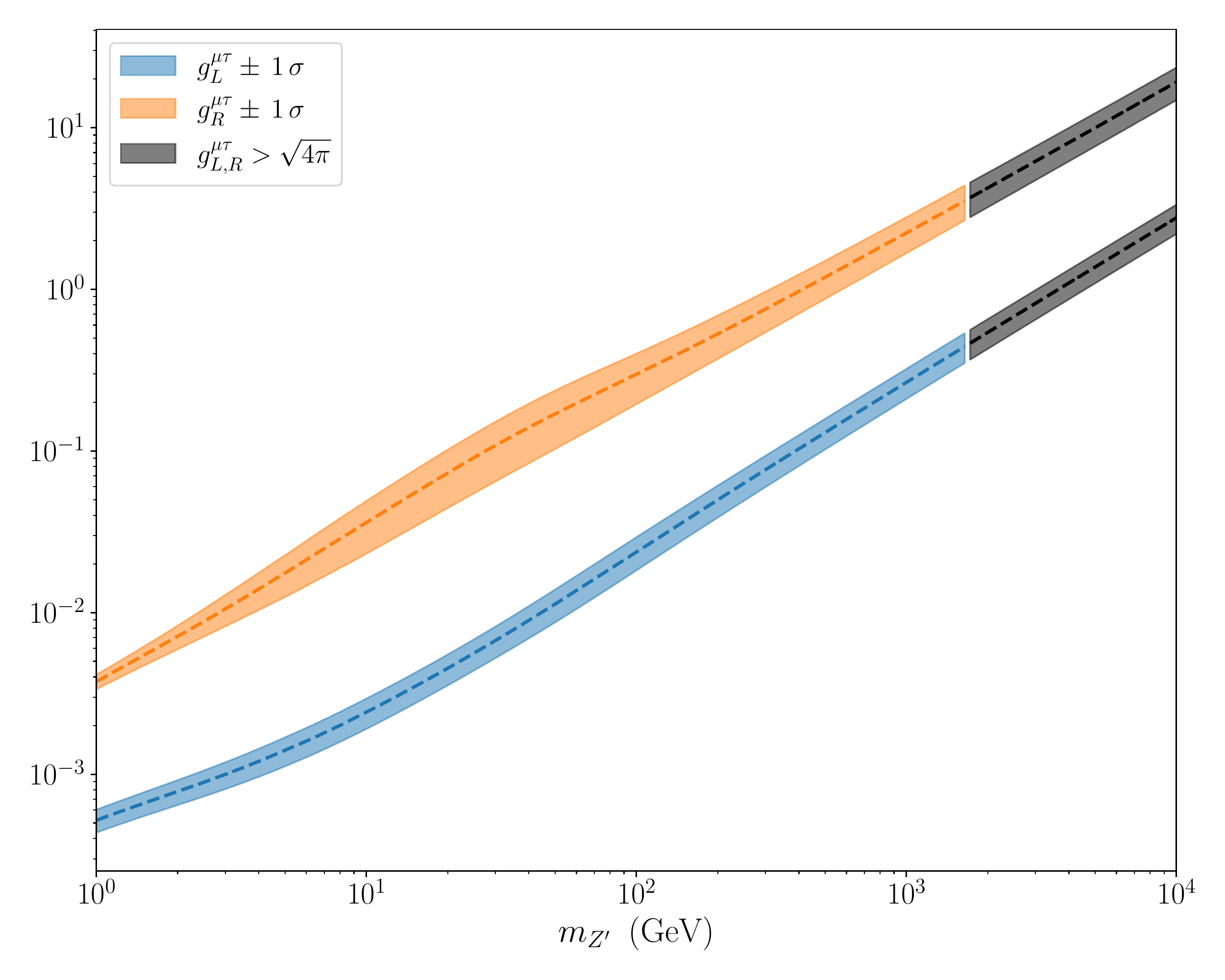}
    \caption{Fits of $g_{L,R}^{\mu\tau}$ as a function of $m_{Z^\prime}$. In blue (orange), $1\,\sigma$ estimate of $g_{L(R)}^{\mu\tau}$; the dashed lines denote the best fit value of the couplings for a given $m_{Z^\prime}$. On the right panel, regimes for which the fit is not satisfactory (non-perturbative couplings) are represented in grey.
    }
    \label{fig:fit_res}
\end{figure}

Another insight on the best-fit regimes of the model's parameter space, favoured by an explanation to $(g-2)_\mu$, is presented in Fig.~\ref{fig:fit_res}, in which $g_{L,R}^{\mu \tau}$ are independently fitted for varying values of  $m_{Z^\prime}$ (for details on the fit procedure, see the beginning of this section).

On the left panel of Fig.~\ref{fig:fit_res}, we present the results of the best-fit for $g_{L,R}^{\mu \tau}$ taking 
$m_{Z^\prime}$ to lie in the range $[2~\text{GeV}, \, 200~\text{GeV}]$.
As can be seen, and despite having taken independent input values for the couplings, the fit leads to a correlation between $g_{L}^{\mu \tau}$ and 
$g_{R}^{\mu \tau}$, strengthening the original findings (and 
underlying assumption leading to Fig.~\ref{fig:g_mu_tau_masses}). 
For completeness, we present in the right panel of Fig.~\ref{fig:fit_res} a similar study, but now including a much wider interval for $m_{Z^\prime}$, up to 10~TeV. However, for mediator masses $\mathcal{O}(1~\text{TeV})$, one clearly runs into a scenario for which no satisfactory fit can be achieved (as the couplings become non-perturbative); in turn this allows to infer a ``soft" limit for the validity of the model, $m_{Z^\prime} \sim \mathcal{O}(1~\text{TeV})$.

\section{Prospects for cLFV probes}\label{sec:cLFVprospects}
\begin{figure}[h!]
    \centering
    \mbox{\includegraphics[width=0.51\textwidth]{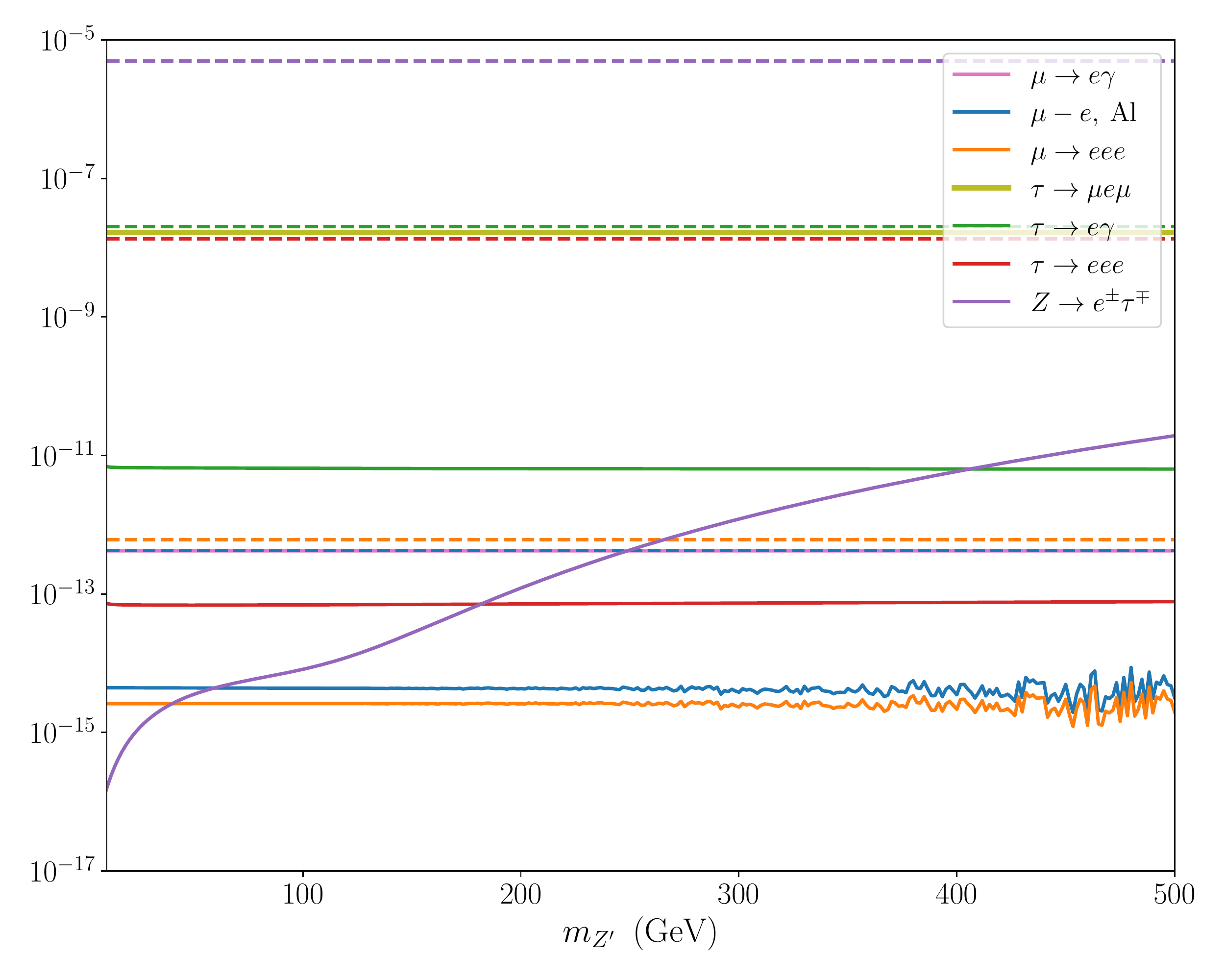}
    \includegraphics[width=0.51\textwidth]{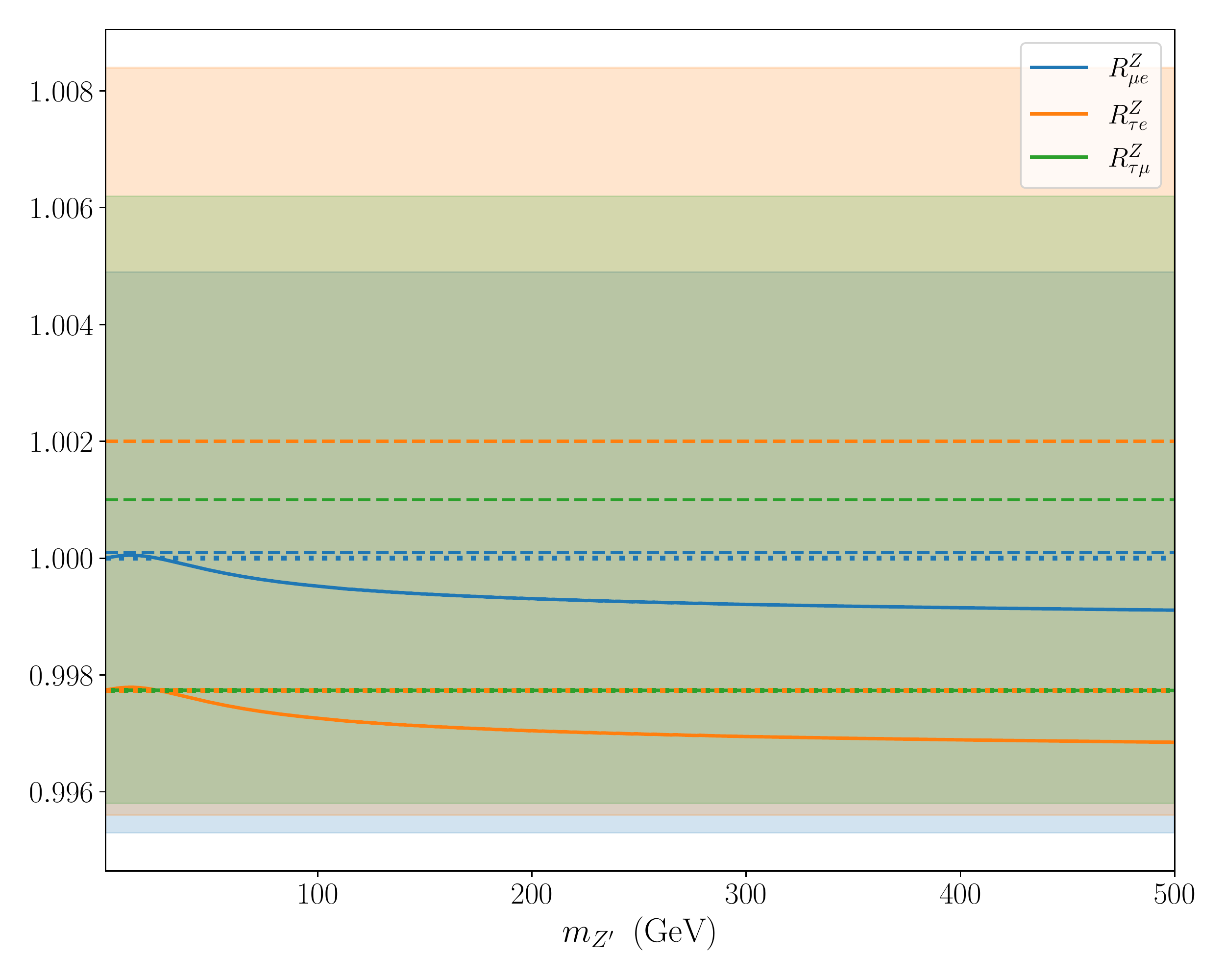}}
    \caption{Illustrative theoretical upper limits for several cLFV decays (left) and predictions for LFU in $Z$ decays (right) for a wide range of $Z^\prime$ masses. On the left, full (dahed) lines denote the predicted upper limits (current experimental bounds) for the distinct cLFV observables. On the right panel, the coloured full (dotted) lines denote the model's expected upper limit (SM prediction), while the dashed lines and shaded regions correspond to the central experimental values and associated $2\, \sigma$ uncertainties.
    }
    \label{fig:upper_limits}
\end{figure}

Interestingly, one can also estimate the (theoretical) upper bounds for several observables, arising from regimes of $g_{L,R}^{\mu\tau}$ and $m_{Z^\prime}$ saturating $\Delta a_\mu$,
while in agreement with all other constraints.
The most relevant predictions are collected in Fig.~\ref{fig:upper_limits}, in which we display the projections for the maximal values of several observables as a function of  
 $m_{Z^\prime}$. The curves are obtained by fitting the $\mu-\tau$ couplings for the different $Z^\prime$ masses, to account for $\Delta a_\mu$ and respecting the constraints from the ratios $R^Z_{\alpha\beta}$ and $R^\tau_{\mu e}$. We then fit the other couplings, $g^{e \tau}_{L,R}$ and $g^{e \mu}_{L,R}$, in order to drive the value of the most stringent cLFV processes (i.e. $\mu \to e \gamma$ and $\tau \to \mu \bar{e} \mu $) to their current experimental upper bounds (see  Table~\ref{tab:cLFVdata}).
 
These illustrative theoretical predictions (which do not take into account effects of operator mixing due to RG evolution for large $m_{Z^\prime}>\Lambda_\text{EW}$) should be compared with the future sensitivity of the associated dedicated searches, see Table~\ref{tab:cLFVdata}.

As can be seen on the left panel\footnote{For larger values of $m_{Z'}$, the behaviour of the predictions for $\mu\to eee$ and muon-electron conversion is due to numerical instabilities of the fit maximising the contributions to $\mu\to e\gamma$ (that is the couplings $g_{L,R}^{e\tau}$) and not a physical effect.} of Fig.~\ref{fig:upper_limits}, 
and while the maximal predictions for the cLFV tau decays\footnote{The maximal predictions for $\tau\to \mu \gamma$ decays appear to be constant (i.e., independent of $m_{Z^\prime}$) since the relevant couplings are those also responsible for $\tau \to \mu \bar e \mu$. Since the latter rate is fixed at its current limit for each value of $m_{Z^\prime}$, one recovers the observed behaviour.} in general lie beyond experimental sensitivity (at best, BR($\tau\to eee$)$\sim \mathcal{O}(10^{-13})$ 
), one has very good prospects for the observation of certain rare muon decays. This is the case of both 
$\mu\to e\gamma$, and muon-electron conversion in nuclei: the former maximal branching ratio is fixed to its current experimental upper limit ($\leq4.2\times 10^{-13}$), and the maximal conversion rate for Aluminium nuclei $\sim  \mathcal{O}(10^{-17,  -16})$, potentially within future experimental sensitivity.
The current experimental limit for $\mathrm{BR}(\mu\to e\gamma)$ also precludes sizeable rates for $\mu\to eee$ (within future sensitivity), such that a future observation of both $\mu\to e\gamma$ and $\mu\to eee$ at comparable rates would falsify $Z^\prime$ models under discussion\footnote{
Notice that the cLFV observables discussed, and in particular their falsifying role, generically probe the presence of a new leptophilic $Z^\prime$ with flavour violating couplings (combinations of  $g^{\mu \tau}$ with either $g^{\mu e}$ or $g^{e \tau}$), and not directly the potential of such constructions to offer an explanation to $\Delta a_\mu$.}.
Also notice that the (full) lines for $\mu \to e\gamma$ and $\tau \to \mu \bar e \mu$ appear superimposed with the corresponding dashed lines denoting the associated experimental bound; this is a consequence of maximising the couplings to be within $1 \sigma$ of the current experimental bounds.

For illustrative purposes, we also display the prospects for the maximal contributions to the cLFV $Z$ decays; as can be seen, and for $m_{Z^\prime} \approx 500$~GeV, one has BR($Z\to e^\pm\tau^\mp$)$\sim 10^{-10}$, marginally at future FCC-ee sensitivity~\cite{Abada:2019lih}.
Although not included here, let us notice that the maximal expected rates for the other cLFV $Z$ decays are 
BR($Z\to \mu^\pm\tau^\mp$)$\sim 10^{-19}$ and BR($Z\to e^\pm\mu^\mp$)$\sim 10^{-16}$.

On the right-handed panel of Fig.~\ref{fig:upper_limits} we display the prospects of the maximal expected deviations in what concerns LFUV ratios of $Z\to \ell \ell$ decay widths, together with the SM predictions and the experimental data. 
It is important to notice that, with the exception of $R_{\tau\mu}^Z$, the $Z$-decay universality ratios are generically predicted to be slightly smaller than in the SM. This is due to sizeable interference effects between the $Z'$-loop contribution and the SM tree-level diagrams.

\mathversion{bold}
\section{A (light) flavour violating $Z^\prime$ at a future muon collider}\label{sec:ZpMuon}
\mathversion{normal}
In addition to the vast array of indirect (and direct, low-energy) searches for a new $Z^\prime$ boson, its presence has also been the object of dedicated programmes at high-energy colliders, from LEP~\cite{ALEPH:2006jhv} to current efforts at the LHC~\cite{ATLAS:2015dva,ATLAS:2019fgd,CMS:2019gwf,CMS:2021ctt,ATLAS:2019erb}. Likewise, extensive work had been done in what concerns the prospects for the discovery of such a NP mediator in the (near) future, in particular at (HL)-LHC~\cite{Iguro:2020qbc,Altmannshofer:2016brv}, Belle II~\cite{Iguro:2020rby} and FCC-ee~\cite{Altmannshofer:2016brv}.
In recent years, an increasing interest for future muon colliders has been put forward by the high-energy particle physics community~\cite{EuropeanStrategyforParticlePhysicsPreparatoryGroup:2019qin, MuonCollider:2022xlm,Aime:2022flm,InternationalMuonCollider:2022qki}, and the prospects for the discovery of a $Z^\prime$ in such facilities have been also considered (see~\cite{Huang:2021nkl}). 

In general, lepton flavour violating $Z^\prime$ bosons can be produced at colliders in processes such as $f\bar f
\to \mu^\pm\mu^\pm\tau^\mp\tau^\mp$, with the striking final state signature of two same-sign lepton  pairs~\cite{Altmannshofer:2016brv,Iguro:2020qbc,Iguro:2020rby}.
In order to complement the discussion, we now address the possible smoking guns of a leptophilic flavour violating $Z^\prime$ at muon colliders. In this section we thus explore possible  off-resonance signatures arising from the presence of a cLFV $Z^\prime$ at $\mu^+\mu^-$ colliders, 
discussing both the associated production cross sections, as well as forward-backward asymmetries.

Here, we neglect the initial muon masses and therefore  Higgs-exchange diagrams.
Furthermore, effects of initial and final state radiation corrections are not taken into account as they are beyond the scope of the present study.
For a detailed discussion of the importance of these effects see e.g.~\cite{Blumlein:2022mrp} and references therein.

\mathversion{bold}
\subsection{cLFV $Z^\prime$ production at a muon collider }
\mathversion{normal}
In what follows, we consider distinct scenarios, both concerning the operating centre of mass energy (i.e. $\sqrt s$) and $m_{Z^\prime}$. In each case, the couplings are determined as to comply with the best-fit regimes for the flavoured observables so far considered: saturating $\Delta a_\mu$, and complying with all the relevant cLFV and LFUV bounds. 
In view of the discussion of the previous sections, the very stringent associated constraints lead to tiny $Z^\prime-e-\ell$ couplings (with $\ell = \mu, \, \tau$); we thus focus on tau-pair production at a future muon collider.  

In the present model, the process $\mu^+\mu^-\to \tau^+\tau^-$ receives contributions from SM $\gamma$ and $Z$ boson exchanges ($s$-channel) as well as from $t$-channel $Z^\prime$ exchange,  
leading to the following matrix element:
\begin{equation}\label{eq:Msum}
    \mathcal M \,= \,\mathcal M_\gamma + \mathcal M_Z - \mathcal M_{Z^\prime}\,,
\end{equation}
in which we highlight the sign difference between the SM ($s$-channel) contributions and the $Z^\prime$ $t$-channel one.
The differential cross section is given by
\begin{eqnarray}
     \dfrac{d\sigma}{d\cos \theta} \, =\, \dfrac{1}{32\pi\, s} \sqrt{1 - 4\,\dfrac{m_\tau^2}{s}} \,|\mathcal{M}|^2 \, ,
\end{eqnarray}
with $\mathcal{M}$ defined in Eq.~(\ref{eq:Msum}), and the details of the computation being collected in Appendix~\ref{app:xsec}.
In the above, $\theta$ is the final state lepton angle with respect to the colliding muon direction in the di-tau centre  of mass frame. 
Concerning the phase space integration, we employ a (realistic) angular cut of $-0.99 \leq \cos\theta \leq 0.99$ to account for the finite detector volume.
We stress here that $\theta$ does not correspond to a reconstructed angle from a detector simulation and $\tau$ identification.
 
\begin{figure}[h!]
    \centering
    \includegraphics[width=0.48\textwidth]{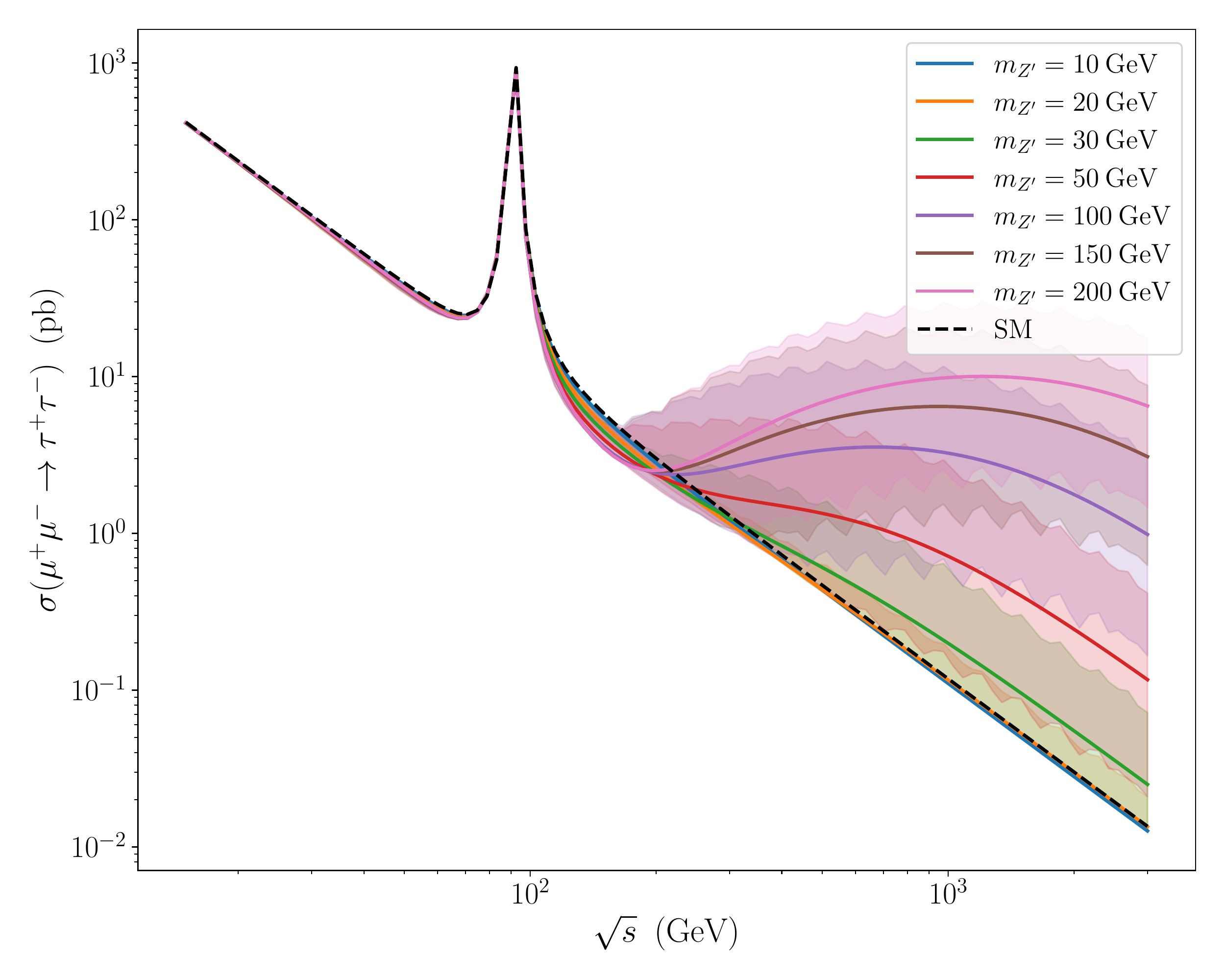}
    \caption{Prospects for $Z^\prime$ production at a muon collider: $\sigma(\mu^+\mu^-\to \tau^+\tau^-)$ cross sections (in pb) as a function of $\sqrt{s}$, for different values of $m_{Z^\prime}$. A black dashed line denotes the SM prediction (at leading order), while the coloured solid lines correspond to distinct values of $m_{Z^\prime}$. The finite width of the coloured bands reflects the uncertainty associated with the fitting of the $g_{L,R}^{\mu\tau}$ couplings (see text for additional details).}
    \label{fig:sigmumutautau_vs_s}
\end{figure}

The results concerning the production cross section 
$\sigma(\mu^+\mu^-\to \tau^+\tau^-)$ as a function of $\sqrt{s}$
are presented in Fig.~\ref{fig:sigmumutautau_vs_s}, for different values of $m_{Z^\prime}$. The SM predictions are separately displayed - we notice that the computation of the latter is done only at leading order and that we neglect the muon mass. 

In all cases, the values of the couplings $g_{L,R}^{\mu\tau}$ are determined by fitting them to the likelihood containing $(g-2)_\mu$, the $Z$-decay LFU ratios $R^Z_{\alpha\beta}$ and the $\tau$-decay LFU ratio $R^{\tau}_{\mu e}$.
The results of these fits, for selected values of $m_{Z^\prime}$ are shown in Table~\ref{tab:couplings_masses}.
Subsequently, for a fixed mass $m_{Z^\prime}$ and $\sqrt{s}$, we sample the couplings around the best fit point according to the likelihood, in order to estimate the uncertainty of our predictions for $\mu^+\mu^-\to \tau^+\tau^-$ processes.
In the following, the shaded regions denote the uncertainty corridor corresponding to the interval between the $16^\text{th}$ and $84^\text{th}$ quantiles of the samples.

While for light $Z^\prime$ mediators (such as those considered in most of the numerical analysis of the previous section) the behaviour of the current BSM construction hardly deviates from the SM expectation, the situation becomes visibly different for heavier states. This is a direct consequence of fitting the $g_{L,R}^{\mu\tau}$ couplings (which thus vary for every value of $m_{Z^\prime}$), and these - as seen from Fig.~\ref{fig:fit_res} - 
significantly increase for $m_{Z^\prime} \gtrsim 30$~GeV, especially the right-handed couplings $g_{R}^{\mu\tau}$ (see Table~\ref{tab:couplings_masses} below). The latter also present the largest uncertainty in the fit, especially for 
$m_{Z^\prime} \lesssim 150$~GeV, as visible from Fig.~\ref{fig:fit_res}, and this accounts for the spread in the different (shaded) regions associated with the coloured curves. 

\renewcommand{\arraystretch}{1.3}
\begin{table}[h!]
    \centering
    \hspace*{-7mm}{\small\begin{tabular}{|c|c|c|}
    \hline
    $m_{Z^\prime}$ (GeV) &$g_L^{\mu\tau}$ & $g_R^{\mu\tau}$  \\
    \hline\hline
    $10$ & \: $0.0024 \pm 0.0005$ \: & \: $0.0360 \pm 0.0129 $ \:  \\
    \hline
    $ 20$ & $0.0045 \pm 0.0009$   & $ 0.0725 \pm 0.0284$  \\
    \hline
    $ 30 $ & $0.0067  \pm 0.0014$   & $0.1076 \pm 0.0431$  \\
    \hline
    $ 50$ & $0.0113  \pm 0.0025$   & $ 0.1699 \pm 0.0667$  \\
    \hline
    $ 100$ & $ 0.0237 \pm 0.0055$   & $0.2969  \pm 0.1025$  \\
    \hline
    $ 150$ & $ 0.0337 \pm 0.0083$   & $ 0.4143 \pm 0.1311$  \\
    \hline
    $ 200 $ & $ 0.0498 \pm 0.0112$   & $0.5288 \pm  0.1587$  \\
    \hline
    \end{tabular}}
    \caption{Best fit values and associated uncertainties of the $\mu-\tau$ couplings for different $Z^\prime$ masses; the fit includes $(g-2)_\mu, \, R^Z_{\alpha\beta}$ and $R^\tau_{\mu e}$ constraints.}
    \label{tab:couplings_masses}
\end{table}
\renewcommand{\arraystretch}{1.}

Studies of heavy $Z^\prime$ resonances ($m_{Z^\prime} \sim \mathcal{O}(\text{TeV})$) at the LHC have been recently pursued, also emphasising the constraining role of the forward-backward asymmetry\cite{Accomando:2015cfa}; however, due to $s$-channel exchange, the studied angular distributions exhibit a resonant behaviour at the pole of the new boson mass,  while deviations for lower $\sqrt{s}$, e.g. at around the $Z$-pole, are negligible.
In contrast, and as we will subsequently discuss in the present study, deviations in the forward-backward asymmetry $A_\text{FB}$ are quite significant already at lower energies, and considerably depart from the SM predictions at large $\sqrt{s}$, due to the $t$-channel exchange of the new boson.
The forward-backward (muon) asymmetry has also been explored in context of lepton flavour violating ($e\mu$) axion-like particles aiming at explaining $\Delta a_\mu$; these light mediators are also responsible for new $t$-channel contributions in $e^+ e^-\to \mu^+\mu^-$ scattering, searched for at the Belle II experiment~\cite{Endo:2020mev}.

\begin{figure}[h!]
    \centering
    \includegraphics[width=0.48\textwidth]{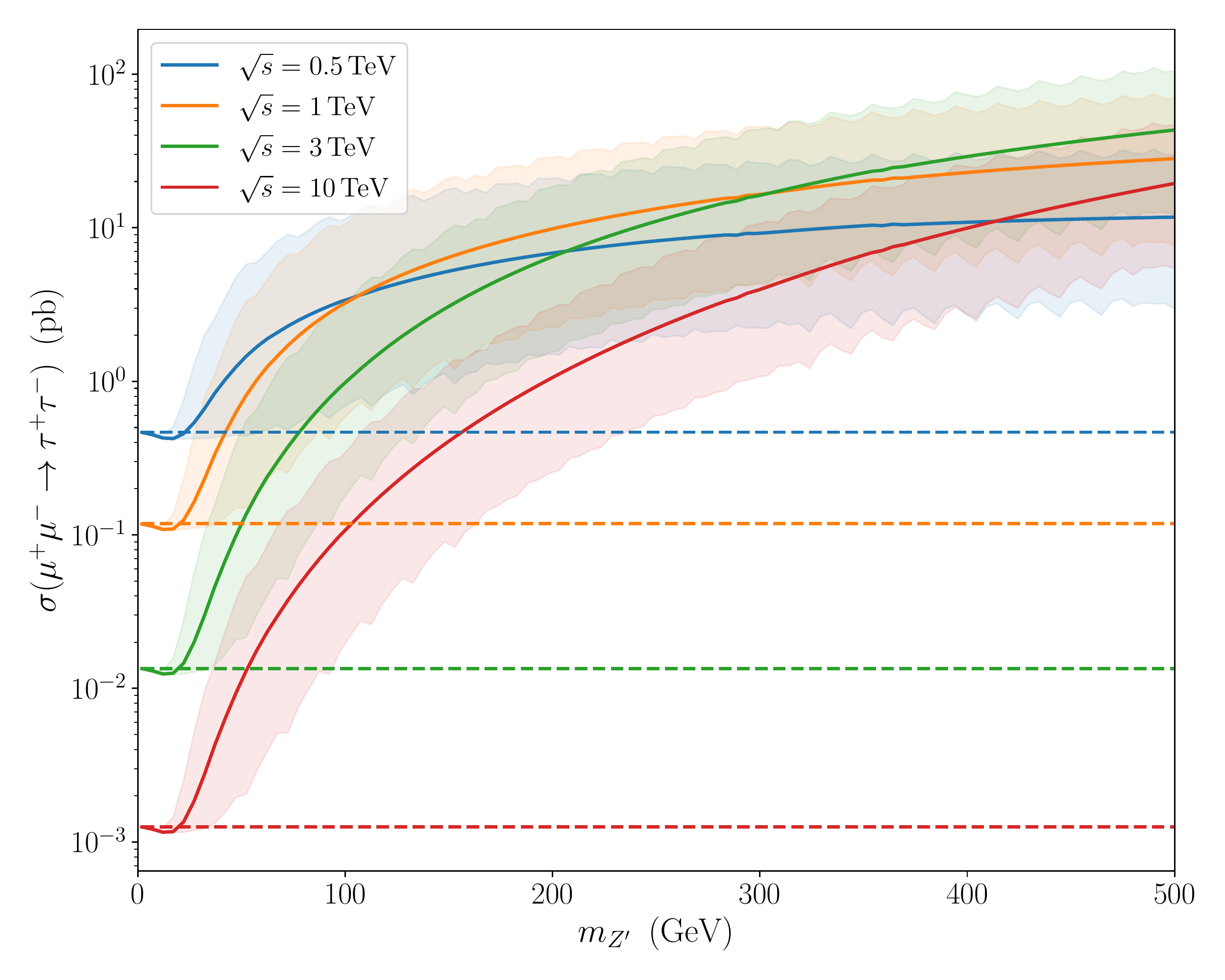}
    \includegraphics[width=0.48\textwidth]{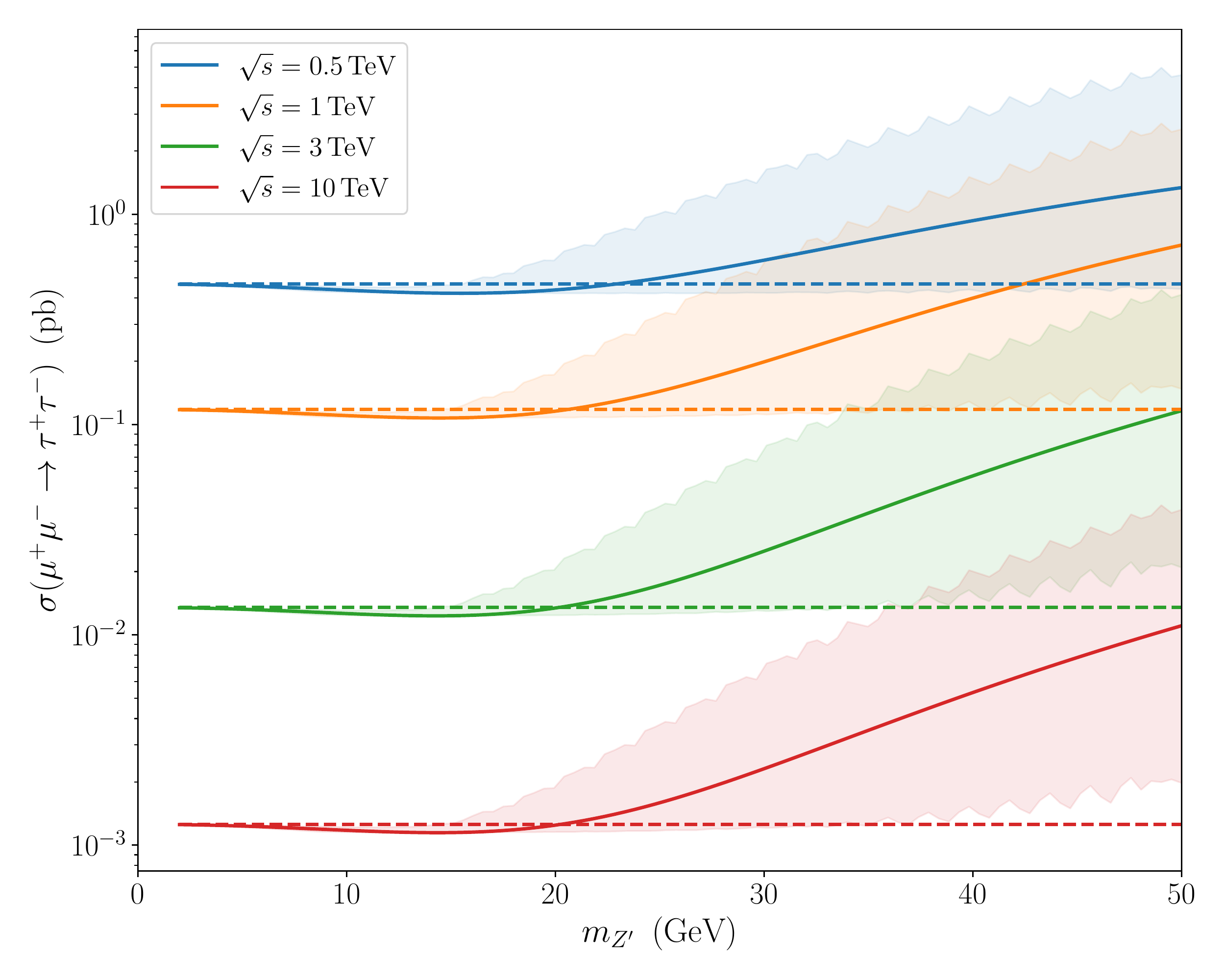}
    \caption{Tau-pair production cross sections, $\sigma(\mu^+\mu^-\to \tau^+\tau^-)$ as a function of $m_{Z^\prime}$ for different values of $\sqrt{s}$. The shaded regions correspond to the fit uncertainty, while the dashed lines denote the corresponding SM predictions (at leading order). The right panel offers a detailed view of the light regime for the $Z^\prime$ ($m_{Z^\prime} \lesssim 50$~GeV).}
    \label{fig:sigmumutautau_vs_mZp}
\end{figure}

Complementary insight can be obtained by considering the dependency of the $\sigma(\mu^+\mu^-\to \tau^+\tau^-)$ production cross section on $m_{Z^\prime}$, for different values of the centre of mass energy, $\sqrt{s}$. 
This is displayed on the panels of Fig.~\ref{fig:sigmumutautau_vs_mZp}, in which we again compare the SM predictions with those of the $Z^\prime$ extension under consideration.

As manifest from Fig.~\ref{fig:sigmumutautau_vs_mZp}, for sufficiently ``heavy" $Z^\prime$ bosons (i.e. $m_{Z^\prime} \gtrsim 30$~GeV) one has a clear departure from the SM prediction; for most choices of $\sqrt{s}$, and as already seen in Fig.~\ref{fig:sigmumutautau_vs_s}, 
$\sigma(\mu^+\mu^-\to \tau^+\tau^-) \gtrsim \mathcal{O}(\text{1~pb})$, rendering this observable highly sensitive to the presence of heavier $Z^\prime$ bosons; even with low statistics, a strong signal in conflict  with the SM prediction could be expected. This information is summarised in Table~\ref{tab:xsec}, in which
we collect the predictions for the $\mu^+\mu^-\to \tau^+\tau^-$ production cross section\footnote{We have also investigated the lepton flavour violating production cross section $\mu^+\mu^-\to \tau^\pm e^\mp$. However, our estimates show that at most one can reach $\mathcal{O}(10^{-9, -8})$~pb for $\sigma(\mu^+\mu^-\to \tau^+e^-)$ for an energy around $\sqrt{s} \sim \mathcal{O}(10^2)$ GeV, thus most likely eluding future observation.} for several values of $m_{Z^\prime}$ and $\sqrt{s}$. 
\renewcommand{\arraystretch}{1.3}
\begin{table}[h!]
    \centering
    \hspace*{-7mm}{\small\begin{tabular}{|c||c|c|c|c|}
     \hline
    \multicolumn{5}{|c|}{Cross section for the process $\mu^+\mu^-\to\tau^+\tau^- $ (pb)} \\
    \hline
    \hline
    $\sqrt{s}$ (TeV)  & $0.5$ & $1$  & $3$ & $10$ \\
    \hline
    SM &\:  $0.47$ \: &\:  $0.12$\:  & \: $\:\,0.014$ \: & \: $0.0013$\: \\
    \hline
    $m_{Z^\prime}=10$ GeV & \: $ 0.43$ \:  & \: $ 0.11 $\:  & \: $ 0.013 $\: & \: $0.0012 $\:  \\
    \hline
     $m_{Z^\prime}=20$ GeV & \: $ 0.44$ \:  & \: $ 0.12 $\:  & \: $ 0.013 $\: & \: $ 0.0012 $\:  \\
    \hline
     $m_{Z^\prime}=30$ GeV & \: $0.60 $ \:  & \: $ 0.20 $\:  & \: $ 0.025 $\: & \: $ 0.0023$\:  \\
    \hline
     $m_{Z^\prime}=50$ GeV & \: $ 1.35$ \:  & \: $  0.72$\:  & \: $ 0.118 $\: & \: $ 0.0112 $\:  \\
    \hline
     $m_{Z^\prime}=100$ GeV & \: $3.38 $ \:  & \: $ 3.24 $\:  & \: $ 0.982 $\: & \: $ 0.1074 $\:  \\
    \hline
     $m_{Z^\prime}=150$ GeV & \: $5.22 $ \:  & \: $ 6.43 $\:  & \: $ 3.080 $\: & \: $0.4075 $\:  \\
    \hline
     $m_{Z^\prime}=200$ GeV & \: $\: 6.84\:$ \:  & \: $\: 9.84 \:$\:  & \: $ 6.488 $\: & \: $ 1.0581$\:  \\
    \hline
    \end{tabular}}
    \caption{Predicted cross sections (pb) for different energies $\sqrt{s}$ and $Z^\prime$ masses, with $-0.99 \leq \cos\theta \leq 0.99$.}
    \label{tab:xsec}
\end{table}
\renewcommand{\arraystretch}{1.}

The benchmark values for $\sqrt{s} = 0.5,\,1,\,3,\,10$~TeV have been chosen in agreement with recent muon collider design proposals~\cite{Delahaye:2019omf, MuonCollider:2022xlm, Aime:2022flm, InternationalMuonCollider:2022qki}.
With the anticipated integrated luminosity of $\mathcal O(\mathrm{ab}^{-1})$~\cite{MuonCollider:2022xlm, Aime:2022flm, InternationalMuonCollider:2022qki}, the $\mu^+\mu^-\to \tau^+\tau^-$ cross section can be expected to be precisely measured at such a facility.

A more extensive study making use of the usual Monte-Carlo ``tool-chain" and detector simulation is left for subsequent studies.
\mathversion{bold}
\subsection{Forward-backward asymmetry $A_\text{FB}$}
\mathversion{normal}
The forward-backward asymmetry, $A_\text{FB}$, has been extensively used a powerful probe for the presence of NP in association with numerous production modes.
At LEP-1 and LEP-2, scans of $\sqrt{s}$ around the $Z$-pole have been performed, in which different final states were studied~\cite{ALEPH:2005ab}. Most of the results have been found to be in excellent agreement with the SM predictions.

Relying on several simplifying assumptions, we have evaluated how the cLFV $Z^\prime$ boson under study could lead to a departure from the SM predictions concerning the forward-backward asymmetry
associated with the $\mu^+\mu^-\to \tau^+\tau^-$ process at a future muon collider. 
We re-iterate that ours is strictly a simple (na\"ive)
tree-level calculation, without taking into account potentially relevant initial and final state radiation (ISR and FSR). Likewise, the results for $A_\text{FB}$ around the $Z$-pole mass are displayed strictly for illustrative purposes.
The forward-backward asymmetry is defined as
\begin{equation}
    A_\text{FB} \, =\, \frac{\sigma_\text{F}\, -\, \sigma_\text{B}}{\sigma_\text{F}\, +\, \sigma_\text{B}}\,,
\end{equation}
with 
\begin{equation}\label{eq:int:dsigma}
    \sigma_\text{F}\, =\, \int^\text{max}_0 
    \dfrac{d\sigma}{d\cos \theta}\, d\cos \theta\,,
    \quad 
    \sigma_\text{B}\, =\, \int^0_\text{min} 
    \dfrac{d\sigma}{d\cos \theta}\, d\cos \theta\,,
\end{equation}
in which the upper (lower) integral limits, ``max (min)", denote 
the extreme values of $\cos \theta$ considered.

\begin{figure}[h!]
    \centering
    \includegraphics[width=0.48\textwidth]{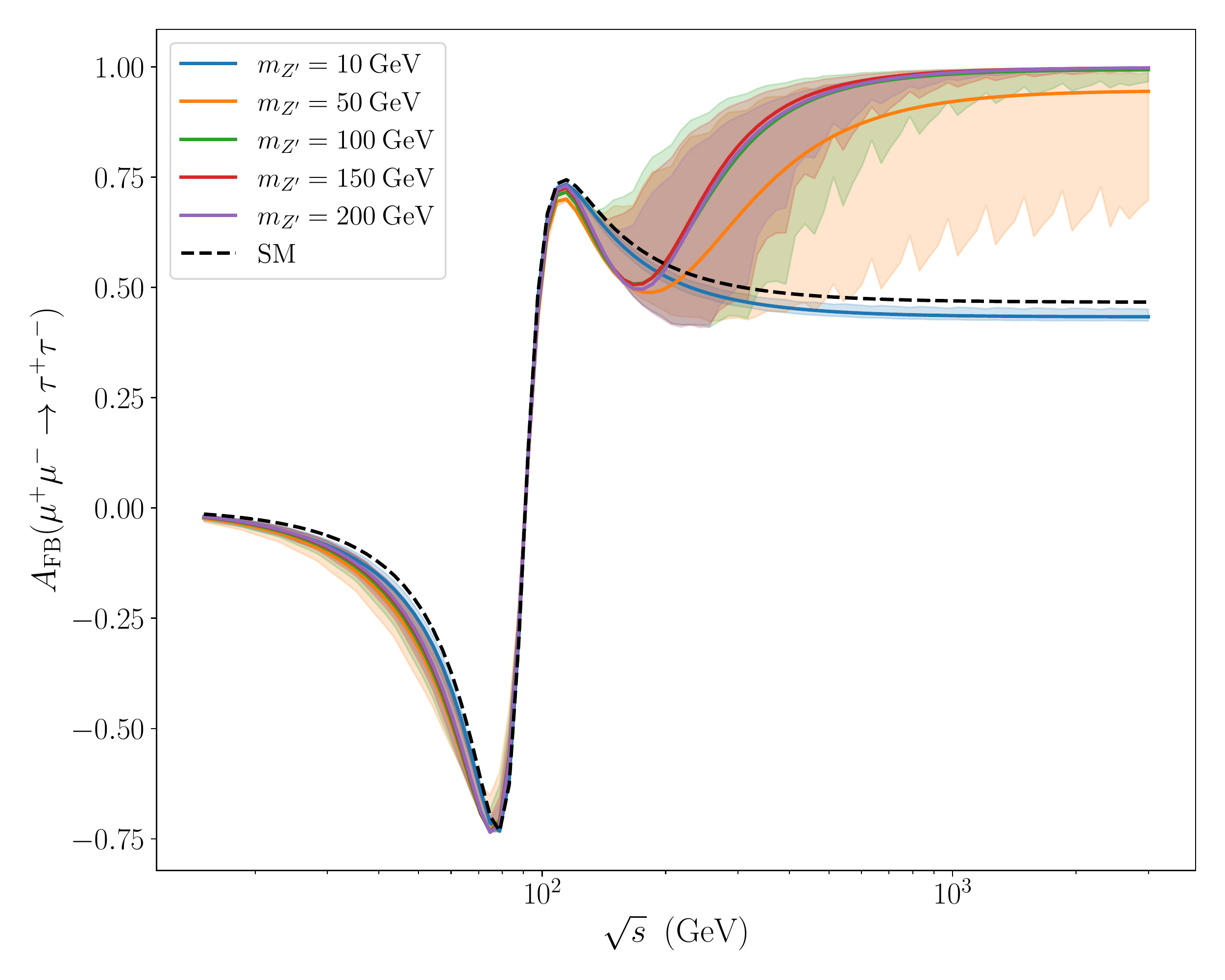}
    \includegraphics[width=0.48\textwidth]{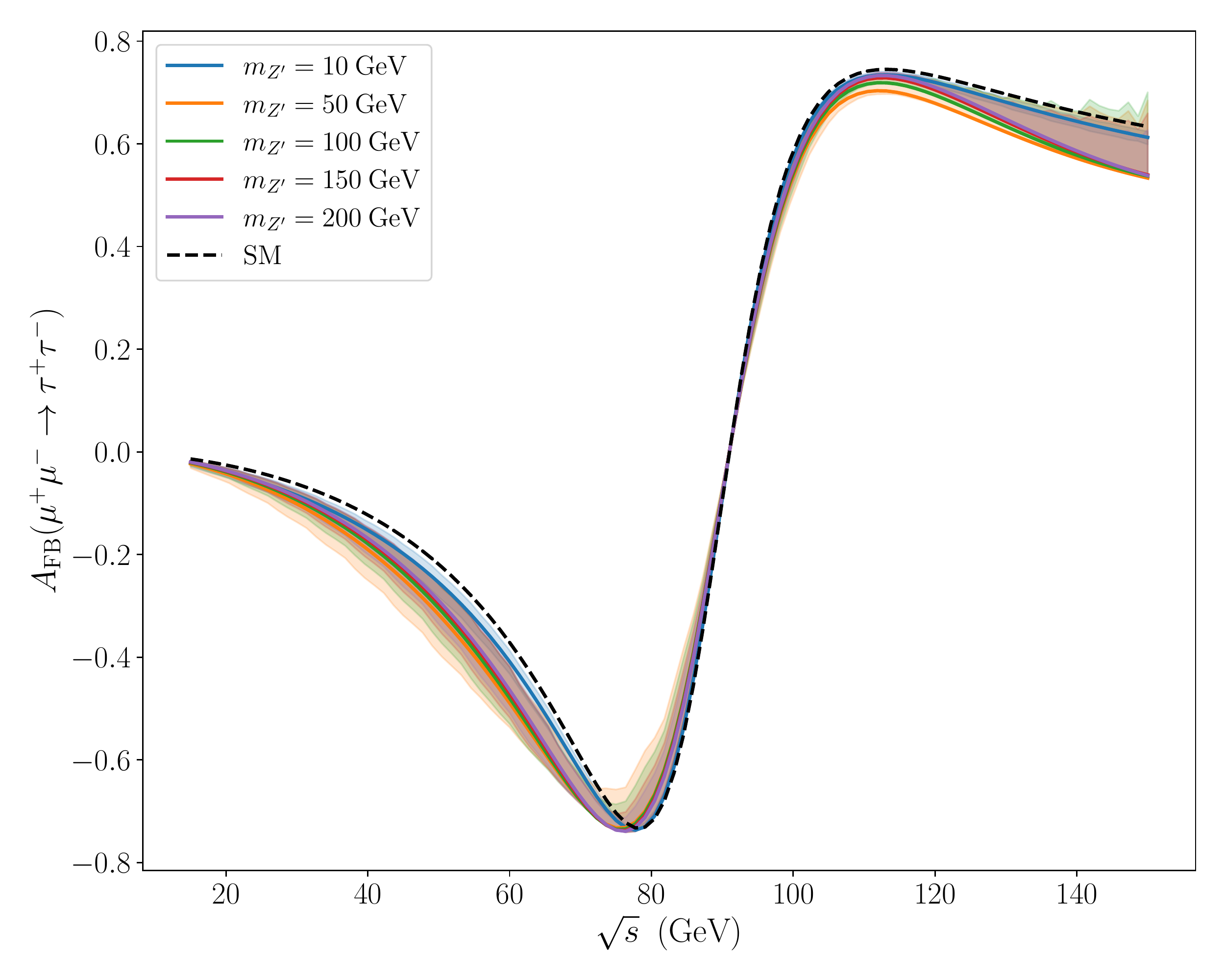}
    \caption{The forward backward asymmetry in $\mu^+\mu^-\to \tau^+\tau^-$ as a function of $\sqrt{s}$ for different values of $m_{Z^\prime}$. On the left panel, general view for $\sqrt s \lesssim 3$~TeV; on the right panel, detailed view around the $Z$-pole. Line code as in Fig.~\ref{fig:sigmumutautau_vs_s}.}
    \label{fig:AFBmumutautau_vs_s}
\end{figure}

The prospects for $A_\text{FB}(\mu^+\mu^-\to \tau^+\tau^-)$ as a function of the centre of mass energy are presented in Fig.~\ref{fig:AFBmumutautau_vs_s}, for several values of 
$m_{Z^\prime}$, and compared to the SM expectation. 
As can be seen, there is a clear difference between the SM and the $Z'$ predictions, increasing with larger $\sqrt{s}$.
Leading to these plots, we have followed the same procedure as 
in the previous subsection - considering for every value of $m_{Z^\prime}$ the best-fit intervals for the couplings $g_{L,R}^{\mu\tau}$ (to comply with cLFV and LFUV bounds, and saturate $\Delta a_\mu$). 
As previously mentioned, in our study we have also imposed (realistic) cuts on the integration limits (which also indirectly reflect finite detector effects), see Eq.~(\ref{eq:int:dsigma}): we have taken $-0.99 \leq \cos \theta \leq 0.99$.

Although $A_\text{FB}(\mu^+\mu^-\to \tau^+\tau^-)$ is significantly larger (and distinguishable from the SM expectation) 
for $\sqrt s \gtrsim 200$~GeV, the predictions for different $m_{Z^\prime}$ values become indistinguishable for 
$\sqrt s \sim 1$~TeV; also recall that, as shown in Fig.~\ref{fig:sigmumutautau_vs_s}, the $Z^\prime$-induced flavour violating rates become very small for high energies (as a consequence of the associated $t$-channel event topology). Further - more detailed - computations might be required in this case.

It is also interesting to notice that there is a clear deviation from the SM expectation for $\sqrt s $ below the $Z$-boson pole: this is manifest from inspection of the right panel of Fig.~\ref{fig:AFBmumutautau_vs_s}, where one can easily have sizeable deviations in such a $\sqrt{s}$ regime.

\begin{figure}[h!]
    \centering
    \includegraphics[width=0.48\textwidth]{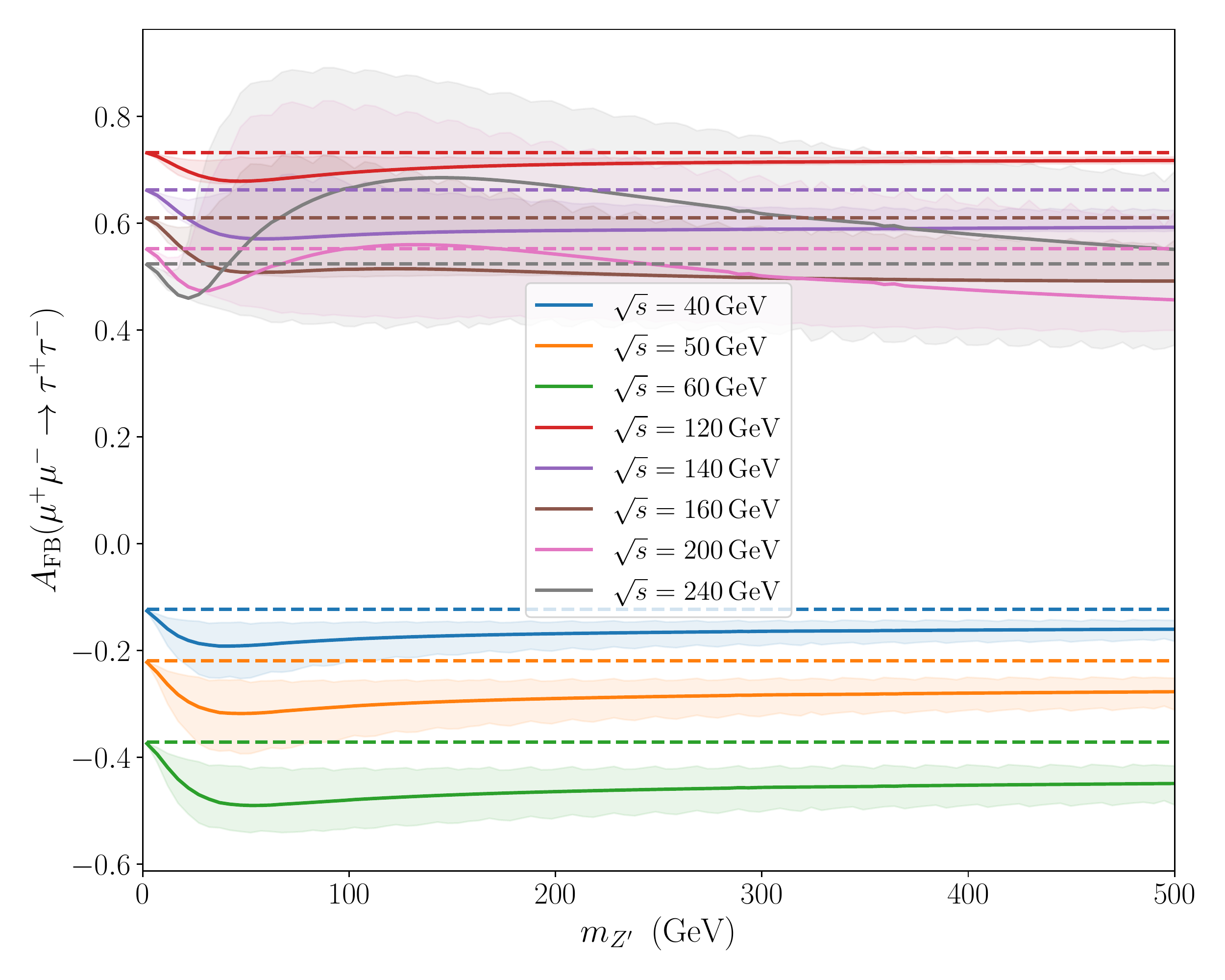}
    \caption{$A_\text{FB}(\mu^+\mu^-\to \tau^+\tau^-)$ as a function of $m_{Z^\prime}$ for different values of $\sqrt{s}$.
    Line code as in Fig.~\ref{fig:sigmumutautau_vs_mZp}.}
    \label{fig:AFBmumutautau_vs_mZp}
\end{figure}

Similarly to what was done in the previous subsection, in Fig.~\ref{fig:AFBmumutautau_vs_mZp} we offer a complementary view of $A_\text{FB}(\mu^+\mu^-\to \tau^+\tau^-)$, now as a function of $m_{Z^\prime}$, considering different values of $\sqrt{s}$. 
As visible, comparatively low values of $\sqrt{s}$, $\mathcal{O}(100\,\text{GeV}-200\,\text{GeV})$ offer the possibility of identifying the presence of the $t$-channel  NP mediator, especially for $m_{Z^\prime} \lesssim 100$~GeV. The expected difference in the asymmetry compared to the SM expectation can be very large, and we recall that for these regimes one expects a sizeable production cross section, see Fig.\ref{fig:sigmumutautau_vs_s}, in general above $1$~pb. 

Notice that for intermediate regimes of $\sqrt{s}$, the 
expected $A_\text{FB}(\mu^+\mu^-\to \tau^+\tau^-)$ comes closer to its SM prediction for large $m_{Z^\prime}$. This can be understood from the $\Delta a_\mu$ driven fit, due to which the couplings do not grow proportionally to the mediator's mass.
Thus, for large  $m_{Z^\prime}$ and comparatively ``small'' $\sqrt{s}$, the new contributions are sufficiently suppressed.

\begin{figure}
    \centering
    \includegraphics[width=0.48\textwidth]{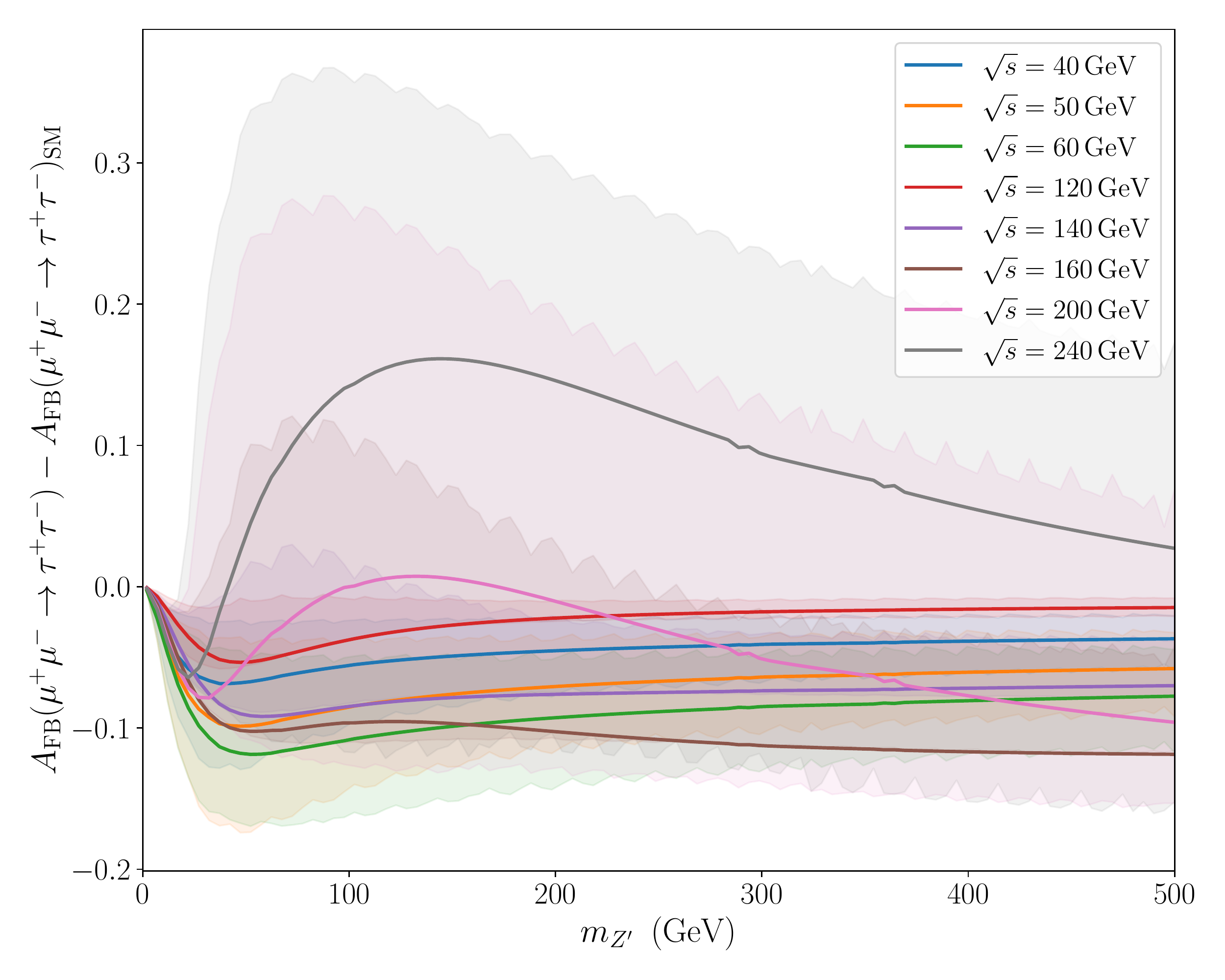}
    \includegraphics[width=0.48\textwidth]{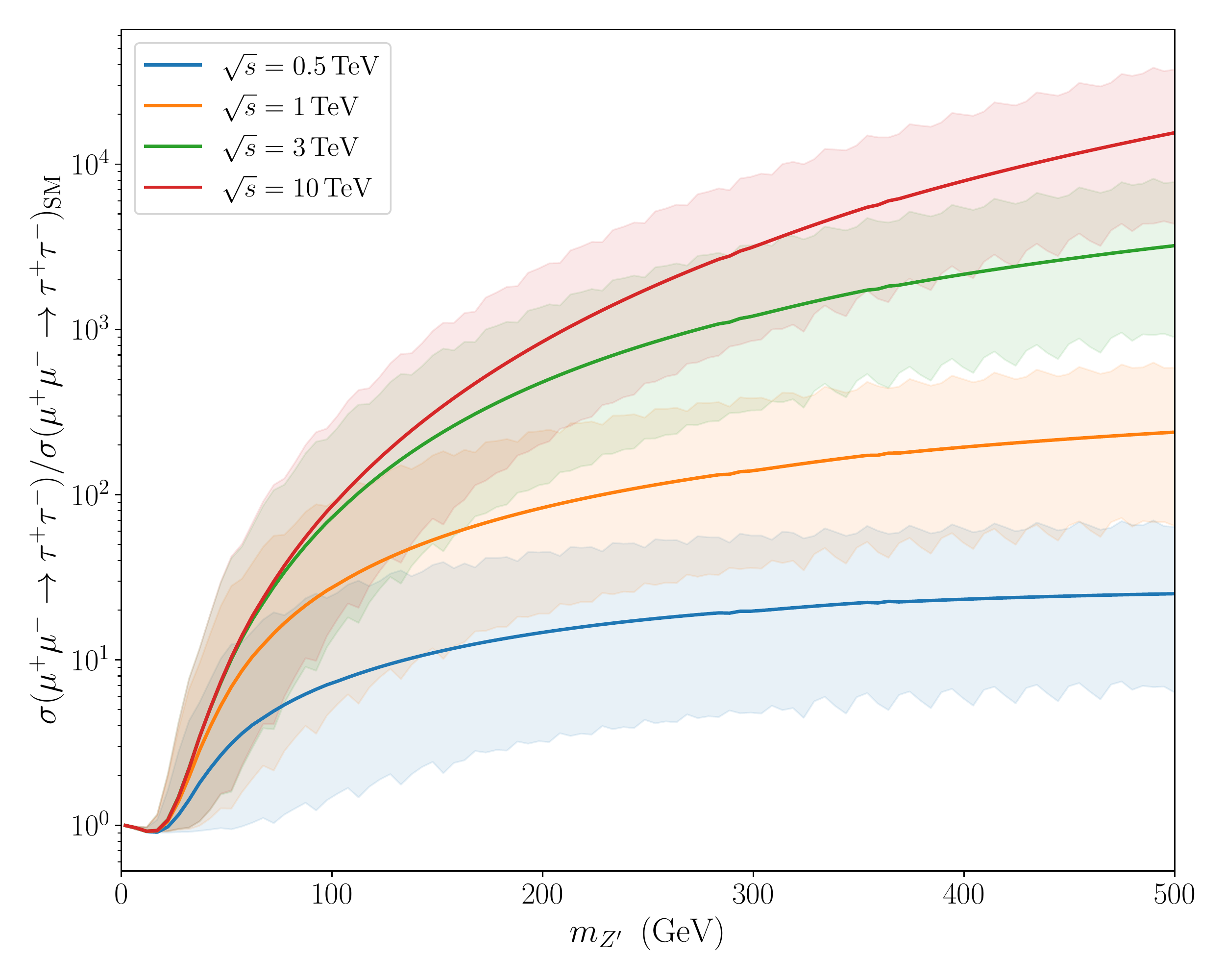}
    \caption{Comparison of $A_\text{FB}(\mu^+\mu^-\to \tau^+\tau^-)$ (left panel) and of the production cross section $\sigma(\mu^+\mu^-\to \tau^+\tau^-)$ (right panel)
    with the SM expectation, as a function of $m_{Z^\prime}$ for different va\-lues of $\sqrt{s}$. }
    \label{fig:sig_AFB_vsSM}
\end{figure}

The striking differences with respect to the SM expectation are summarised in Fig.~\ref{fig:sig_AFB_vsSM}, in which we compare the prospects for the production cross section and $A_\text{FB}$, considering
\begin{equation}
   \dfrac{\sigma(\mu^+\mu^-\to \tau^+\tau^-)}{\sigma(\mu^+\mu^-\to \tau^+\tau^-)_\text{SM}} \,\quad \text{and} \quad
    A_\text{FB}(\mu^+\mu^-\to \tau^+\tau^-) - 
    A_\text{FB}(\mu^+\mu^-\to \tau^+\tau^-)_\text{SM}\,.
\end{equation}
As can be seen, our results show that even for comparatively small $\sqrt{s}$ the relevant regions of the model's parameter space can be falsified as an explanation to $\Delta a_\mu$ at a future muon collider.
At larger $\sqrt{s}$, precise measurements of the $\mu^+\mu^-\to \tau^+\tau^-$ production cross section may be used to constrain the mediator mass, even in the absence of a resonance in the spectrum. 

\section{Further discussion and outlook}
Motivated by a minimal framework  to address the tensions between theory and experiment in the anomalous magnetic moments of (light) charged leptons, 
we have considered a SM extension via a $Z^\prime$ model. These 
well-motivated constructions have been extensively investigated in the literature (also in view of their potential interest in relation with the so-called LFUV anomalies in $B$-meson decays), and their flavour conserving couplings to fermions are very strongly constrained by a vast array of experimental measurements and searches. 
In our study, we have thus considered an ad-hoc (toy) model of a 
leptophilic light $Z^\prime$, which only couples in a flavour-violating manner to leptons. 

As we have argued here, $\Delta a_e^\text{Rb}$ and 
$\Delta a_e^\text{Cs}$ can be separately accounted for (for example, respectively relying on a same-sign or opposite-sign combination of the $g_{L,R}^{e\tau}$ couplings); however, a simultaneous explanation of $\Delta a_e$ (either $\Delta a_e^\text{Rb}$ or $\Delta a_e^\text{Cs}$) and $\Delta a_\mu$ is precluded due to having excessive contributions to numerous cLFV (and also LFUV) observables.  
In particular, the most stringent bounds emerge from
cLFV $\tau \to \mu \bar e \mu$ decays (for $g_{L,R}^{e\mu}$), and 
from  $\mu \to e \gamma$ decays ($g_{L,R}^{e\tau}$); for $g_{L,R}^{\mu\tau}$ couplings, the LFU-probing ratios, $R^Z_{\alpha \beta}$ ($\alpha \neq \beta$) and $R^\tau_{e \mu}$ play the most constraining roles. 

Thus, and focusing on an explanation to the current tension in the muon anomalous magnetic moment, one is led to $\Delta a_\mu$-favoured regimes, both for the relevant couplings and for the mass of the new mediator. In all cases, the latter regimes - and the strong constraints from cLFV and LFUV observables - strongly suggest that in this class of models one expects SM-compatible values for $\Delta a_e$. For the tau magnetic moment, one predicts values compatible with 0, typically 
$\Delta a_\tau \sim [- 10^{-6}, -10^{-7}]$.

In what concerns cLFV observables, and in spite of their very constraining role, one has good prospects for the observation of certain rare muon decays in the regimes favoured by an explanation of $\Delta a_\mu$, in particular for $\mu\to e\gamma$, and muon-electron conversion in nuclei.
Interestingly, in these $\Delta a_\mu$-favoured regimes, the stringent constraint imposed by the current experimental limit on $\mathrm{BR}(\mu\to e\gamma)$ leads to a theoretical upper limit for $\mathrm{BR}(\mu\to eee)$ that is smaller than the sensitivities of next-generation experiments.
Thus, a future observation of the latter two cLFV observables would contribute to falsify these $Z'$ models.

We have also emphasised the role of a future muon collider in probing the model's regimes which are preferred by $\Delta a_\mu$, for a wide range of $Z^\prime$ masses. Relying on a simple estimation of the production cross-section, we have pointed out that one should have sizeable $\sigma(\mu^+\mu^-\to \tau^+\tau^-)$, for various choices of the centre of mass energy. In addition to deviations from the SM regarding $\sigma(\mu^+\mu^-\to \tau^+\tau^-)$, the forward-backward asymmetry $A_\text{FB}(\mu^+\mu^-\to \tau^+\tau^-)$ is also expected to exhibit distinctive features, which can be used to put this minimal construction to the test.

In the future, a more precise determination of $a_\mu^\text{exp}$, possible in view of the expected reduction in the FNAL measurement uncertainty  (ten-fold improvement in statistics~\cite{Tewsley-Booth:2022umi}), together with improved SM predictions, will further allow to constrain the present model's parameter space. Likewise, clarifying the situation in what concerns the electron $(g-2)$ would also contribute to further test (or falsify) the model.

The toy model here considered - and despite not succeeding in providing a simultaneous explanation to the apparent tensions in 
$\Delta a_{e,\mu}$ - offers an interesting starting point to the analysis of complete SM extensions via new $U(1)^\prime$ mediators. As pointed out, new fields should be added - as is the case of a new scalar responsible for symmetry breaking and for giving a mass to $Z^\prime$ boson, and additional (heavy) fermions to properly cancel gauge anomalies and provide a realistic origin of the (strictly) flavour violating couplings. 
In turn, one expects that new interactions of the scalar with SM matter (i.e. the new Yukawa couplings) will then contribute to the many processes described above; one 
must then explore if in this case it is possible to account for  both  $(g-2)_{e,\mu}$ without conflict with the relevant cLFV and LFUV observables which so far preclude an simultaneously explanation of the tensions.
Further extensions are also possible concerning the fermion (matter) sector, such as introducing right-handed neutrinos, and considering potential neutrino mass generation mechanisms. 
Likewise, one can also envisage considering the role of the new fields and couplings in what concerns an explanation to the dark matter problem.

\section*{Acknowledgements}
This project has received support from the European Union's Horizon 2020 research and innovation programme under the Marie Sk\l{}odowska-Curie grant agreement No.~860881 (HIDDe$\nu$ network) and from the IN2P3 (CNRS) Master Project, ``Flavour probes: lepton sector and beyond'' (16-PH-169).

\appendix
\section{Appendix}
\mathversion{bold}
\subsection{New contributions to LFC $Z\to \ell^+ \ell^-$ decays}
\label{app:lfcZ}
\mathversion{normal}
In this appendix we summarise the most relevant details concerning the new contributions to the $Z \to \ell_\alpha^+\ell_\alpha^-$ decay (same-flavour charged lepton pair), arising from a $Z^\prime$ in the loop.
The matrix element can be written as 
\begin{eqnarray}
  \mathcal{M}_{Z \to \ell_\alpha^+\ell_\alpha^-}^{Z^\prime} &=& u_\alpha \left[ \gamma^\mu \, \left(\mathcal{C}_L^V\,   P_L + \mathcal{C}_R^V\,   P_R \right) +  \sigma^{\mu \nu}\, q_\nu \, \left(\mathcal{C}_L^T \,P_L  + \mathcal{C}_R^T \,P_R \right) \right] v_\alpha \, \epsilon_\mu^*(q) \, , 
\end{eqnarray} 
where $u_\alpha, \, v_\alpha$ denote the $\ell_\alpha$ spinors, $\epsilon_\mu(q)$ the $Z$-boson polarisation vector with $q$ the $Z$ momentum, and $\mathcal{C}_X^{V, T}$ are the vector and tensor coefficients (computed from the decay triangle diagram). For Hermitian (real in our case)  $Z^\prime$ couplings, these are given by:
\begin{eqnarray}
\mathcal{C}^T_L &=& -\frac{i}{16 \pi^2  m_{Z^\prime}^2}
\left\{2 \left[ g_L^{i \alpha}  g_L^{i \alpha,*}  g_L^Z m_\alpha +  g_R^{i \alpha}  g_R^{i \alpha,*}  g_R^Z m_\alpha -  g_L^{i \alpha}  g_R^{i \alpha,*} ( g_L^Z +  g_R^Z) m_i \right] m_{Z^\prime}^2 C_{0} \right. \nonumber\\
&+& \left[4  g_L^{i \alpha}  g_L^{i \alpha,*}  g_L^Z m_\alpha m_{Z^\prime}^2 +  g_R^{i \alpha}  g_R^{i \alpha,*}  g_R^Z m_\alpha (m_\alpha^2 - m_i^2 + 2 m_{Z^\prime}^2) \right. \nonumber\\
&& \hspace{2mm}\left. +  g_L^{i \alpha}  g_R^{i \alpha,*} m_i (-2  g_L^Z m_{Z^\prime}^2 +  g_R^Z (-m_\alpha^2 + m_i^2 - 2 m_{Z^\prime}^2))\right] C_{1} \nonumber\\
&+& \left[4  g_R^{i \alpha}  g_R^{i \alpha,*}  g_R^Z m_\alpha m_{Z^\prime}^2 +  g_L^{i \alpha}  g_L^{i \alpha,*}  g_L^Z m_\alpha (m_\alpha^2 - m_i^2 + 2 m_{Z^\prime}^2) \right.
\nonumber\\
&& \hspace{2mm}\left.+  g_L^{i \alpha}  g_R^{i \alpha,*} m_i (-2  g_R^Z m_{Z^\prime}^2 +  g_L^Z (-m_\alpha^2 + m_i^2 - 2 m_{Z^\prime}^2))\right] C_{2} \nonumber\\
&+& m_\alpha \left[ g_R^{i \alpha}  g_R^Z m_\alpha ( g_R^{i \alpha,*} m_\alpha -  g_L^{i \alpha,*} m_i) +  g_L^{i \alpha} (-( g_R^{i \alpha,*}  g_R^Z m_\alpha m_i) +  g_L^{i \alpha,*}  g_R^Z m_i^2 + 2  g_L^{i \alpha,*}  g_L^Z m_{Z^\prime}^2)\right] C_{11}
\nonumber\\
&+& m_\alpha \left[ g_L^{i \alpha}  g_L^Z m_\alpha ( g_L^{i \alpha,*} m_\alpha -  g_R^{i \alpha,*} m_i) +  g_R^{i \alpha} (-( g_L^{i \alpha,*}  g_L^Z m_\alpha m_i) +  g_R^{i \alpha,*}  g_L^Z m_i^2 + 2  g_R^{i \alpha,*}  g_R^Z m_{Z^\prime}^2)\right] C_{22} \nonumber\\
&+&
m_\alpha \left[-( g_L^{i \alpha,*}  g_R^{i \alpha} ( g_L^Z +  g_R^Z) m_\alpha m_i) -  g_L^{i \alpha}  g_R^{i \alpha,*} ( g_L^Z +  g_R^Z) m_\alpha m_i +  g_L^{i \alpha}  g_L^{i \alpha,*} ( g_R^Z m_i^2 +  g_L^Z (m_\alpha^2 + 2 m_{Z^\prime}^2)) \right.\nonumber\\ 
&& \hspace{2mm} \left.\left.+  g_R^{i \alpha}  g_R^{i \alpha,*} ( g_L^Z m_i^2 +  g_R^Z (m_\alpha^2 + 2 m_{Z^\prime}^2))\right] C_{12}
\right\}\,,
\end{eqnarray}
\begin{eqnarray}
\mathcal{C}^V_L &=& \frac{1}{16 \pi^2 m_{Z^\prime}^2}
\left\{
-4  g_L^{i \alpha}  g_L^{i \alpha,*}  g_L^Z m_{Z^\prime}^2 B_{0}^\alpha  \right.
\nonumber \\
&-& \left[ g_R^{i \alpha}  g_R^Z m_\alpha ( g_R^{i \alpha,*} m_\alpha -  g_L^{i \alpha,*} m_i) +  g_L^{i \alpha} (-( g_R^{i \alpha,*}  g_R^Z m_\alpha m_i) +  g_L^{i \alpha,*}  g_R^Z m_i^2 - 2  g_L^{i \alpha,*}  g_L^Z m_{Z^\prime}^2)\right] B_{0}^q \nonumber \\
&+& \left[ g_R^{i \alpha}g_R^{i \alpha,*} g_R^Z m_\alpha^2 m_{Z^\prime}^2
-  g_R^{i \alpha} g_L^{i \alpha,*} (2  g_L^Z +  g_R^Z) m_i m_\alpha m_{Z^\prime}^2 - g_L^{i \alpha} g_R^{i \alpha,*} (2  g_L^Z +  g_R^Z) m_\alpha m_i m_{Z^\prime}^2
\right. \nonumber\\
&&\hspace{2mm} \left. 
+  g_L^{i \alpha} g_L^{i \alpha,*} \left( g_R^Z m_i^2 m_{Z^\prime}^2 
+  g_L^Z (m_\alpha^4 + m_i^4 - 4 m_i^2 m_{Z^\prime}^2 - 2 m_\alpha^2 (m_i^2 - 2 m_{Z^\prime}^2) + 2 m_{Z^\prime}^2 (m_{Z^\prime}^2 + m_Z^2))\right)\right] C_{0} 
\nonumber \\
&+& m_\alpha \left[4  g_R^{i \alpha}  g_R^{i \alpha,*}  g_R^Z m_\alpha m_{Z^\prime}^2 + 2  g_L^{i \alpha}  g_L^{i \alpha,*}  g_L^Z m_\alpha (m_\alpha^2 - m_i^2 + 4 m_{Z^\prime}^2) 
\right. \nonumber\\
&&\hspace{2mm} \left.  +  \left(g_L^{i \alpha,*}  g_R^{i \alpha} + g_L^{i \alpha}  g_R^{i \alpha,*} \right)m_i (-2  g_R^Z m_{Z^\prime}^2 +  g_L^Z (-m_\alpha^2 + m_i^2 - 2 m_{Z^\prime}^2))\right] \left[C_{1} + C_{2} \right] 
\nonumber \\
&+& m_\alpha^2 \left[-( g_L^{i \alpha,*}  g_R^{i \alpha} ( g_L^Z +  g_R^Z) m_\alpha m_i) -  g_L^{i \alpha}  g_R^{i \alpha,*} ( g_L^Z +  g_R^Z) m_\alpha m_i +  g_L^{i \alpha}  g_L^{i \alpha,*} ( g_R^Z m_i^2 +  g_L^Z (m_\alpha^2 + 2 m_{Z^\prime}^2)) 
\right. \nonumber\\
&&\hspace{2mm} \left. +  g_R^{i \alpha}  g_R^{i \alpha,*} ( g_L^Z m_i^2 +  g_R^Z (m_\alpha^2 + 2 m_{Z^\prime}^2))\right] \left[C_{11} + C_{22} + 2 C_{12}\right] 
\nonumber \\
&+& \left. 2 \left[ g_R^{i \alpha}  g_R^Z m_\alpha ( g_R^{i \alpha,*} m_\alpha -  g_L^{i \alpha,*} m_i) +  g_L^{i \alpha} (-( g_R^{i \alpha,*}  g_R^Z m_\alpha m_i) +  g_L^{i \alpha,*}  g_R^Z m_i^2 + 2  g_L^{i \alpha,*}  g_L^Z m_{Z^\prime}^2)\right] C_{00}
\right\}\,,
\end{eqnarray}
In the above, $B_0^\alpha \equiv B_0( m_\alpha^2, m_{Z^\prime}^2, m_i^2)$, $B_0^q \equiv B_0( q^2, m_i^2, m_i^2)$ and $C_{PV} \equiv C_{PV}\left(m_\alpha^2,q^2,m_\alpha^2,m_{Z^\prime}^2,m_i^2,m_i^2\right)$  denote the Passarino-Veltman functions (with $PV \, = \, 0,1,2,00,11,12,22$) and $m_i$ the mass of the internal lepton. Notice that the dimensions of the vector and tensor coefficients are respectively $\left[ \mathcal{C}_L^{V} \right] = m^0$ and $\left[ \mathcal{C}_L^{T} \right] = m^{-1}$ (recall that certain Passarino-Veltman functions are dimensionful quantities). Finally, the RH coefficients can be obtained by exchanging $L\leftrightarrow R$ in the expressions for 
$\mathcal{C}_L^{V, T}$.
We further set $q^2 = m_Z^2$ for the on-shell decays.
Moreover, and as already discussed in the main part of this work, we treat the divergences appearing in the $B_0$ and $C_{00}$ functions in the $\overline{\text{MS}}$-scheme and set the 't Hooft scale to $\mu^2 = m_Z^2$.

The contributions to the leptonic $Z$ decay width (SM-like, $Z^\prime$ and interference terms) can be cast as
\begin{equation}
    \Gamma (Z\to \ell^+\ell^-) \,\simeq\, \Gamma^{\text{SM}_\text{full}} \,+\, \Gamma^{\text{SM}_\text{tree}-Z^\prime} \,+\, \Gamma^{Z^\prime}\,,
\end{equation}
with 
\begin{eqnarray}
  \Gamma^{\text{SM}_\text{tree}-Z^\prime} &=&  \dfrac{\lambda^{1/2}(m_Z,m_\alpha, m_\alpha)}{16\pi m_Z^3} \frac{2}{3} 
  \left\{ 2\left( m_Z^2 - m_\alpha^2 \right)\left(\mathcal{C}^V_L\, g_L^Z + \mathcal{C}^V_R \, g_R^Z \right) + 6m_\alpha^2 \left(\mathcal{C}^V_L\, g_L^R + \mathcal{C}^V_R \, g_L^Z \right) \right. \nonumber \\
  &-& \left. 3i m_\alpha m_Z^2 \left[ \mathcal{C}^T_L \left(g_L^Z + g_L^R \right) + \mathcal{C}^T_R \left(g_L^Z + g_L^R \right) \right]\right\}\, ,\\
\Gamma^{Z^\prime} &=& \dfrac{\lambda^{1/2}(m_Z,m_\alpha, m_\alpha)}{16\pi m_Z^3} \frac{1}{3}  
\left\{ 2\left(m_Z^2 - m_\alpha^2 \right)\left(|\mathcal{C}^V_L|^2 + |\mathcal{C}^V_R|^2 \right) + 6m_\alpha^2\left( \mathcal{C}^{V}_L \mathcal{C}^{V,*}_R + \mathcal{C}^{V}_R \mathcal{C}^{V,*}_L\right)\right. \nonumber \\
&+& \left(m_Z^4 + 2 m_\alpha^2 m_Z^2 \right)\left(|\mathcal{C}^T_L|^2 + |\mathcal{C}^T_R|^2 \right) + 6m_\alpha^2m_Z^2 \left( \mathcal{C}^{T}_L \mathcal{C}^{T,*}_R + \mathcal{C}^{T}_R \mathcal{C}^{T,*}_L\right) \nonumber \\
&+& \left. 3i m_\alpha m_Z^2\left[ \mathcal{C}^{T,*}_L \left( \mathcal{C}^V_L + \mathcal{C}^V_R \right) + \mathcal{C}^{T,*}_R \left( \mathcal{C}^V_L + \mathcal{C}^V_R \right) - \text{c.c.} \right] \right\}\,,
\end{eqnarray}
and $\Gamma^{\text{SM}_\text{full}}$ is given in~\cite{Freitas:2014hra} at 2-loop accuracy.
In the above, the K\"all\'en-function is defined as
\begin{equation}\label{eq:kallen}
    \lambda(a,b,c) = (a^2 - b^2 - c^2)^2 - 4b^2 c^2\,.
\end{equation}

\mathversion{bold}
\subsection{Lepton flavour violating $Z$ decay amplitude and coefficients}\label{app:LFV_Zdecays}
\mathversion{normal}
We now consider the effects arising from the presence of the new $Z^\prime$ boson in what concerns cLFV $Z \to \ell_\alpha^+ \ell_\beta^-$ decays. Analogously to the previous appendix, the matrix element can be cast as:
\begin{eqnarray}
   \mathcal{M}_{Z \to \ell_\alpha^+\ell_\beta^-} &=& u_\beta \left[ \gamma^\mu \, \left(\mathcal{K}_L^V\,   P_L + \mathcal{K}_R^V\,   P_R \right) +  \sigma^{\mu \nu}\, q_\nu \, \left(\mathcal{K}_L^T \,P_L  + \mathcal{K}_R^T \,P_R \right) \right] v_\alpha \, \epsilon_\mu^*(q) \, , 
\end{eqnarray}
already explicitly cast in terms of the vector and tensor coefficients, $\mathcal{C}_L^{V, T}$. 
In the above, $u_\beta, \, v_\alpha$ denote the $\ell_\beta, \, \ell_\alpha$ spinors respectively, $\epsilon_\mu(q)$ is the $Z$-boson polarisation vector, with $q$ the $Z$ momentum. As before, $\mathcal{K}_X^{V, T}$ can be computed from the decay triangle diagram and are given, for hermitian (real) $Z^\prime$ couplings, by:
\begin{eqnarray}
\mathcal{K}_L^T &=& - \dfrac{i}{16 \pi^2 m_{Z^\prime}^2}
\left\{
2 \left[ g_L^{i \alpha, *}  g_L^{i \beta}  g^Z_L m_\alpha +  g_R^{i \alpha, *}  g_R^{i \beta}  g^Z_R m_\beta -  g_L^{i \beta}  g_R^{i \alpha, *} ( g^Z_L +  g^Z_R) m_{i}\right] m_{Z^\prime}^2 C_0 \right.\nonumber \\
&+& \left[4  g_L^{i \alpha, *}  g_L^{i \beta}  g^Z_L m_\alpha m_{Z^\prime}^2 +  g_R^{i \alpha, *}  g_R^{i \beta}  g^Z_R m_\beta (m_\alpha^2 - m_{i}^2 + 2 m_{Z^\prime}^2) 
\right. \nonumber \\
&& \left. \hspace{2mm}+  g_L^{i \beta}  g_R^{i \alpha, *} m_{i} (-2  g^Z_L m_{Z^\prime}^2 +  g^Z_R (-m_\alpha^2 + m_{i}^2 - 2 m_{Z^\prime}^2))\right] C_1 \nonumber \\
&+& \left[4  g_R^{i \alpha, *}  g_R^{i \beta}  g^Z_R m_\beta m_{Z^\prime}^2 +  g_L^{i \alpha, *}  g_L^{i \beta}  g^Z_L m_\alpha (m_\beta^2 - m_{i}^2 + 2 m_{Z^\prime}^2) 
\right. \nonumber \\
&& \left. \hspace{2mm}
+  g_L^{i \beta}  g_R^{i \alpha, *} m_{i} (-2  g^Z_R m_{Z^\prime}^2 +  g^Z_L (-m_\beta^2 + m_{i}^2 - 2 m_{Z^\prime}^2))\right] C_2 \nonumber \\ 
&+& m_\alpha \left[ g_R^{i \alpha, *}  g^Z_R m_\alpha ( g_R^{i \beta} m_\beta -  g_L^{i \beta} m_{i}) +  g_L^{i \alpha, *} (-( g_R^{i \beta}  g^Z_R m_\beta m_{i}) +  g_L^{i \beta}  g^Z_R m_{i}^2 + 2  g_L^{i \beta}  g^Z_L m_{Z^\prime}^2)\right] C_{11} \nonumber \\
&+& m_\beta \left[ g_L^{i \alpha, *}  g^Z_L m_\alpha ( g_L^{i \beta} m_\beta -  g_R^{i \beta} m_{i}) +  g_R^{i \alpha, *} (-( g_L^{i \beta}  g^Z_L m_\beta m_{i}) +  g_R^{i \beta}  g^Z_L m_{i}^2 + 2  g_R^{i \beta}  g^Z_R m_{Z^\prime}^2)\right] C_{22}\nonumber \\ 
&+&  \left[- g_L^{i \alpha, *}  g_R^{i \beta} ( g^Z_L +  g^Z_R) m_\alpha m_\beta m_{i} -  g_L^{i \beta}  g_R^{i \alpha, *} ( g^Z_R m_\alpha^2 +  g^Z_L m_\beta^2) m_{i} +  g_R^{i \alpha, *}  g_R^{i \beta} m_\beta ( g^Z_L m_{i}^2 +  g^Z_R (m_\alpha^2 + 2 m_{Z^\prime}^2)) \right. \nonumber \\
&& \left.\left. \hspace{2mm}
+  g_L^{i \alpha, *}  g_L^{i \beta} m_\alpha ( g^Z_R m_{i}^2 +  g^Z_L (m_\beta^2 + 2 m_{Z^\prime}^2))\right] C_{12} 
\right\}\,,
\end{eqnarray}
\begin{eqnarray}
\mathcal{K}_L^V &=&  \dfrac{1}{16 \pi^2 m_{Z^\prime}^2}
\left\{ -2  g_L^{i \alpha, *}  g_L^{i \beta}  g^Z_L m_{Z^\prime}^2 \left[
B_0(m_\alpha^2,m_{Z^\prime}^2,m_i^2)+ B_0(m_\beta^2,m_{Z^\prime}^2,m_i^2)
 \right] \right. \nonumber \\
&+& \left[ g_R^{i \alpha, *}  g^Z_R m_\alpha (-( g_R^{i \beta} m_\beta) +  g_L^{i \beta} m_{i}) +  g_L^{i \alpha, *} ( g_R^{i \beta}  g^Z_R m_\beta m_{i} -  g_L^{i \beta}  g^Z_R m_{i}^2 + 2  g_L^{i \beta}  g^Z_L m_{Z^\prime}^2)\right] B_0(q^2, m_i^2,m_i^2) 
\nonumber \\
&+& \dfrac{g^Z_L m_{i}}{m_\alpha^2 - m_\beta^2}\left[  -2  g_R^{i \alpha, *}  g_R^{i \beta} m_\alpha m_\beta m_{i} -  g_L^{i \alpha, *}  g_L^{i \beta} (m_\alpha^2 + m_\beta^2) m_{i} +  g_R^{i \beta}  g_L^{i \alpha, *} m_\beta (m_\alpha^2 + m_{i}^2 - 4 m_{Z^\prime}^2) \right.\nonumber\\
&&\hspace{2mm}\left.
+  g_R^{i \alpha, *}  g_L^{i \beta} m_\alpha (m_\beta^2 + m_{i}^2 - 4 m_{Z^\prime}^2) \right] \left[
B_0(m_\alpha^2, m_i^2 ,m_{Z^\prime}^2) - B_0(m_\beta^2, m_i^2 ,m_{Z^\prime}^2)
\right] \nonumber \\
&-& \dfrac{g^Z_L m_\alpha}{m_\alpha^2 - m_\beta^2}\left[ -2  g_R^{i \beta}  g_L^{i \alpha, *} m_\alpha m_\beta m_{i} -  g_R^{i \alpha, *}  g_L^{i \beta} (m_\alpha^2 + m_\beta^2) m_{i} +  g_R^{i \alpha, *}  g_R^{i \beta} m_\beta (m_\alpha^2 + m_{i}^2 + 2 m_{Z^\prime}^2) \right.
\nonumber\\
&&\hspace{2mm} \left. +  g_L^{i \alpha, *}  g_L^{i \beta} m_\alpha (m_\beta^2 + m_{i}^2 + 2 m_{Z^\prime}^2) \right]B_1(m_\alpha^2, m_i^2 ,m_{Z^\prime}^2)
\nonumber \\
&+& \dfrac{g^Z_L m_\beta }{m_\alpha^2 - m_\beta^2}\left[-2  g_R^{i \alpha, *}  g_L^{i \beta} m_\alpha m_\beta m_{i} -  g_R^{i \beta}  g_L^{i \alpha, *} (m_\alpha^2 + m_\beta^2) m_{i} +  g_L^{i \alpha, *}  g_L^{i \beta} m_\beta (m_\alpha^2 + m_{i}^2 + 2 m_{Z^\prime}^2) \right.
\nonumber\\
&&\hspace{2mm} \left.+  g_R^{i \alpha, *}  g_R^{i \beta} m_\alpha (m_\beta^2 + m_{i}^2 + 2 m_{Z^\prime}^2)\right] B_1(m_\beta^2, m_i^2 ,m_{Z^\prime}^2)
\nonumber \\
&+& \left[ g_R^{i \alpha, *} m_\alpha ( g_R^{i \beta}  g^Z_R m_\beta -  g_L^{i \beta} ( g^Z_R + 2  g^Z_L) m_{i}) m_{Z^\prime}^2 
- g_L^{i \alpha, *}  g_R^{i \beta} ( g^Z_R + 2  g^Z_L) m_\beta m_{i} m_{Z^\prime}^2
+ g_L^{i \alpha, *}  g_L^{i \beta}  g^Z_R m_{i}^2 m_{Z^\prime}^2 
\right.
\nonumber\\
&&\hspace{2mm} \left.
+ g_L^{i \alpha, *}  g_L^{i \beta}  g^Z_L (m_{i}^4 - 4 m_{i}^2 m_{Z^\prime}^2 + 2 m_{Z^\prime}^4 - m_\beta^2 (m_{i}^2 - 2 m_{Z^\prime}^2) + m_\alpha^2 (m_\beta^2 - m_{i}^2 + 2 m_{Z^\prime}^2) + 2 m_{Z^\prime}^2 m_{Z}^2))\right] C_0 \nonumber \\
&+&\left[4  g_R^{i \alpha, *}  g_R^{i \beta}  g^Z_R m_\alpha m_\beta m_{Z^\prime}^2 +  g_R^{i \beta}  g_L^{i \alpha, *} m_\beta m_{i} (-2  g^Z_R m_{Z^\prime}^2 +  g^Z_L (-m_\alpha^2 + m_{i}^2 - 2 m_{Z^\prime}^2)) 
\right.
\nonumber\\
&&\hspace{2mm} \left.
+  g_R^{i \alpha, *}  g_L^{i \beta} m_\alpha m_{i} (-2  g^Z_R m_{Z^\prime}^2 +  g^Z_L (-m_\beta^2 + m_{i}^2 - 2 m_{Z^\prime}^2)) 
\right.
\nonumber\\
&&\hspace{2mm} \left.+  g_L^{i \alpha, *}  g_L^{i \beta}  g^Z_L (-(m_\beta^2 (m_{i}^2 - 2 m_{Z^\prime}^2)) + m_\alpha^2 (2 m_\beta^2 - m_{i}^2 + 6 m_{Z^\prime}^2))\right] C_1
\nonumber \\
&+& \left[4  g_R^{i \alpha, *}  g_R^{i \beta}  g^Z_R m_\alpha m_\beta m_{Z^\prime}^2 +  g_R^{i \beta}  g_L^{i \alpha, *} m_\beta m_{i} (-2  g^Z_R m_{Z^\prime}^2 +  g^Z_L (-m_\alpha^2 + m_{i}^2 - 2 m_{Z^\prime}^2)) 
\right.
\nonumber\\
&&\hspace{2mm} \left. +  g_R^{i \alpha, *}  g_L^{i \beta} m_\alpha m_{i} (-2  g^Z_R m_{Z^\prime}^2 +  g^Z_L (-m_\beta^2 + m_{i}^2 - 2 m_{Z^\prime}^2)) 
\right.
\nonumber\\
&&\hspace{2mm} \left.+  g_L^{i \alpha, *}  g_L^{i \beta}  g^Z_L (-(m_\beta^2 (m_{i}^2 - 6 m_{Z^\prime}^2)) + m_\alpha^2 (2 m_\beta^2 - m_{i}^2 + 2 m_{Z^\prime}^2))\right] C_2
\nonumber \\
&+& m_\alpha \left[-( g_R^{i \beta}  g_L^{i \alpha, *} ( g^Z_R +  g^Z_L) m_\alpha m_\beta m_{i}) -  g_R^{i \alpha, *}  g_L^{i \beta} ( g^Z_R m_\alpha^2 +  g^Z_L m_\beta^2) m_{i} 
\right.
\nonumber\\
&&\hspace{2mm} \left.+  g_R^{i \alpha, *}  g_R^{i \beta} m_\beta ( g^Z_L m_{i}^2 +  g^Z_R (m_\alpha^2 + 2 m_{Z^\prime}^2)) +  g_L^{i \alpha, *}  g_L^{i \beta} m_\alpha ( g^Z_R m_{i}^2 +  g^Z_L (m_\beta^2 + 2 m_{Z^\prime}^2))\right] C_{11}
\nonumber \\
&+& m_\beta \left[-( g_R^{i \alpha, *}  g_L^{i \beta} ( g^Z_R +  g^Z_L) m_\alpha m_\beta m_{i}) -  g_R^{i \beta}  g_L^{i \alpha, *} ( g^Z_L m_\alpha^2 +  g^Z_R m_\beta^2) m_{i} 
\right.
\nonumber\\
&&\hspace{2mm} \left.+  g_L^{i \alpha, *}  g_L^{i \beta} m_\beta ( g^Z_R m_{i}^2 +  g^Z_L (m_\alpha^2 + 2 m_{Z^\prime}^2)) +  g_R^{i \alpha, *}  g_R^{i \beta} m_\alpha ( g^Z_L m_{i}^2 +  g^Z_R (m_\beta^2 + 2 m_{Z^\prime}^2))\right] C_{22}
\nonumber \\
&+& \left[-( g_R^{i \beta}  g_L^{i \alpha, *} m_\beta (2  g^Z_L m_\alpha^2 +  g^Z_R (m_\alpha^2 + m_\beta^2)) m_{i}) -  g_R^{i \alpha, *}  g_L^{i \beta} m_\alpha (2  g^Z_L m_\beta^2 +  g^Z_R (m_\alpha^2 + m_\beta^2)) m_{i} 
\right.
\nonumber\\
&&\hspace{2mm} \left.+  g_R^{i \alpha, *}  g_R^{i \beta} m_\alpha m_\beta (2  g^Z_L m_{i}^2 +  g^Z_R (m_\alpha^2 + m_\beta^2 + 4 m_{Z^\prime}^2)) 
\right.
\nonumber\\
&&\hspace{2mm} \left.+  g_L^{i \alpha, *}  g_L^{i \beta} ( g^Z_R (m_\alpha^2 + m_\beta^2) m_{i}^2 + 2  g^Z_L (m_\beta^2 m_{Z^\prime}^2 + m_\alpha^2 (m_\beta^2 + m_{Z^\prime}^2)))\right] C_{12}
\nonumber \\
&+& \left. 2 \left[ g_R^{i \alpha, *}  g^Z_R m_\alpha ( g_R^{i \beta} m_\beta -  g_L^{i \beta} m_{i}) +  g_L^{i \alpha, *} (-( g_R^{i \beta}  g^Z_R m_\beta m_{i}) +  g_L^{i \beta}  g^Z_R m_{i}^2 + 2  g_L^{i \beta}  g^Z_L m_{Z^\prime}^2)\right] C_{00}
\right\}\,.
\end{eqnarray}
As in the previous appendix, $C_{PV} \equiv C_{PV}(m_\alpha^2, q^2, m_\beta^2, m_{Z^\prime}^2, m_i^2, m_i^2)$, are the Passarino-Veltman functions, 
$\ell_i$ is the internal lepton, and the RH coefficients are obtained by exchanging ($L \leftrightarrow R$). As before, the dimensions of the vector and tensor coefficients are respectively $\left[ \mathcal{K}_L^{V} \right] = m^0$ and $\left[ \mathcal{K}_L^{T} \right] = m^{-1}$.
As before, we treat the divergences appearing in the $B_{0,1}$ and $C_{00}$ functions in the $\overline{\text{MS}}$-scheme and set the 't Hooft scale to $\mu^2 = m_Z^2$.

Finally, the partial width for the decay $Z \to \ell_\alpha^+ \ell_\beta^-$ reads:
\begin{eqnarray}
\Gamma &=& \dfrac{\lambda^{1/2}(m_Z,m_\alpha, m_\beta)}{16\pi m_Z^3}
\frac{1}{3 m_Z^2} \left\{ 
\left[2 m_Z^4 - m_Z^2 \left(m_\alpha^2 + m_\beta^2\right) -\left(m_\alpha^2-m_\beta^2\right)^2\right] \left(|\mathcal{K}^{V}_L|^2+|\mathcal{K}^{V}_R|^2\right)
\right.\nonumber \\
&+& \left[m_Z^6 + m_Z^4 \left(m_\alpha^2+m_\beta^2\right) - 2 m_Z^2 \left(m_\alpha^2 - m_\beta^2\right)^2 \right] \left(|\mathcal{K}^{T}_L|^2+|\mathcal{K}^{T}_R|^2\right) + 6 m_Z^2 m_\alpha m_\beta \left(\mathcal{K}^{V}_L \mathcal{K}^{V, *}_R + \text{c.c.} \right) \nonumber\\
&+& 6 m_Z^4 m_\alpha m_\beta \left(\mathcal{K}^{T}_L \mathcal{K}^{T, *}_R + \text{c.c.} \right) + 3i  m_Z^2 m_\beta \left[m_\beta^2 - m_\alpha^2 -  m_Z^2 \right] \left(\mathcal{K}^{V}_L \mathcal{K}^{T,*}_L + \mathcal{K}^{V}_R \mathcal{K}^{T,*}_R  - \text{c.c.}\right) \nonumber \\
&+& \left. 3i m_Z^2 m_\alpha \left[ m_\alpha^2 -  m_\beta^2 -  m_Z^2 \right] \left(\mathcal{K}^{V}_L \mathcal{K}^{T,*}_R + \mathcal{K}^{V}_R \mathcal{K}^{T,*}_L  - \text{c.c.} \right) 
\right\}\, ,
\end{eqnarray}
in which "c.c." denotes the complex conjugate.
\mathversion{bold}
\subsection{Tau-pair production at a muon collider: matrix elements for $ \mu^- \mu^+ \to \tau^- \tau^+$}\label{app:xsec}
\mathversion{normal}
The matrix element for the $ \mu^- \mu^+ \to \tau^- \tau^+$ process, considering $s$-channel $Z$ and $\gamma$ exchange, as well as $t$-channel $Z^\prime$ contributions (and neglecting the muon mass) is given by 
\begin{equation}
    |\mathcal M|^2 \,= \,|\mathcal M_\gamma + \mathcal M_Z - \mathcal M_{Z^\prime}|^2\,, 
\end{equation}
as introduced in Eq.~(\ref{eq:Msum}), with 
\begin{eqnarray}
    \mathcal{M}_{Z^\prime} &=& \dfrac{1}{t - m_{Z^\prime}^2 + i\Gamma_{Z^\prime} m_{Z^\prime}} \left[v_\mu \, \gamma_\alpha\,\left(   g_L^{\mu\tau}\,  P_L+ g_R^{\mu\tau}\,  P_R \right)v_\tau\right] \left[ u_\tau \, \gamma^\alpha \,\left(  g_L^{\mu\tau, *} \, P_L + g_R^{\mu\tau, *} \, P_R \right) u_\mu \right] \nonumber\\
    &-& \dfrac{m_\tau^2}{m_{Z^\prime}^2 (t - m_{Z^\prime}^2 + i\Gamma_{Z^\prime} m_{Z^\prime})} \left[ v_\mu\left(   g_{R}^{\mu\tau}\,  P_L +  g_{L}^{\mu\tau}\,  P_R \right) v_\tau\right]  \left[ u_\tau \left(  g_L^{\mu\tau, *} \, P_L + g_R^{\mu\tau, *} \, P_R \right) u_\mu \right]\, ,\\
    \mathcal{M}_{Z} &=& \dfrac{g}{c_w \,(s - m_{Z}^2 + i\Gamma_{Z} m_{Z})} 
   \left[ v_\mu \, \gamma_\alpha\,  \left(g_V - g_A \,\gamma_5\right) u_\mu \right] \left[ u_\tau  \gamma^\alpha \, \left(g_V - g_A \,\gamma_5\right)  v_\tau \right] \, ,\\
    \mathcal{M}_{\gamma} &=& \dfrac{e^2}{s} 
    \left[ v_\mu \,\gamma_\alpha \,  u_\mu \right]  \left[u_\tau \, \gamma^\alpha \, v_\tau \right]  \, .
\end{eqnarray}
The matrix elements squared (and interference terms) are accordingly given by:
\begin{eqnarray}
    |\mathcal{M}_{Z^\prime}|^2 &=& \frac{4 |g_L^{\mu\tau}|^4 \left(m_\tau^2-u\right)^2+8 |g_L^{\mu\tau}|^2  |g_R^{\mu\tau}|^2  s \left(s-2 m_\tau^2\right)+4 |g_R^{\mu\tau}|^4 \left(m_\tau^2-u\right)^2}{4 \left(\Gamma_{Z^\prime}^2 m_{Z^\prime}^2+\left(m_{Z^\prime}^2-t\right)^2\right)} \nonumber\\
   &+& \frac{ (|g_L^{\mu\tau}|^2 +|g_R^{\mu\tau}|^2 )^2 \left(4 m_\tau^4 s + m_\tau^4 \left(m_\tau^2-t\right)^2 \right)}{4 m_{Z^\prime}^2 \left(\Gamma_{Z^\prime}^2 m_{Z^\prime}^2+\left(m_{Z^\prime}^2-t\right)^2\right)}\, ,
\\
    |\mathcal{M}_{Z}|^2 &=& \dfrac{2 g^4}{c_w^4 \left(\Gamma_{Z}^2 m_Z^2+\left(m_Z^2-s\right)^2\right)} \Big\{ g_A^4 \left(2 m_\tau^4-2 m_\tau^2 (s+t+u)+t^2+u^2\right) \nonumber\\
    &+&  2  g_A^2 g_V^2 \left(2 m_\tau^4+2 m_\tau^2 (t-3 u)-t^2+3 u^2\right) \nonumber\\
    &+&  g_V^4 \left(2 m_\tau^4+2 m_\tau^2 (s-t-u)+t^2+u^2\right) \Big\} ,
\\
    |\mathcal{M}_\gamma|^2 &=& \frac{2 e^4 \left(2 m_\tau^4+2 m_\tau^2 (s-t-u)+t^2+u^2\right)}{s^2}\, ,
\\
    2\mathrm{Re}\mathcal{M}_{Z^\prime}\mathcal{M}_\gamma^\dag &=& \frac{e^2 \left(m_{Z^\prime}^2-t\right) (|g_L^{\mu\tau}|^2+|g_R^{\mu\tau}|^2)}{m_{Z^\prime}^2 s \left(\Gamma_{Z^\prime}^2 m_{Z^\prime}^2+\left(m_{Z^\prime}^2-t\right)^2\right)} \nonumber \\
    & \times&
    \Big[m_\tau^6+m_\tau^4 \left(2 m_{Z^\prime}^2+s-2 t\right)+m_\tau^2 \left(2 m_{Z^\prime}^2 (s-2 u)+t^2\right)+2 m_{Z^\prime}^2 u^2\Big]\, ,
\\
   2\mathrm{Re}\mathcal{M}_{Z} \mathcal{M}_\gamma^\dag  &=& \frac{4 e^2 g^2 \left(s-m_Z^2\right) }{c_w^2 s \left(\Gamma_{Z}^2 m_Z^2+\left(m_Z^2-s\right)^2\right)}
   \nonumber \\
    & \times& \Big[g_A^2 (t-u) \left(2 m_\tau^2-t-u\right)+g_V^2 \left(2 m_\tau^4+2 m_\tau^2 (s-t-u)+t^2+u^2\right)\Big]\, ,
\\
  2\mathrm{Re}\mathcal{M}_{Z^\prime} \mathcal{M}_{Z}^\dag &=& -\frac{g^2 \left(\Gamma_{Z} \Gamma_{Z^\prime} m_Z m_{Z^\prime}+\left(m_Z^2-s\right) \left(m_{Z^\prime}^2-t\right)\right)}{c_w^2 m_{Z^\prime}^2 \left(\Gamma_{Z}^2 m_Z^2+\left(m_Z^2-s\right)^2\right) \left(\Gamma_{Z^\prime}^2 m_{Z^\prime}^2+\left(m_{Z^\prime}^2-t\right)^2\right)} \nonumber \\
  &\times&
\Big\{g_V^2 (|g_L^{\mu\tau}|^2+|g_R^{\mu\tau}|^2) \left[m_\tau^6+m_\tau^4 \left(2 m_{Z^\prime}^2+s-2 t\right)+m_\tau^2 \left(2 m_{Z^\prime}^2 (s-2 u)+t^2\right)+2 m_{Z^\prime}^2 u^2\right]
  \nonumber \\
&-& g_A^2 (|g_L^{\mu\tau}|^2+|g_R^{\mu\tau}|^2) \left[m_\tau^6-m_\tau^4 \left(2 m_{Z^\prime}^2+s+2 t\right)+m_\tau^2 \left(2 m_{Z^\prime}^2 (s+2 u)+t^2\right)-2 m_{Z^\prime}^2 u^2\right] \nonumber \\
    &+&2 g_A g_V (|g_L^{\mu\tau}|^2-|g_R^{\mu\tau}|^2) \left[m_\tau^4 \left(2 m_{Z^\prime}^2+s\right)-4 m_\tau^2 m_{Z^\prime}^2 u+2 m_{Z^\prime}^2 u^2\right]
 \Big\}\, ,
\end{eqnarray}
where $e$ is the electric charge,  $\frac{g}{c_w}g_{A(V)}$ are the axial (vector) couplings of the $Z$ boson to charged leptons, $\Gamma_V$ denotes the width of the vector boson, with $V=Z, Z^\prime$ (see Appendix~\ref{app:Zpwidth} for details). 
The Mandelstam variables, $s, t, u$, can be expressed in function of the energy $\sqrt{s}$ and the angle $\theta$ (the final state lepton angle with respect to the colliding muon direction in the di-tau centre  of mass frame) as:
\begin{eqnarray}
  t \, = \, -\frac{s}{2}\left[ 1 - \sqrt{1 - 4\frac{m_\tau^2}{s}}\cos \theta - 2\frac{m_\tau^2}{s}\right]\, , \quad\quad u \, = \, -\frac{s}{2}\left[ 1 + \sqrt{1 - 4\frac{m_\tau^2}{s}}\cos \theta - 2\frac{m_\tau^2}{s}\right]\,.
\end{eqnarray}

\mathversion{bold}
\subsection{Width of the $Z^\prime$}\label{app:Zpwidth}
\mathversion{normal}
The $Z^\prime$ boson width can be estimated from its tree-level decays to charged leptons and neutrinos.
The decay rate of a vector boson $V$ to two fermions $F_{1,2}$ is given by (see, for example~\cite{Abada:2013aba}):
\begin{eqnarray}
  \Gamma_{VFF} &=& \dfrac{\lambda^{1/2}(m_V, m_{F_1}, m_{F_2})}{48\pi \,m_V^3}\nonumber\\
            &\times& \left[(|b_L|^2 + |b_R|^2)\left(2 m_V^2 - m_{F_1}^2 - m_{F_2}^2 - \dfrac{(m_{F_1}^2 - m_{F_2}^2)^2}{m_V^2}\right) + 12\, m_{F_1}\,m_{F_2}\,\mathrm{Re}(b_L\, b_R^\ast)\right]\,,
\end{eqnarray}
in which $\lambda(a,b,c)$ is the K\"all\'en-function already defined in Eq.~(\ref{eq:kallen}).
For $Z^\prime\to \ell_\alpha\ell_\beta$ decays, the couplings $b_{L,R}$ are given by $b_{L,R} = g_{L,R}^{\alpha\beta}$, whereas for $Z^\prime\to \nu_\alpha\nu_\beta$ decays one has $b_{L} = g_{L}^{\alpha\beta}$ (with $b_R = 0$).
The total width $\Gamma_{Z^\prime}$ is then determined by the sum over all possible leptonic final states, neutral and charged, flavour conserving and/or flavour violating.


\begin{thebibliography}{99}
{\small
\bibitem{Muong-2:2021ojo}
B.~Abi \textit{et al.} [Muon g-2],
Phys. Rev. Lett. \textbf{126} (2021) no.14, 141801
[arXiv:2104.03281 [hep-ex]].

\bibitem{Muong-2:2006rrc}
G.~W.~Bennett \textit{et al.} [Muon g-2],
Phys. Rev. D \textbf{73} (2006), 072003
[arXiv:hep-ex/0602035 [hep-ex]].

\bibitem{Borsanyi:2020mff}
S.~Borsanyi, Z.~Fodor, J.~N.~Guenther, C.~Hoelbling, S.~D.~Katz, L.~Lellouch, T.~Lippert, K.~Miura, L.~Parato and K.~K.~Szabo, \textit{et al.}
Nature \textbf{593} (2021) no.7857, 51-55
[arXiv:2002.12347 [hep-lat]].

\bibitem{Aoyama:2020ynm}
T.~Aoyama, N.~Asmussen, M.~Benayoun, J.~Bijnens, T.~Blum, M.~Bruno, I.~Caprini, C.~M.~Carloni Calame, M.~C\`e and G.~Colangelo, \textit{et al.}
Phys. Rept. \textbf{887} (2020), 1-166
[arXiv:2006.04822 [hep-ph]].

\bibitem{Aoyama:2020ynm}
T.~Aoyama, N.~Asmussen, M.~Benayoun, J.~Bijnens, T.~Blum, M.~Bruno, I.~Caprini, C.~M.~Carloni Calame, M.~C\`e and G.~Colangelo, \textit{et al.}
Phys. Rept. \textbf{887} (2020), 1-166
[arXiv:2006.04822 [hep-ph]].

\bibitem{Aoyama:2012wk}
T.~Aoyama, M.~Hayakawa, T.~Kinoshita and M.~Nio,
Phys. Rev. Lett. \textbf{109} (2012), 111808
[arXiv:1205.5370 [hep-ph]].

\bibitem{Aoyama:2019ryr}
T.~Aoyama, T.~Kinoshita and M.~Nio,
Atoms \textbf{7} (2019) no.1, 28

\bibitem{Czarnecki:2002nt}
A.~Czarnecki, W.~J.~Marciano and A.~Vainshtein,
Phys. Rev. D \textbf{67} (2003), 073006
[erratum: Phys. Rev. D \textbf{73} (2006), 119901]
[arXiv:hep-ph/0212229 [hep-ph]].

\bibitem{Gnendiger:2013pva}
C.~Gnendiger, D.~St\"ockinger and H.~St\"ockinger-Kim,
Phys. Rev. D \textbf{88} (2013), 053005
[arXiv:1306.5546 [hep-ph]].

\bibitem{Davier:2017zfy}
M.~Davier, A.~Hoecker, B.~Malaescu and Z.~Zhang,
Eur. Phys. J. C \textbf{77} (2017) no.12, 827
[arXiv:1706.09436 [hep-ph]].

\bibitem{Keshavarzi:2018mgv}
A.~Keshavarzi, D.~Nomura and T.~Teubner,
Phys. Rev. D \textbf{97} (2018) no.11, 114025
[arXiv:1802.02995 [hep-ph]].

\bibitem{Colangelo:2018mtw}
G.~Colangelo, M.~Hoferichter and P.~Stoffer,
JHEP \textbf{02} (2019), 006
[arXiv:1810.00007 [hep-ph]].

\bibitem{Hoferichter:2019mqg}
M.~Hoferichter, B.~L.~Hoid and B.~Kubis,
JHEP \textbf{08} (2019), 137
[arXiv:1907.01556 [hep-ph]].

\bibitem{Davier:2019can}
M.~Davier, A.~Hoecker, B.~Malaescu and Z.~Zhang,
Eur. Phys. J. C \textbf{80} (2020) no.3, 241
[erratum: Eur. Phys. J. C \textbf{80} (2020) no.5, 410]
[arXiv:1908.00921 [hep-ph]].

\bibitem{Keshavarzi:2019abf}
A.~Keshavarzi, D.~Nomura and T.~Teubner,
Phys. Rev. D \textbf{101} (2020) no.1, 014029
[arXiv:1911.00367 [hep-ph]].

\bibitem{Kurz:2014wya}
A.~Kurz, T.~Liu, P.~Marquard and M.~Steinhauser,
Phys. Lett. B \textbf{734} (2014), 144-147
[arXiv:1403.6400 [hep-ph]].

\bibitem{Melnikov:2003xd}
K.~Melnikov and A.~Vainshtein,
Phys. Rev. D \textbf{70} (2004), 113006
[arXiv:hep-ph/0312226 [hep-ph]].

\bibitem{Masjuan:2017tvw}
P.~Masjuan and P.~Sanchez-Puertas,
Phys. Rev. D \textbf{95} (2017) no.5, 054026
[arXiv:1701.05829 [hep-ph]].

\bibitem{Colangelo:2017fiz}
G.~Colangelo, M.~Hoferichter, M.~Procura and P.~Stoffer,
JHEP \textbf{04} (2017), 161
[arXiv:1702.07347 [hep-ph]].

\bibitem{Hoferichter:2018kwz}
M.~Hoferichter, B.~L.~Hoid, B.~Kubis, S.~Leupold and S.~P.~Schneider,
JHEP \textbf{10} (2018), 141
[arXiv:1808.04823 [hep-ph]].

\bibitem{Gerardin:2019vio}
A.~G\'erardin, H.~B.~Meyer and A.~Nyffeler,
Phys. Rev. D \textbf{100} (2019) no.3, 034520
[arXiv:1903.09471 [hep-lat]].

\bibitem{Bijnens:2019ghy}
J.~Bijnens, N.~Hermansson-Truedsson and A.~Rodr\'\i{}guez-S\'anchez,
Phys. Lett. B \textbf{798} (2019), 134994
[arXiv:1908.03331 [hep-ph]].

\bibitem{Colangelo:2019uex}
G.~Colangelo, F.~Hagelstein, M.~Hoferichter, L.~Laub and P.~Stoffer,
JHEP \textbf{03} (2020), 101
[arXiv:1910.13432 [hep-ph]].

\bibitem{Blum:2019ugy}
T.~Blum, N.~Christ, M.~Hayakawa, T.~Izubuchi, L.~Jin, C.~Jung and C.~Lehner,
Phys. Rev. Lett. \textbf{124} (2020) no.13, 132002
[arXiv:1911.08123 [hep-lat]].

\bibitem{Colangelo:2014qya}
G.~Colangelo, M.~Hoferichter, A.~Nyffeler, M.~Passera and P.~Stoffer,
Phys. Lett. B \textbf{735} (2014), 90-91
[arXiv:1403.7512 [hep-ph]].


\bibitem{Athron:2021iuf}
P.~Athron, C.~Bal\'azs, D.~H.~J.~Jacob, W.~Kotlarski, D.~St\"ockinger and H.~St\"ockinger-Kim,
JHEP \textbf{09} (2021), 080
[arXiv:2104.03691 [hep-ph]].

\bibitem{Parker:2018vye}
R.~H.~Parker, C.~Yu, W.~Zhong, B.~Estey and H.~M\"uller,
Science \textbf{360} (2018), 191
[arXiv:1812.04130 [physics.atom-ph]].

\bibitem{Yu:2019gdh}
C.~Yu, W.~Zhong, B.~Estey, J.~Kwan, R.~H.~Parker and H.~M\"uller,
Annalen Phys. \textbf{531} (2019) no.5, 1800346.

\bibitem{Morel:2020dww}
L.~Morel, Z.~Yao, P.~Clad\'e and S.~Guellati-Kh\'elifa,
Nature \textbf{588} (2020) no.7836, 61-65.

\bibitem{Giudice:2012ms}
G.~F.~Giudice, P.~Paradisi and M.~Passera,
JHEP \textbf{11} (2012), 113
[arXiv:1208.6583 [hep-ph]].

\bibitem{Davoudiasl:2018fbb}
H.~Davoudiasl and W.~J.~Marciano,
Phys. Rev. D \textbf{98} (2018) no.7, 075011
[arXiv:1806.10252 [hep-ph]].

\bibitem{Kahn:2016vjr}
Y.~Kahn, G.~Krnjaic, S.~Mishra-Sharma and T.~M.~P.~Tait,
JHEP \textbf{05} (2017), 002
[arXiv:1609.09072 [hep-ph]].

\bibitem{Crivellin:2018qmi}
A.~Crivellin, M.~Hoferichter and P.~Schmidt-Wellenburg,
Phys. Rev. D \textbf{98} (2018) no.11, 113002
[arXiv:1807.11484 [hep-ph]].

\bibitem{Liu:2018xkx}
J.~Liu, C.~E.~M.~Wagner and X.~P.~Wang,
JHEP \textbf{03} (2019), 008
[arXiv:1810.11028 [hep-ph]].

\bibitem{Dutta:2018fge}
B.~Dutta and Y.~Mimura,
Phys. Lett. B \textbf{790} (2019), 563-567
[arXiv:1811.10209 [hep-ph]].

\bibitem{Han:2018znu}
X.~F.~Han, T.~Li, L.~Wang and Y.~Zhang,
Phys. Rev. D \textbf{99} (2019) no.9, 095034
[arXiv:1812.02449 [hep-ph]].

\bibitem{Endo:2019bcj}
M.~Endo and W.~Yin,
JHEP \textbf{08} (2019), 122
[arXiv:1906.08768 [hep-ph]].

\bibitem{Kawamura:2019rth}
J.~Kawamura, S.~Raby and A.~Trautner,
Phys. Rev. D \textbf{100} (2019) no.5, 055030
[arXiv:1906.11297 [hep-ph]].

\bibitem{Abdullah:2019ofw}
M.~Abdullah, B.~Dutta, S.~Ghosh and T.~Li,
Phys. Rev. D \textbf{100} (2019) no.11, 115006
[arXiv:1907.08109 [hep-ph]].

\bibitem{Badziak:2019gaf}
M.~Badziak and K.~Sakurai,
JHEP \textbf{10} (2019), 024
[arXiv:1908.03607 [hep-ph]].

\bibitem{CarcamoHernandez:2019ydc}
A.~E.~C\'arcamo Hern\'andez, S.~F.~King, H.~Lee and S.~J.~Rowley,
Phys. Rev. D \textbf{101} (2020) no.11, 115016
[arXiv:1910.10734 [hep-ph]].

\bibitem{Hiller:2019mou}
G.~Hiller, C.~Hormigos-Feliu, D.~F.~Litim and T.~Steudtner,
Phys. Rev. D \textbf{102} (2020) no.7, 071901
[arXiv:1910.14062 [hep-ph]].

\bibitem{Cornella:2019uxs}
C.~Cornella, P.~Paradisi and O.~Sumensari,
JHEP \textbf{01} (2020), 158
[arXiv:1911.06279 [hep-ph]].

\bibitem{CarcamoHernandez:2020pxw}
A.~E.~C\'arcamo Hern\'andez, Y.~Hidalgo Vel\'asquez, S.~Kovalenko, H.~N.~Long, N.~A.~P\'erez-Julve and V.~V.~Vien,
Eur. Phys. J. C \textbf{81} (2021) no.2, 191
[arXiv:2002.07347 [hep-ph]].

\bibitem{Haba:2020gkr}
N.~Haba, Y.~Shimizu and T.~Yamada,
PTEP \textbf{2020} (2020) no.9, 093B05
[arXiv:2002.10230 [hep-ph]].

\bibitem{Bigaran:2020jil}
I.~Bigaran and R.~R.~Volkas,
Phys. Rev. D \textbf{102} (2020) no.7, 075037
[arXiv:2002.12544 [hep-ph]].

\bibitem{Jana:2020pxx}
S.~Jana, V.~P.~K. and S.~Saad,
Phys. Rev. D \textbf{101} (2020) no.11, 115037
[arXiv:2003.03386 [hep-ph]].

\bibitem{Calibbi:2020emz}
L.~Calibbi, M.~L.~L\'opez-Ib\'a\~nez, A.~Melis and O.~Vives,
JHEP \textbf{06} (2020), 087
[arXiv:2003.06633 [hep-ph]].

\bibitem{Chen:2020jvl}
C.~H.~Chen and T.~Nomura,
Nucl. Phys. B \textbf{964} (2021), 115314
[arXiv:2003.07638 [hep-ph]].

\bibitem{Yang:2020bmh}
J.~L.~Yang, T.~F.~Feng and H.~B.~Zhang,
J. Phys. G \textbf{47} (2020) no.5, 055004
[arXiv:2003.09781 [hep-ph]].

\bibitem{Dorsner:2020aaz}
I.~Dor\v{s}ner, S.~Fajfer and S.~Saad,
Phys. Rev. D \textbf{102} (2020) no.7, 075007
[arXiv:2006.11624 [hep-ph]].

\bibitem{Hati:2020fzp}
C.~Hati, J.~Kriewald, J.~Orloff and A.~M.~Teixeira,
JHEP \textbf{07} (2020), 235
[arXiv:2005.00028 [hep-ph]].

\bibitem{DelleRose:2020oaa}
L.~Delle Rose, S.~Khalil and S.~Moretti,
Phys. Lett. B \textbf{816} (2021), 136216
[arXiv:2012.06911 [hep-ph]].

\bibitem{Hernandez:2021tii}
A.~E.~C.~Hern\'andez, S.~F.~King and H.~Lee,
Phys. Rev. D \textbf{103} (2021) no.11, 115024
[arXiv:2101.05819 [hep-ph]].

\bibitem{Bodas:2021fsy}
A.~Bodas, R.~Coy and S.~J.~D.~King,
Eur. Phys. J. C \textbf{81} (2021) no.12, 1065
[arXiv:2102.07781 [hep-ph]].

\bibitem{Han:2021gfu}
X.~F.~Han, T.~Li, H.~X.~Wang, L.~Wang and Y.~Zhang,
Phys. Rev. D \textbf{104} (2021) no.11, 115001
[arXiv:2104.03227 [hep-ph]].

\bibitem{Hernandez:2021iss}
A.~E.~C.~Hern\'andez, S.~Kovalenko, M.~Maniatis and I.~Schmidt,
JHEP \textbf{10} (2021), 036
[arXiv:2104.07047 [hep-ph]].

\bibitem{Borah:2021khc}
D.~Borah, M.~Dutta, S.~Mahapatra and N.~Sahu,
Phys. Rev. D \textbf{105} (2022) no.1, 015029
[arXiv:2109.02699 [hep-ph]].

\bibitem{Hue:2021xzl}
L.~T.~Hue, A.~E.~C\'arcamo Hern\'andez, H.~N.~Long and T.~T.~Hong,
``Heavy singly charged Higgs bosons and inverse seesaw neutrinos as origins of large $(g-2)_{e,\mu}$ in two higgs doublet models,''
arXiv:2110.01356 [hep-ph].

\bibitem{Langacker:2008yv}
P.~Langacker,
Rev. Mod. Phys. \textbf{81} (2009), 1199-1228
[arXiv:0801.1345 [hep-ph]].

\bibitem{Altmannshofer:2016jzy}
W.~Altmannshofer, S.~Gori, S.~Profumo and F.~S.~Queiroz,
JHEP \textbf{12} (2016), 106
[arXiv:1609.04026 [hep-ph]].

\bibitem{Correia:2016xcs}
F.~C.~Correia and S.~Fajfer,
Phys. Rev. D \textbf{94} (2016) no.11, 115023
[arXiv:1609.00860 [hep-ph]].

\bibitem{Correia:2019pnn}
F.~C.~Correia and S.~Fajfer,
JHEP \textbf{10} (2019), 278
[arXiv:1905.03867 [hep-ph]].

\bibitem{Correia:2019woz}
F.~C.~Correia and S.~Fajfer,
JHEP \textbf{10} (2019), 279
[arXiv:1905.03872 [hep-ph]].

\bibitem{Altmannshofer:2014cfa}
W.~Altmannshofer, S.~Gori, M.~Pospelov and I.~Yavin,
Phys. Rev. D \textbf{89} (2014), 095033
[arXiv:1403.1269 [hep-ph]].

\bibitem{Crivellin:2015mga}
A.~Crivellin, G.~D'Ambrosio and J.~Heeck,
Phys. Rev. Lett. \textbf{114} (2015), 151801
[arXiv:1501.00993 [hep-ph]].

\bibitem{Crivellin:2015lwa}
A.~Crivellin, G.~D'Ambrosio and J.~Heeck,
Phys. Rev. D \textbf{91} (2015) no.7, 075006
[arXiv:1503.03477 [hep-ph]].

\bibitem{Sierra:2015fma}
D.~Aristizabal Sierra, F.~Staub and A.~Vicente,
Phys. Rev. D \textbf{92} (2015) no.1, 015001
[arXiv:1503.06077 [hep-ph]].

\bibitem{Crivellin:2015era}
A.~Crivellin, L.~Hofer, J.~Matias, U.~Nierste, S.~Pokorski and J.~Rosiek,
Phys. Rev. D \textbf{92} (2015) no.5, 054013
[arXiv:1504.07928 [hep-ph]].

\bibitem{Celis:2015ara}
A.~Celis, J.~Fuentes-Martin, M.~Jung and H.~Serodio,
Phys. Rev. D \textbf{92} (2015) no.1, 015007
[arXiv:1505.03079 [hep-ph]].

\bibitem{Bhatia:2017tgo}
D.~Bhatia, S.~Chakraborty and A.~Dighe,
JHEP \textbf{03} (2017), 117
[arXiv:1701.05825 [hep-ph]].

\bibitem{Kamenik:2017tnu}
J.~F.~Kamenik, Y.~Soreq and J.~Zupan,
Phys. Rev. D \textbf{97} (2018) no.3, 035002
[arXiv:1704.06005 [hep-ph]].

\bibitem{Chen:2017usq}
C.~H.~Chen and T.~Nomura,
Phys. Lett. B \textbf{777} (2018), 420-427
[arXiv:1707.03249 [hep-ph]].

\bibitem{Camargo-Molina:2018cwu}
J.~E.~Camargo-Molina, A.~Celis and D.~A.~Faroughy,
Phys. Lett. B \textbf{784} (2018), 284-293
[arXiv:1805.04917 [hep-ph]].

\bibitem{Darme:2018hqg}
L.~Darm\'e, K.~Kowalska, L.~Roszkowski and E.~M.~Sessolo,
JHEP \textbf{10} (2018), 052
[arXiv:1806.06036 [hep-ph]].

\bibitem{Baek:2018aru}
S.~Baek and C.~Yu,
JHEP \textbf{11} (2018), 054
[arXiv:1806.05967 [hep-ph]].

\bibitem{Biswas:2019twf}
A.~Biswas and A.~Shaw,
JHEP \textbf{05} (2019), 165
[arXiv:1903.08745 [hep-ph]].

\bibitem{Allanach:2019iiy}
B.~C.~Allanach and J.~Davighi,
Eur. Phys. J. C \textbf{79} (2019) no.11, 908
[arXiv:1905.10327 [hep-ph]].

\bibitem{Crivellin:2020oup}
A.~Crivellin, C.~A.~Manzari, M.~Alguero and J.~Matias,
Phys. Rev. Lett. \textbf{127} (2021) no.1, 011801
[arXiv:2010.14504 [hep-ph]].

\bibitem{Crivellin:2022obd}
A.~Crivellin, C.~A.~Manzari, W.~Altmannshofer, G.~Inguglia, P.~Feichtinger and J.~M.~Camalich,
``Towards excluding a light $Z^\prime$ explanation of $b\to s\ell^+\ell^-$,''
arXiv:2202.12900 [hep-ph].

\bibitem{Raggi:2015noa}
M.~Raggi [NA48/2],
Nuovo Cim. C \textbf{38} (2016) no.4, 132
[arXiv:1508.01307 [hep-ex]].

\bibitem{Anastasi:2015qla}
A.~Anastasi, D.~Babusci, G.~Bencivenni, M.~Berlowski, C.~Bloise, F.~Bossi, P.~Branchini, A.~Budano, L.~Caldeira Balkest\r{a}hl and B.~Cao, \textit{et al.}
Phys. Lett. B \textbf{750} (2015), 633-637
[arXiv:1509.00740 [hep-ex]].

\bibitem{Androic:2018kni}
D.~Androi\'c \textit{et al.} [Qweak],
Nature \textbf{557} (2018) no.7704, 207-211
[arXiv:1905.08283 [nucl-ex]].

\bibitem{Baek:2001kca}
S.~Baek, N.~G.~Deshpande, X.~G.~He and P.~Ko,
Phys. Rev. D \textbf{64} (2001), 055006
[arXiv:hep-ph/0104141 [hep-ph]].

\bibitem{Ma:2001md}
E.~Ma, D.~P.~Roy and S.~Roy,
Phys. Lett. B \textbf{525} (2002), 101-106
[arXiv:hep-ph/0110146 [hep-ph]].

\bibitem{Amaral:2021rzw}
D.~W.~P.~Amaral, D.~G.~Cerdeno, A.~Cheek and P.~Foldenauer,
Eur. Phys. J. C \textbf{81} (2021) no.10, 861
[arXiv:2104.03297 [hep-ph]].

\bibitem{Heeck:2011wj}
J.~Heeck and W.~Rodejohann,
Phys. Rev. D \textbf{84} (2011), 075007
[arXiv:1107.5238 [hep-ph]].

\bibitem{Foldenauer:2018zrz}
P.~Foldenauer,
Phys. Rev. D \textbf{99} (2019) no.3, 035007
[arXiv:1808.03647 [hep-ph]].

\bibitem{Gninenko:2018tlp}
S.~N.~Gninenko and N.~V.~Krasnikov,
Phys. Lett. B \textbf{783} (2018), 24-28
[arXiv:1801.10448 [hep-ph]].

\bibitem{Holst:2021lzm}
I.~Holst, D.~Hooper and G.~Krnjaic,
Phys. Rev. Lett. \textbf{128} (2022) no.14, 141802
[arXiv:2107.09067 [hep-ph]].

\bibitem{Huang:2021nkl}
G.~y.~Huang, F.~S.~Queiroz and W.~Rodejohann,
Phys. Rev. D \textbf{103} (2021) no.9, 095005
[arXiv:2101.04956 [hep-ph]].

\bibitem{Banerjee:2020zvi}
H.~Banerjee, B.~Dutta and S.~Roy,
JHEP \textbf{03} (2021), 211
[arXiv:2011.05083 [hep-ph]].

\bibitem{Kamada:2018zxi}
A.~Kamada, K.~Kaneta, K.~Yanagi and H.~B.~Yu,
JHEP \textbf{06} (2018), 117
[arXiv:1805.00651 [hep-ph]].

\bibitem{Araki:2017wyg}
T.~Araki, S.~Hoshino, T.~Ota, J.~Sato and T.~Shimomura,
Phys. Rev. D \textbf{95} (2017) no.5, 055006
[arXiv:1702.01497 [hep-ph]].

\bibitem{Araki:2014ona}
T.~Araki, F.~Kaneko, Y.~Konishi, T.~Ota, J.~Sato and T.~Shimomura,
Phys. Rev. D \textbf{91} (2015) no.3, 037301
[arXiv:1409.4180 [hep-ph]].

\bibitem{Patra:2016shz}
S.~Patra, S.~Rao, N.~Sahoo and N.~Sahu,
Nucl. Phys. B \textbf{917} (2017), 317-336
[arXiv:1607.04046 [hep-ph]].

\bibitem{Harigaya:2013twa}
K.~Harigaya, T.~Igari, M.~M.~Nojiri, M.~Takeuchi and K.~Tobe,
JHEP \textbf{03} (2014), 105
[arXiv:1311.0870 [hep-ph]].

\bibitem{Frank:2021nkq}
M.~Frank, Y.~Hi\c{c}y\i{}lmaz, S.~Mondal, \"O.~\"Ozdal and C.~S.~\"Un,
JHEP \textbf{10} (2021), 063
[arXiv:2107.04116 [hep-ph]].

\bibitem{Bhattacharyya:2018evo}
G.~Bhattacharyya, M.~Dutra, Y.~Mambrini and M.~Pierre,
Phys. Rev. D \textbf{98} (2018) no.3, 035038
[arXiv:1806.00016 [hep-ph]].

\bibitem{Arcadi:2017jqd}
G.~Arcadi, P.~Ghosh, Y.~Mambrini, M.~Pierre and F.~S.~Queiroz,
JCAP \textbf{11} (2017), 020
[arXiv:1706.04198 [hep-ph]].

\bibitem{Arcadi:2013qia}
G.~Arcadi, Y.~Mambrini, M.~H.~G.~Tytgat and B.~Zaldivar,
JHEP \textbf{03} (2014), 134
[arXiv:1401.0221 [hep-ph]].

\bibitem{Dudas:2009uq}
E.~Dudas, Y.~Mambrini, S.~Pokorski and A.~Romagnoni,
JHEP \textbf{08} (2009), 014
[arXiv:0904.1745 [hep-ph]].

\bibitem{Chun:2010ve}
E.~J.~Chun, J.~C.~Park and S.~Scopel,
JHEP \textbf{02} (2011), 100
[arXiv:1011.3300 [hep-ph]].

\bibitem{Frandsen:2011cg}
M.~T.~Frandsen, F.~Kahlhoefer, S.~Sarkar and K.~Schmidt-Hoberg,
JHEP \textbf{09} (2011), 128
[arXiv:1107.2118 [hep-ph]].

\bibitem{Alves:2013tqa}
A.~Alves, S.~Profumo and F.~S.~Queiroz,
JHEP \textbf{04} (2014), 063
[arXiv:1312.5281 [hep-ph]].

\bibitem{Foot:1994vd}
R.~Foot, X.~G.~He, H.~Lew and R.~R.~Volkas,
Phys. Rev. D \textbf{50} (1994), 4571-4580
[arXiv:hep-ph/9401250 [hep-ph]].

\bibitem{Heeck:2016xkh}
J.~Heeck,
Phys. Lett. B \textbf{758} (2016), 101-105
[arXiv:1602.03810 [hep-ph]].

\bibitem{Altmannshofer:2016brv}
W.~Altmannshofer, C.~Y.~Chen, P.~S.~Bhupal Dev and A.~Soni,
Phys. Lett. B \textbf{762} (2016), 389-398
[arXiv:1607.06832 [hep-ph]].

\bibitem{Buras:2021btx}
A.~J.~Buras, A.~Crivellin, F.~Kirk, C.~A.~Manzari and M.~Montull,
JHEP \textbf{06} (2021), 068
[arXiv:2104.07680 [hep-ph]].

\bibitem{Crivellin:2013hpa}
A.~Crivellin, S.~Najjari and J.~Rosiek,
JHEP \textbf{04} (2014), 167
[arXiv:1312.0634 [hep-ph]].

\bibitem{EuropeanStrategyforParticlePhysicsPreparatoryGroup:2019qin}
R.~K.~Ellis, B.~Heinemann, J.~de Blas, M.~Cepeda, C.~Grojean, F.~Maltoni, A.~Nisati, E.~Petit, R.~Rattazzi and W.~Verkerke, \textit{et al.}
``Physics Briefing Book: Input for the European Strategy for Particle Physics Update 2020,''
arXiv:1910.11775 [hep-ex].

\bibitem{Delahaye:2019omf}
J.~P.~Delahaye, M.~Diemoz, K.~Long, B.~Mansouli\'e, N.~Pastrone, L.~Rivkin, D.~Schulte, A.~Skrinsky and A.~Wulzer,
``Muon Colliders,''
arXiv:1901.06150 [physics.acc-ph].

\bibitem{Long:2020wfp}
K.~Long, D.~Lucchesi, M.~Palmer, N.~Pastrone, D.~Schulte and V.~Shiltsev,
Nature Phys. \textbf{17} (2021) no.3, 289-292
[arXiv:2007.15684 [physics.acc-ph]].

\bibitem{MuonCollider:2022xlm}
J.~De Blas \textit{et al.} [Muon Collider],
``The physics case of a 3 TeV muon collider stage,''
arXiv:2203.07261 [hep-ph].

\bibitem{NA64:2018lsq}
D.~Banerjee \textit{et al.} [NA64],
Phys. Rev. Lett. \textbf{120} (2018) no.23, 231802
[arXiv:1803.07748 [hep-ex]].

\bibitem{NA64:2019auh}
D.~Banerjee \textit{et al.} [NA64],
Phys. Rev. D \textbf{101} (2020) no.7, 071101
[arXiv:1912.11389 [hep-ex]].

\bibitem{BaBar:2014zli}
J.~P.~Lees \textit{et al.} [BaBar],
Phys. Rev. Lett. \textbf{113} (2014) no.20, 201801
[arXiv:1406.2980 [hep-ex]].

\bibitem{SLACE158:2005uay}
P.~L.~Anthony \textit{et al.} [SLAC E158],
Phys. Rev. Lett. \textbf{95} (2005), 081601
[arXiv:hep-ex/0504049 [hep-ex]].

\bibitem{Deniz:2009mu}
M.~Deniz \textit{et al.} [TEXONO],
Phys. Rev. D \textbf{81} (2010), 072001
[arXiv:0911.1597 [hep-ex]].

\bibitem{Vilain:1993kd}
P.~Vilain \textit{et al.} [CHARM-II],
Phys. Lett. B \textbf{302} (1993), 351-355.

\bibitem{CCFR:1991lpl}
S.~R.~Mishra \textit{et al.} [CCFR],
Phys. Rev. Lett. \textbf{66} (1991), 3117-3120.

\bibitem{Aoki:1982ed}
K.~I.~Aoki, Z.~Hioki, M.~Konuma, R.~Kawabe and T.~Muta,
Prog. Theor. Phys. Suppl. \textbf{73} (1982), 1-225.

\bibitem{Espriu:2002xv}
D.~Espriu, J.~Manzano and P.~Talavera,
Phys. Rev. D \textbf{66} (2002), 076002
[arXiv:hep-ph/0204085 [hep-ph]].

\bibitem{Denner:1991kt}
A.~Denner,
Fortsch. Phys. \textbf{41} (1993), 307-420
[arXiv:0709.1075 [hep-ph]].

\bibitem{Freitas:2014hra}
A.~Freitas,
JHEP \textbf{04} (2014), 070
[arXiv:1401.2447 [hep-ph]].

\bibitem{ParticleDataGroup:2020ssz}
P.~A.~Zyla \textit{et al.} [Particle Data Group],
PTEP \textbf{2020} (2020) no.8, 083C01.

\bibitem{Patel:2015tea}
H.~H.~Patel,
Comput. Phys. Commun. \textbf{197} (2015), 276-290
[arXiv:1503.01469 [hep-ph]].

\bibitem{ATLAS:2014vur}
G.~Aad \textit{et al.} [ATLAS],
Phys. Rev. D \textbf{90} (2014) no.7, 072010
[arXiv:1408.5774 [hep-ex]].

\bibitem{OPAL:1995grn}
R.~Akers \textit{et al.} [OPAL],
Z. Phys. C \textbf{67} (1995), 555-564.

\bibitem{TheMEG:2016wtm}
A.~M.~Baldini \textit{et al.} [MEG],
Eur. Phys. J. C \textbf{76} (2016) no.8, 434
[arXiv:1605.05081 [hep-ex]].

\bibitem{Baldini:2018nnn}
A.~M.~Baldini \textit{et al.} [MEG II],
Eur. Phys. J. C \textbf{78} (2018) no.5, 380
[arXiv:1801.04688 [physics.ins-det]].

\bibitem{Aubert:2009ag}
B.~Aubert \textit{et al.} [BaBar],
Phys. Rev. Lett. \textbf{104} (2010), 021802
[arXiv:0908.2381 [hep-ex]].

\bibitem{Kou:2018nap}
E.~Kou \textit{et al.} [Belle-II],
PTEP \textbf{2019} (2019) no.12, 123C01
[erratum: PTEP \textbf{2020} (2020) no.2, 029201]
[arXiv:1808.10567 [hep-ex]].

\bibitem{Bellgardt:1987du}
U.~Bellgardt \textit{et al.} [SINDRUM],
Nucl. Phys. B \textbf{299} (1988), 1-6.

\bibitem{Blondel:2013ia}
A.~Blondel, A.~Bravar, M.~Pohl, S.~Bachmann, N.~Berger, M.~Kiehn, A.~Schoning, D.~Wiedner, B.~Windelband and P.~Eckert, \textit{et al.}
``Research Proposal for an Experiment to Search for the Decay $\mu \to eee$,''
arXiv:1301.6113 [physics.ins-det].

\bibitem{Hayasaka:2010np}
K.~Hayasaka, K.~Inami, Y.~Miyazaki, K.~Arinstein, V.~Aulchenko, T.~Aushev, A.~M.~Bakich, A.~Bay, K.~Belous and V.~Bhardwaj, \textit{et al.}
Phys. Lett. B \textbf{687} (2010), 139-143
[arXiv:1001.3221 [hep-ex]].

\bibitem{Abada:2019lih}
A.~Abada \textit{et al.} [FCC],
Eur. Phys. J. C \textbf{79} (2019) no.6, 474.

\bibitem{Bertl:2006up}
W.~H.~Bertl \textit{et al.} [SINDRUM II],
Eur. Phys. J. C \textbf{47} (2006), 337-346.

\bibitem{Nguyen:2015vkk}
T.~M.~Nguyen [DeeMe],
PoS \textbf{FPCP2015} (2015), 060.

\bibitem{Krikler:2015msn}
B.~E.~Krikler [COMET],
``An Overview of the COMET Experiment and its Recent Progress,''
arXiv:1512.08564 [physics.ins-det].

\bibitem{KunoESPP19}
      Y.~Kuno,
 Presentation at the Flavour Session of the CERN Council Open Symposium on the Update of the European Strategy for Particle Physics. Granada, 13-16 May 2019.

\bibitem{Adamov:2018vin}
R.~Abramishvili \textit{et al.} [COMET],
PTEP \textbf{2020} (2020) no.3, 033C01
[arXiv:1812.09018 [physics.ins-det]].

\bibitem{Bartoszek:2014mya}
L.~Bartoszek \textit{et al.} [Mu2e],
``Mu2e Technical Design Report,''
arXiv:1501.05241 [physics.ins-det].

\bibitem{Aad:2014bca}
G.~Aad \textit{et al.} [ATLAS],
Phys. Rev. D \textbf{90} (2014) no.7, 072010
[arXiv:1408.5774 [hep-ex]].

\bibitem{Akers:1995gz}
R.~Akers \textit{et al.} [OPAL],
Z. Phys. C \textbf{67} (1995), 555-564.

\bibitem{Abada:2014kba}
A.~Abada, M.~E.~Krauss, W.~Porod, F.~Staub, A.~Vicente and C.~Weiland,
JHEP \textbf{11} (2014), 048
[arXiv:1408.0138 [hep-ph]].

\bibitem{Hahn:1998yk}
T.~Hahn and M.~Perez-Victoria,
Comput. Phys. Commun. \textbf{118} (1999), 153-165
[arXiv:hep-ph/9807565 [hep-ph]].

\bibitem{Passarino:1978jh}
G.~Passarino and M.~J.~G.~Veltman,
Nucl. Phys. B \textbf{160} (1979), 151-207.

\bibitem{Hisano:1995cp}
J.~Hisano, T.~Moroi, K.~Tobe and M.~Yamaguchi,
Phys. Rev. D \textbf{53} (1996), 2442-2459
[arXiv:hep-ph/9510309 [hep-ph]].

\bibitem{Kitano:2002mt}
R.~Kitano, M.~Koike and Y.~Okada,
Phys. Rev. D \textbf{66} (2002), 096002
[erratum: Phys. Rev. D \textbf{76} (2007), 059902]
[arXiv:hep-ph/0203110 [hep-ph]].

\bibitem{Fukuyama:2021iyw}
T.~Fukuyama, Y.~Mimura and Y.~Uesaka,
Phys. Rev. D \textbf{105} (2022) no.1, 015026
[arXiv:2108.10736 [hep-ph]].

\bibitem{Willmann:1998gd}
L.~Willmann, P.~V.~Schmidt, H.~P.~Wirtz, R.~Abela, V.~Baranov, J.~Bagaturia, W.~H.~Bertl, R.~Engfer, A.~Grossmann and V.~W.~Hughes, \textit{et al.}
Phys. Rev. Lett. \textbf{82} (1999), 49-52
[arXiv:hep-ex/9807011 [hep-ex]].

\bibitem{ARGUS:1991zhv}
H.~Albrecht \textit{et al.} [ARGUS],
Z. Phys. C \textbf{53} (1992), 367-374.

\bibitem{CLEO:1996oro}
A.~Anastassov \textit{et al.} [CLEO],
Phys. Rev. D \textbf{55} (1997), 2559-2576
[erratum: Phys. Rev. D \textbf{58} (1998), 119904].

\bibitem{BaBar:2009lyd}
B.~Aubert \textit{et al.} [BaBar],
Phys. Rev. Lett. \textbf{105} (2010), 051602
[arXiv:0912.0242 [hep-ex]].

\bibitem{Pich:2009zza}
A.~Pich, I.~Boyko, D.~Dedovich and I.~I.~Bigi,
Int. J. Mod. Phys. A \textbf{24S1} (2009), 715-737.

\bibitem{HFLAV:2019otj}
Y.~S.~Amhis \textit{et al.} [HFLAV],
Eur. Phys. J. C \textbf{81} (2021) no.3, 226
[arXiv:1909.12524 [hep-ex]].

\bibitem{Sirlin:1980nh}
A.~Sirlin,
Phys. Rev. D \textbf{22} (1980), 971-981.

\bibitem{Lindner:2016bgg}
M.~Lindner, M.~Platscher and F.~S.~Queiroz,
Phys. Rept. \textbf{731} (2018), 1-82
[arXiv:1610.06587 [hep-ph]].

\bibitem{Jegerlehner:2009ry}
F.~Jegerlehner and A.~Nyffeler,
Phys. Rept. \textbf{477} (2009), 1-110
[arXiv:0902.3360 [hep-ph]].

\bibitem{Leveille:1977rc}
J.~P.~Leveille,
Nucl. Phys. B \textbf{137} (1978), 63-76.

\bibitem{Iguro:2020rby}
S.~Iguro, Y.~Omura and M.~Takeuchi,
JHEP \textbf{09} (2020), 144
[arXiv:2002.12728 [hep-ph]].

\bibitem{Abada:2021zcm}
A.~Abada, J.~Kriewald and A.~M.~Teixeira,
Eur. Phys. J. C \textbf{81} (2021) no.11, 1016
[arXiv:2107.06313 [hep-ph]].

\bibitem{ALEPH:2006jhv}
S.~Schael \textit{et al.} [ALEPH],
Eur. Phys. J. C \textbf{49} (2007), 411-437
[arXiv:hep-ex/0609051 [hep-ex]].

\bibitem{ATLAS:2015dva}
G.~Aad \textit{et al.} [ATLAS],
Phys. Rev. Lett. \textbf{115} (2015) no.3, 031801
[arXiv:1503.04430 [hep-ex]].

\bibitem{ATLAS:2019fgd}
G.~Aad \textit{et al.} [ATLAS],
JHEP \textbf{03} (2020), 145
[arXiv:1910.08447 [hep-ex]].

\bibitem{CMS:2019gwf}
A.~M.~Sirunyan \textit{et al.} [CMS],
JHEP \textbf{05} (2020), 033
[arXiv:1911.03947 [hep-ex]].

\bibitem{CMS:2021ctt}
A.~M.~Sirunyan \textit{et al.} [CMS],
JHEP \textbf{07} (2021), 208
[arXiv:2103.02708 [hep-ex]].

\bibitem{ATLAS:2019erb}
G.~Aad \textit{et al.} [ATLAS],
Phys. Lett. B \textbf{796} (2019), 68-87
[arXiv:1903.06248 [hep-ex]].

\bibitem{Iguro:2020qbc}
S.~Iguro, K.~A.~Mohan and C.~P.~Yuan,
Phys. Rev. D \textbf{101} (2020) no.7, 075011
[arXiv:2001.09079 [hep-ph]].

\bibitem{Aime:2022flm}
C.~Aim\`e, A.~Apyan, M.~Attia Mahmoud, N.~Bartosik, F.~Batsch, A.~Bertolin, M.~Bonesini, D.~Buttazzo, M.~Casarsa and M.~G.~Catanesi, \textit{et al.}
``Muon Collider Physics Summary,''
arXiv:2203.07256 [hep-ph].

\bibitem{InternationalMuonCollider:2022qki}
D.~Stratakis \textit{et al.} [International Muon Collider],
``A Muon Collider Facility for Physics Discovery,''
arXiv:2203.08033 [physics.acc-ph].

\bibitem{Blumlein:2022mrp}
J.~Bl\"umlein and K.~Sch\"onwald,
Mod. Phys. Lett. A \textbf{37} (2022) no.07, 2230004
[arXiv:2202.08476 [hep-ph]].

\bibitem{Accomando:2015cfa}
E.~Accomando, A.~Belyaev, J.~Fiaschi, K.~Mimasu, S.~Moretti and C.~Shepherd-Themistocleous,
JHEP \textbf{01} (2016), 127
[arXiv:1503.02672 [hep-ph]].

\bibitem{Endo:2020mev}
M.~Endo, S.~Iguro and T.~Kitahara,
JHEP \textbf{06} (2020), 040
[arXiv:2002.05948 [hep-ph]].

\bibitem{ALEPH:2005ab}
S.~Schael \textit{et al.} [ALEPH, DELPHI, L3, OPAL, SLD, LEP Electroweak Working Group, SLD Electroweak Group and SLD Heavy Flavour Group],
Phys. Rept. \textbf{427} (2006), 257-454
[arXiv:hep-ex/0509008 [hep-ex]].

\bibitem{Tewsley-Booth:2022umi}
A.~Tewsley-Booth [Muon g-2],
``The First Measurement of the Muon Anomalous Magnetic Moment from the Fermilab Muon $g-2$ Collaboration,''
arXiv:2202.11172 [hep-ex].

\bibitem{Abada:2013aba}
A.~Abada, A.~M.~Teixeira, A.~Vicente and C.~Weiland,
JHEP \textbf{02} (2014), 091
[arXiv:1311.2830 [hep-ph]].

}
\end{thebibliography}
\end{document}